\documentclass[12pt]{iopart}
\usepackage{graphicx}
\usepackage{epsfig}
\usepackage{ulem}
\usepackage{amssymb}
\usepackage{color}

\begin{document}

\title[Wire and extended ladder model predict THz oscillations...]{Wire and extended ladder model predict THz oscillations in DNA monomers, dimers and trimers}

\author{K. Lambropoulos, K. Kaklamanis, A. Morphis, M. Tassi, R. Lopp$^{a}$, G. Georgiadis, M. Theodorakou, M. Chatzieleftheriou$^{b}$, and C. Simserides}

\address{National and Kapodistrian University of Athens, Department of Physics, Department of Solid State Physics,
Panepistimiopolis, 15784 Zografos, Athens, Greece}

\ead{csimseri@phys.uoa.gr}

\vspace{2pc}

\address{$^{a}$ Current Affiliation: Georg-August-Universit\"{a}t G\"{o}ttingen, Fakult\"{a}t f\"{u}r Physik, Friedrich-Hund-Platz 1
D-37077 G\"{o}ttingen, Germany}

\address{$^{b}$ Current Affiliation: University of Copenhagen, Niels Bohr Institute, Blegdamsvej 17, DK-2100 Copenhagen, Denmark}

\vspace{10pt}

\begin{indented}
\item[] \today
\end{indented}

\begin{abstract}
We call \textit{monomer} a B-DNA base pair and study, analytically and numerically, electron or hole oscillations in \textit{monomers}, \textit{dimers} and \textit{trimers}.
We employ two Tight Binding (TB) approaches:
(I) at the base-pair level, using the on-site energies of the base pairs and the hopping parameters between successive base pairs i.e. \textit{a wire model}, and
(II) at the single-base level, using the on-site energies of the bases and the hopping parameters between neighbouring bases, specifically between
(a) two successive bases in the same strand,
(b) complementary bases that define a base pair, and
(c) diagonally located bases of successive base pairs, i.e. \textit{an extended ladder model} since it also includes the diagonal hoppings (c).
For \textit{monomers}, with TB II, we predict periodic carrier oscillations with frequency $f \approx$ 50-550 THz.
For \textit{dimers},   with TB I,  we predict periodic carrier oscillations with           $f \approx$ 0.25-100 THz.
For \textit{trimers made of identical monomers}, with TB I, we predict periodic carrier oscillations with $f \approx$ 0.5-33 THz.
In other cases, either with TB I or TB II, the oscillations may be not strictly periodic, but Fourier analysis shows similar frequency content.
For dimers and trimers, TB I and TB II are successfully compared giving complementary aspects of the oscillations.
\end{abstract}





\vspace{2pc}
%
\submitto{\JPCM}
%
%
%

\section{Introduction}   
\label{sec:introduction} 
Charge transfer (CT) in biological molecules attracts recently considerable interest among the physical, chemical, biological and medical communities. It also attracts a broad spectrum of interdisciplinary scientists and engineers. This is because CT constitutes the basis of many biological processes e.g. in various proteins~\cite{Page:2003} including metalloproteins~\cite{GrayWinkler:2010} and enzymes~\cite{Moser:2010} with medical and bioengineering applications \cite{Artes:2014,Kannan:2009}.
CT plays a central role in DNA damage and repair \cite{Dandliker:1997,Rajski:2000,Giese:2006} and it, also, might be an indicator to discriminate between pathogenic and non-pathogenic mutations at an early stage \cite{Shih:2011}.
At least for twenty years, there have been many experimental attempts to recognize the electronic properties of DNA cf. e.g.~\cite{Storm:2001,Porath:2000,Cohen:2005,Yoo:2001,FinkSchoenenberger:1999,Xu:2004}.
Today we know that many factors related to environment (the aqueous solution, the concentration of counterions), extraction process, conducts, purity, base-pair sequence, geometry etc influence CT in DNA. These factors can be categorized into intrinsic and extrinsic. In this work we focus on maybe the most important of the intrinsic factors, i.e. the effect of alternating the base (or base-bair) sequence, which affects the overlaps across the $\pi$-stack.  Additionally, we have to discriminate between the words transport (usually implying the use of electrodes), transfer, migration (a transfer over rather long distances). The carriers (electrons or holes) can be inserted via electrodes, generated by UV irradiation or by reduction and oxidation. The nature of charge transport, transfer, migration along the DNA double helix is important for many scientific fields like physics, chemistry, biology, medicine and engineering.
Although unbiased charge transfer in DNA nearly vanishes after 10 to 20 nm~\cite{Simserides:2014, LChMKTS:2015}, DNA still remains a promising candidate as an electronic component in molecular electronics, e.g. as a short molecular wire~\cite{Wohlgamuth:2013}. Favouring geometries and base sequences have still to be explored e.g. incorporation of sequences serving as molecular rectifiers, using non-natural bases or using the triplet acceptor anthraquinone for hole injection~\cite{LewisWasielewski:2013}. Structural fluctuations could be another important factor which influences quantum transport
through DNA molecular wires~\cite{Gutierrez:2010}. The contact of DNA segments with experimentally involved surfaces and interfaces is another research direction e.g. for bio-sensoric applications. For example, charge transfer on the contact of DNA with gold nanoparticles~\cite{Abouzar:2012} and  polyelectrolyte multilayers \cite{Poghossian:2013} has been recently investigated.

During the last decade, the scientific literature has been enriched with works studying carrier oscillations within ``molecular'' systems.
Real-Time Time-Dependent Density Functional Theory (RT-TDDFT)~\cite{RT-TDDFT} simulations predicted oscillations ($\approx$ 0.1-10 PHz)  within p-nitroaniline and FTC chromophore~\cite{Takimoto:2007}, as well as within zinc porphyrin, green fluorescent protein chromophores and the adenine-thymine base pair~\cite{LopataGovind:2011}.
It was shown that in a simplified single-stranded helix of 101 bases, a collinear uniform electric field induces THz Bloch oscillations~\cite{Malyshev:2009}.
Single and multiple charge transfer within a typical DNA dimer in connection to a bosonic bath has been studied, too~\cite{Tornow:2010}.
Each base pair was approximated by a single site, as in our Tight Binding (TB) approach at the base-pair level denoted in the present article as TB I [cf. Section~\ref{sec:general} and Appendix A]. In the subspace of single charge transfer between base pairs, having initially placed the charge at the donor site and having used a ``typical hopping matrix element'' of 0.2 eV, the authors obtained a period slightly greater than 10 fs. Let us call $T$ the period and $f$ the frequency.
Applying our equation~\cite{Simserides:2014,LKGS:2014} $f= \frac{1}{T} = \frac{\sqrt{(2t)^2 + \Delta^2}}{h}$, with $ t = $ 0.2 eV for the ``typical hopping matrix element'' and difference of the on-site energies $\Delta = 0$ for identical dimers i.e. as in Ref.~\cite{Tornow:2010},
we obtain $T \approx$ 10.34 fs, which agrees splendidly with the dotted line in Fig.~4 of Ref.~\cite{Tornow:2010}.

Recently, we studied B-DNA dimers, trimers and polymers with TB I~\cite{Simserides:2014,LChMKTS:2015,LKGS:2014}.
This approach allowed us to readily determine the spatiotemporal evolution of holes or electrons along a $N$ base-pair DNA segment. With TB I, we have already shown~\cite{Simserides:2014,LKGS:2014} that for all dimers and for trimers made of identical monomers the carrier movement is periodic with frequencies in the mid- and far-infrared i.e. approximately in the THz domain~\cite{ISO20473}. This part of the electromagnetic (EM) spectrum is significant for biological sciences because it can be used to extract complementary to traditional spectroscopic measurements information e.g. on low-frequency bond vibrations, hydrogen bond stretching, bond torsions in liquids and gases etc and because it is relatively non-invasive compared to higher-frequency regions of the EM spectrum~\cite{Yin:2012}. Within TB I, we also showed that, generally, increasing the number of monomers above three, periodicity is lost~\cite{Simserides:2014,LKGS:2014}. Even for the simplest tetramer, the carrier movement is not periodic~\cite{Lambropoulos:2014}.
For periodic cases, we defined~\cite{Simserides:2014,LKGS:2014} the maximum transfer percentage $p$, e.g. the maximum probability to find the carrier at the last monomer having placed it initially at the first monomer and the pure maximum transfer rate $\frac{p}{T}=pf$.
For all cases, either periodic or not, the pure mean transfer rate $k$ (cf.~Eq.~\ref{meantransferrateN(I)} in Appendix A) and the speed $u = kd$, where $d = (N-1) \times $ 3.4 {\AA} is the charge transfer distance, can be used to characterize the system. Within TB I, our analytical calculations and numerical results showed that for dimers $k=2\frac{p}{T}$ and for trimers made of identical monomers $k \approx 1.3108 \frac{p}{T}$.
Using $k$ to evaluate the easiness of charge transfer, we calculated the inverse decay length $\beta$ for exponential fits $k(d)$ and the exponent $\eta$ for power law fits $k(N)$. Studying B-DNA polymers and segments taken from experiments~\cite{Simserides:2014,LChMKTS:2015}, we determined the ranges of values of $\beta$ and $\eta$. Our TB I was used~\cite{Simserides:2014} to reproduce theoretical and experimental findings for various B-DNA segments~\cite{WangLewisSankey:2004,Giese:2001,Murphy:1993,Arkin:1996,Giese:1999}. For monomer-polymers and dimer-polymers, we studied HOMO and LUMO eigenspectra, mean over time probabilities to find the carrier at a particular monomer and mean transfer rates, we illustrated how increasing the number of different parameters involved in TB I, the fall of $k(d)$ or $k(N)$ becomes steeper and we circumscribed the range covered by $\beta$ and $\eta$ ~\cite{LChMKTS:2015}. Finally, both for the time-independent and the time-dependent problem, we analyzed the {\it palindromicity} and the {\it degree of eigenspectrum dependence} of the probabilities to find the carrier at a particular monomer~\cite{LChMKTS:2015}.

As in Refs.~\cite{Simserides:2014,LChMKTS:2015,LKGS:2014,Lambropoulos:2014}, here we call \textit{monomer} a B-DNA base pair and study carrier oscillations in \textit{dimers} and \textit{trimers}. Moreover, for the first time we study \textit{monomers}.
However, here we employ two Tight Binding (TB) approaches:
(I) at the base-pair level, using the on-site energies of the base pairs and the hopping parameters between successive base pairs which is \textit{a wire model}~\cite{CMRR:2007:Chapter} (this was the approach employed in our previous works~\cite{Simserides:2014,LChMKTS:2015,LKGS:2014,Lambropoulos:2014})
and
(II) at the single-base level, using the on-site energies of the bases and the hopping parameters between neighbouring bases, specifically between
(a) two successive bases in the same strand,
(b) complementary bases that define a base pair, and
(c) diagonally located bases of successive base pairs, which is \textit{an extended ladder model} since it also includes the diagonal hoppings (c). Hence, it is more elaborate than the usual ladder model~\cite{CMRR:2007:Chapter}. Extrinsic effects, such as the aqueousness and the presence of counterions can also be taken into account in TB models which then can be renormalized via a decimation procedure to obtain either wire or ladder models, like the ones presented here \cite{Macia:2009}.
Inclusion of the diagonal hoppings is essential e.g. for dimers made of identical monomers with crosswise purines cf. Section 4.
Additionally, TB II allows us to study charge oscillations within monomers, which is not possible for TB I since there a site is a base pair.
A few preliminary results with TB II have been included in Ref.~\cite{PIERS:2015} and in Ref.~\cite{Kaklamanis:2015}.
We assume that isolation of a few consecutive B-DNA base pairs is possible,
e.g. by connecting at the boundaries moieties with very small transfer integrals with the segment of interest.
The TB parameters that we use can be found in Refs.~\cite{Simserides:2014,HKS:2010-2011,MehrezAnantram:2005}.
We solve analytically and numerically, with the eigenvalue method, a system of (I) $N$ or (II) $2N$ coupled differential equations to determine the spatiotemporal evolution of an extra carrier (electron or hole) along a $N$ base-pair DNA segment.
Carriers move either between the HOMOs or between the LUMOs of the relevant sites [(I) base pairs, (II) bases].

A legitimate critique to our theoretical predictions could be relative to a possible comparison with the experiment.
As far as we know, such experiments in so short DNA segments (monomers, dimers, trimers) do not exist.
However, nowadays a variety of experimental techniques can be used to probe CT in biological molecules including optical, electrochemical and scanning probe techniques such as Electrochemical STM and Conductive AFM \cite{Artes:2014}.
Another method is the femtosecond transient absorption spectroscopy, with which one can measure the transfer rates of photoinduced carriers (holes or electrons) in small molecular systems \cite{Berera:2009,Gorczak:2015}. This method has been successfully employed in order to study single and double-stranded DNA oligonucleotides, slightly larger than the systems studied here \cite{CrespoHernandez:2005}.
In our case, we can imagine the isolation of an oligomer by putting at its end moieties with very small hopping integral with the ends of the oligomer under investigation and subsequent use of one of these techniques. The techniques must be in the position to probe oscillations with frequency content in the 0.1 to 1000 THz regime i.e. time scales of 10 ps to 1 fs.
Additionally, our method TB I has already been used to successfully reproduce experimental results with longer DNA segments relative to transfer rates and mean occupation probabilities~\cite{Simserides:2014}.

In this work we show that for monomers TB II predicts periodic carrier oscillations with frequency $f \approx$ 50-550 THz (but with very small transfer percentages), while TB I predicts periodic carrier oscillations with $f \approx$ 0.25-100 THz for dimers and with $f \approx$ 0.5-33 THz for trimers made of identical monomers. In other cases, either with TB I or TB II, oscillations are not strictly periodic, but Fourier analysis shows similar frequency content.
For dimers and trimers TB I and TB II give complementary aspects of the oscillations.
For dimers made of identical monomers we have large carrier transfer and the occupation probability is equally shared between the two monomers which constitute the dimer. For dimers made of identical monomers, if purines are crosswise to purines, interstrand carrier transfer dominates, i.e. we have significant diagonal transfer uncovered by TB II which includes \textit{diagonal hoppings}; if purines are on the same strand, intrastrand carrier transfer dominates. For dimers made of different monomers, TB II basically shows intrastrand carrier transfer (but in small percentage).
In this work we show that THz oscillations in DNA monomers, dimers and trimers exist and we study the frequency content, the maximum transfer percentages, the transfer rates between sites and the mean probabilities to find the carrier at a site. Hence, one could imagine the future built of a source or receiver of EM radiation in the range 0.1 THz to 1000 THz made of tiny DNA segments.

The rest of the article is organized as follows:
In Sec.~\ref{sec:general} we outline our TB I and TB II approaches in general terms skipping technical details, which are presented in Appendix A.
Our results for monomers, dimers and trimers are presented in Sections~\ref{sec:ResultsMonomers}, \ref{sec:ResultsDimers} and \ref{sec:ResultsTrimers}, respectively. In Sec.~\ref{sec:Conclusion} we state our conclusions.

\section{General}    
\label{sec:general} 
We begin with our notation. By YX we denote two successive base pairs, according to the convention
\begin{eqnarray*}
   &\vdots& \\
5' &      & 3' \\
\textrm{Y}  &   -  & \textrm{Y}_{\textrm{compl}} \\
\textrm{X}  &   -  & \textrm{X}_{\textrm{compl}} \\
3' &      & 5' \\
   &\vdots&
\label{bpdimer}
\end{eqnarray*}
for the DNA strands orientation.
We denote by X, X$_{\textrm{compl}}$, Y, Y$_{\textrm{compl}}$ DNA bases, where
X$_{\textrm{compl}}$ (Y$_{\textrm{compl}}$) is the complementary base of X (Y).
In other words, the notation YX means that the bases Y and X of
two successive base pairs are located at the same strand in the direction $5'-3'$.
X-X$_{\textrm{compl}}$ is the one base pair and
Y-Y$_{\textrm{compl}}$ is the other base pair,
separated and twisted by 3.4 {\AA} and $36^{\circ}$, respectively,
relatively to the first base pair, along the growth axis of the nucleotide chain.
For example, the notation GT denotes that one strand contains G and T in the direction $5'-3'$ and the complementary
strand contains C and A in the direction $3'-5'$.
In the sense explained above, we can talk for equivalent dimers i.e.
$\textrm{YX} \equiv \textrm{X}_{\textrm{compl}}\textrm{Y}_{\textrm{compl}}$
and expand the notion of equivalency to $N$-mers.

Furthermore, we suppose that an extra hole or electron inserted in a DNA segment travels through HOMOs or LUMOs, respectively. Hence, for each base pair or {\it monomer},
the highest occupied molecular orbital (HOMO) and the lowest unoccupied molecular orbital (LUMO) play a key role.

\begin{figure} [h!]
\centering
\includegraphics[width=7.5cm]{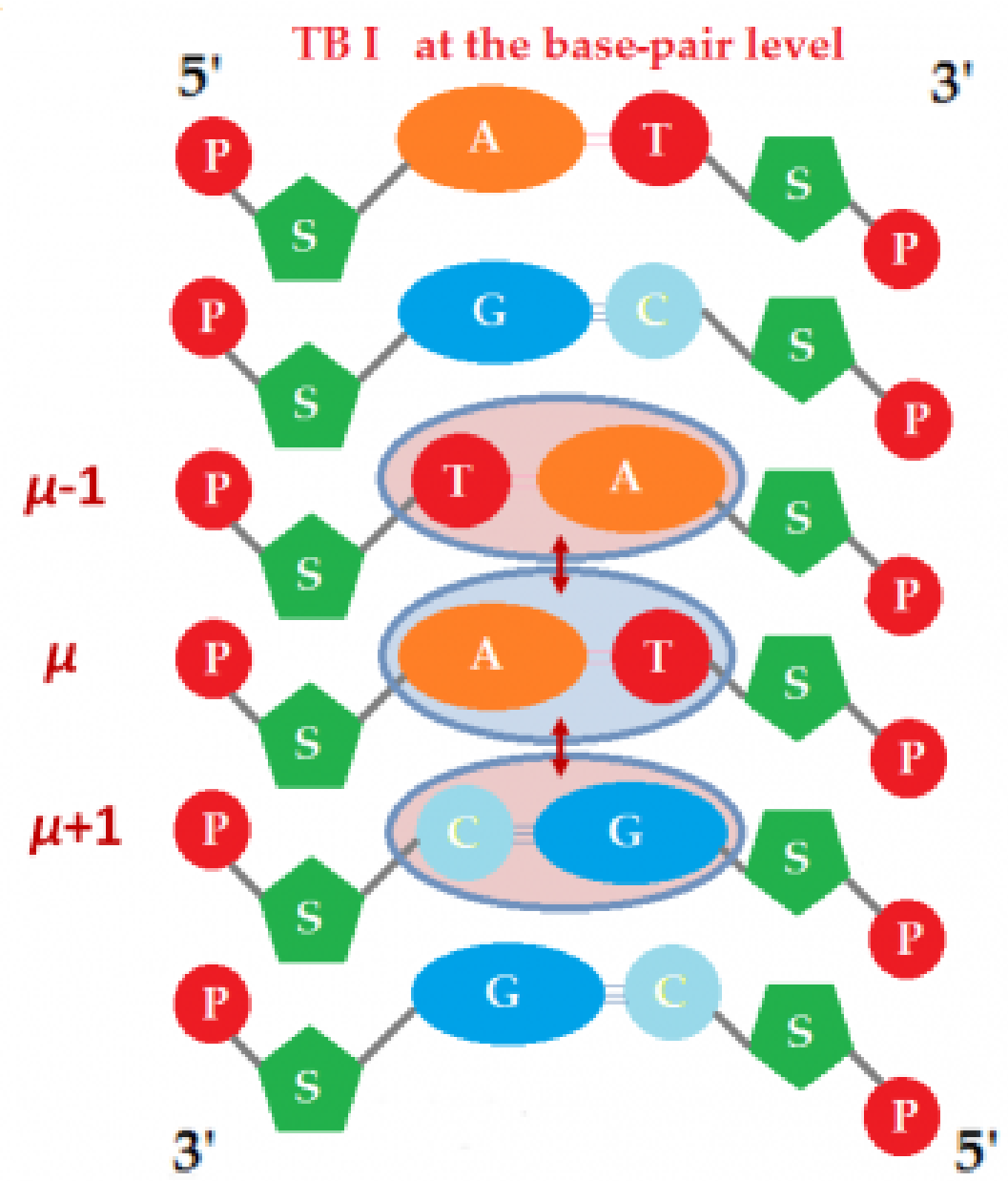}
\includegraphics[width=7.5cm]{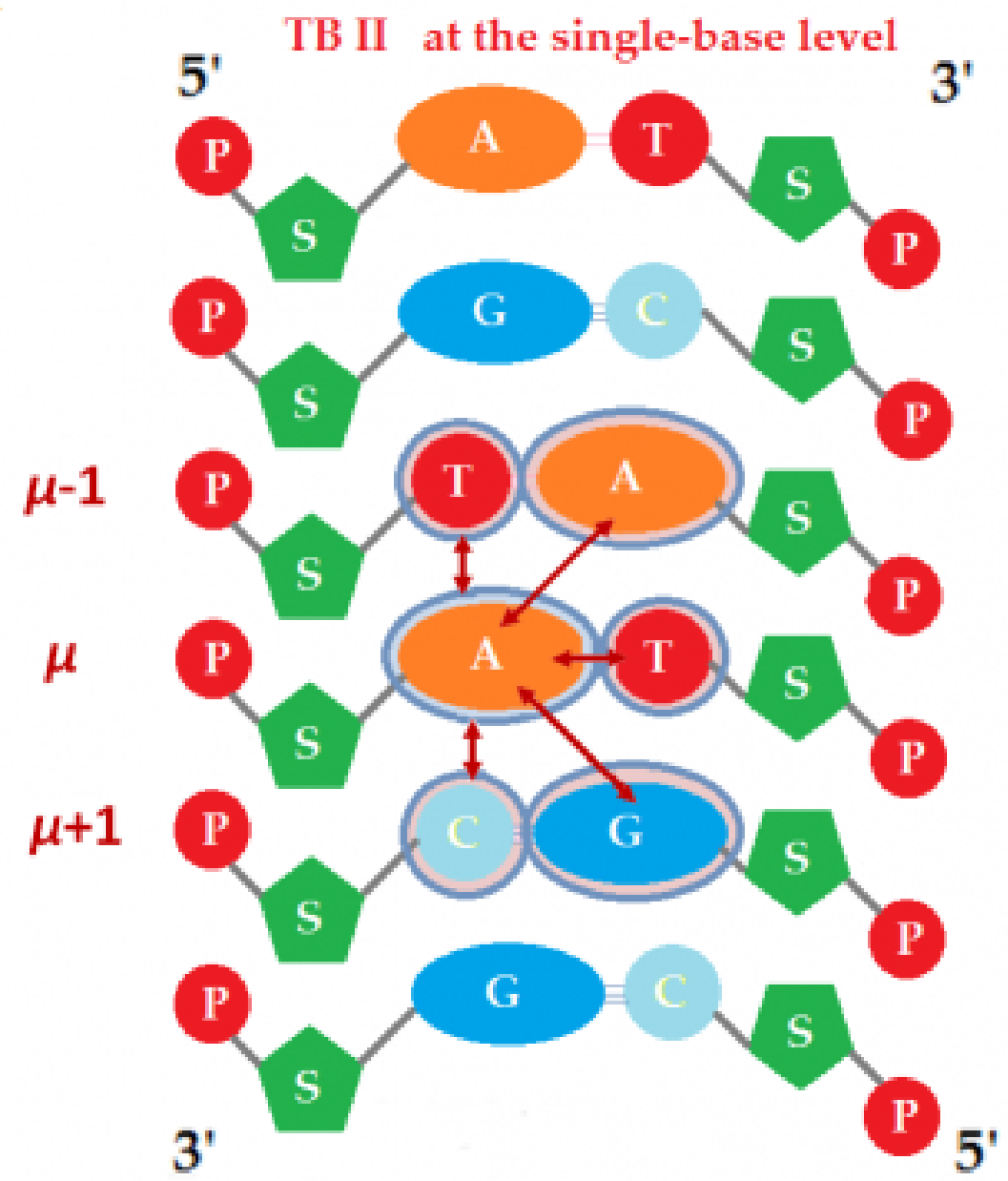}
\caption{Our Tight-Binding approaches: TB I (left) and TB II (right).}
\label{fig:TBITBII}
\end{figure}

We utilize two Tight-Binding approximations, the main points of which are explained below. The mathematical details are given in Appendix A. Here we merely describe them in general terms to make the manuscript more accessible for readers from different areas of the rather wide DNA science.
In TB I the carrier is located at a base pair and it can move to the next or to the previous base pair. Hence, this is a \textit{wire model}~\cite{Albuquerque:2014,CMRR:2007}. In TB II the carrier is located at a base and it can move (1) to its complementary base of the same base pair or (2) to the next or the previous base of the same strand or (3) to the diagonally located base of the other strand of the next or the previous base pair in the $5'-5'$ or in the $3'-3'$ direction, respectively. If we ignore (3), this would be a \textit{ladder model}~\cite{Albuquerque:2014,CMRR:2007}. Here, since we also include diagonal hoppings we call it an \textit{extended ladder model}~\cite{Albuquerque:2014,CMRR:2007}.
It will become evident below that in some cases (e.g. in dimers when purines are crosswise to purines) interstrand carrier transfer dominates, in other words we have significant diagonal transfer, which justifies the inclusion of diagonal hoppings in our model TB II. The two TB models are explained schematically in Fig.~\ref{fig:TBITBII}.
In Sections \ref{sec:ResultsMonomers}, \ref{sec:ResultsDimers}, and \ref{sec:ResultsTrimers} we show that THz oscillations in DNA monomers, dimers and trimers exist. We study the frequency content of these oscillations; where necessary, we employ Fourier analysis. For periodic cases, we study the maximum carrier transfer percentage $p$ from an initial site to a final site, as well as the relevant maximum transfer rate $pf$ showing not only how fast the transfer is, but also what the maximum carrier transfer percentage is. For all cases, either periodic or not, we use the mean transfer rate $k_{ij}$, which shows not only how fast the transfer from site $i$ to site $j$ is, but also what the mean carrier transfer percentage from site $i$ to site $j$ is (cf. Eq.~\ref{meantransferrateN}).

\section{Monomers}          
\label{sec:ResultsMonomers} 
\begin{figure} [h!]
\includegraphics[width=7.8cm]{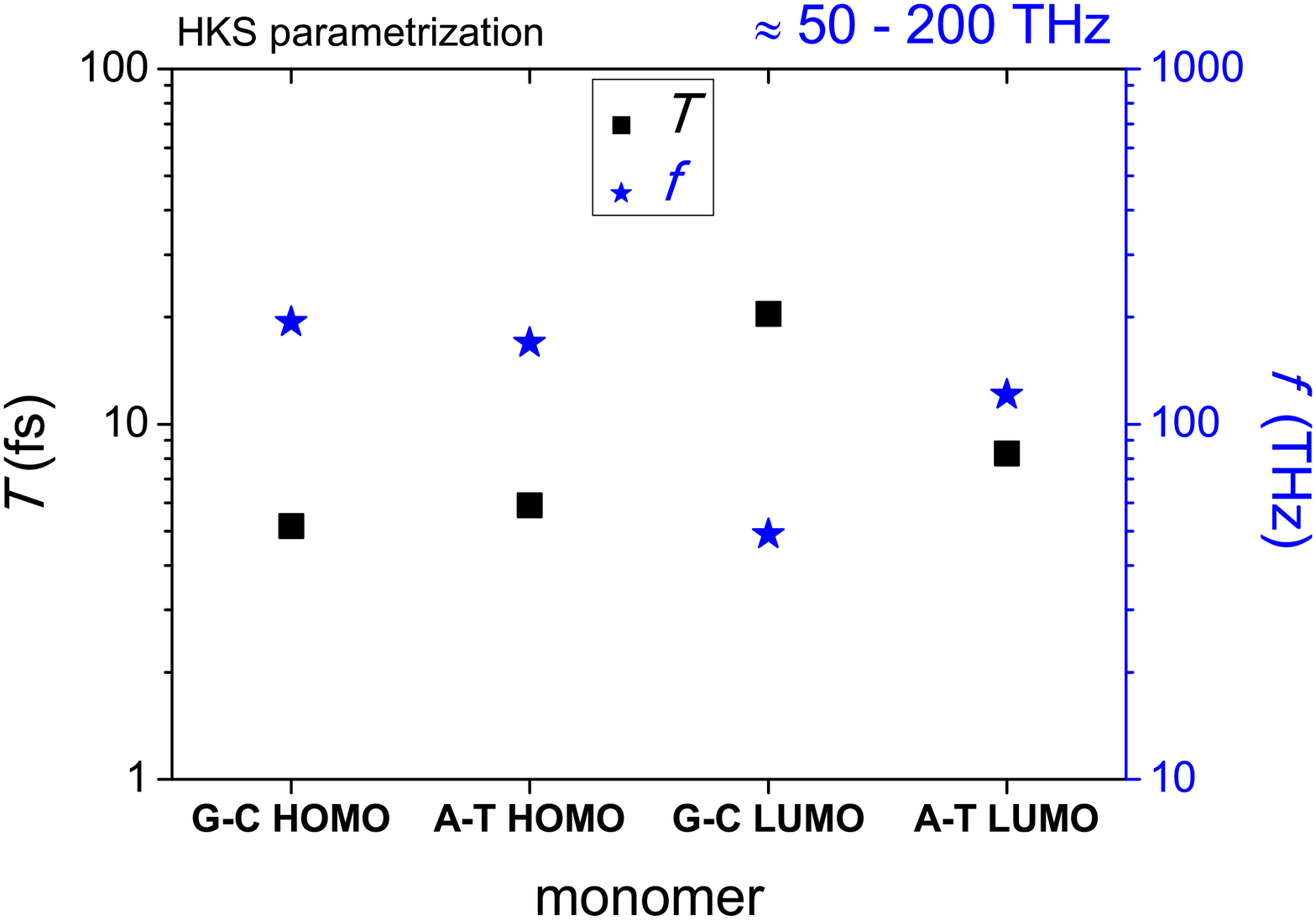}
\includegraphics[width=7.8cm]{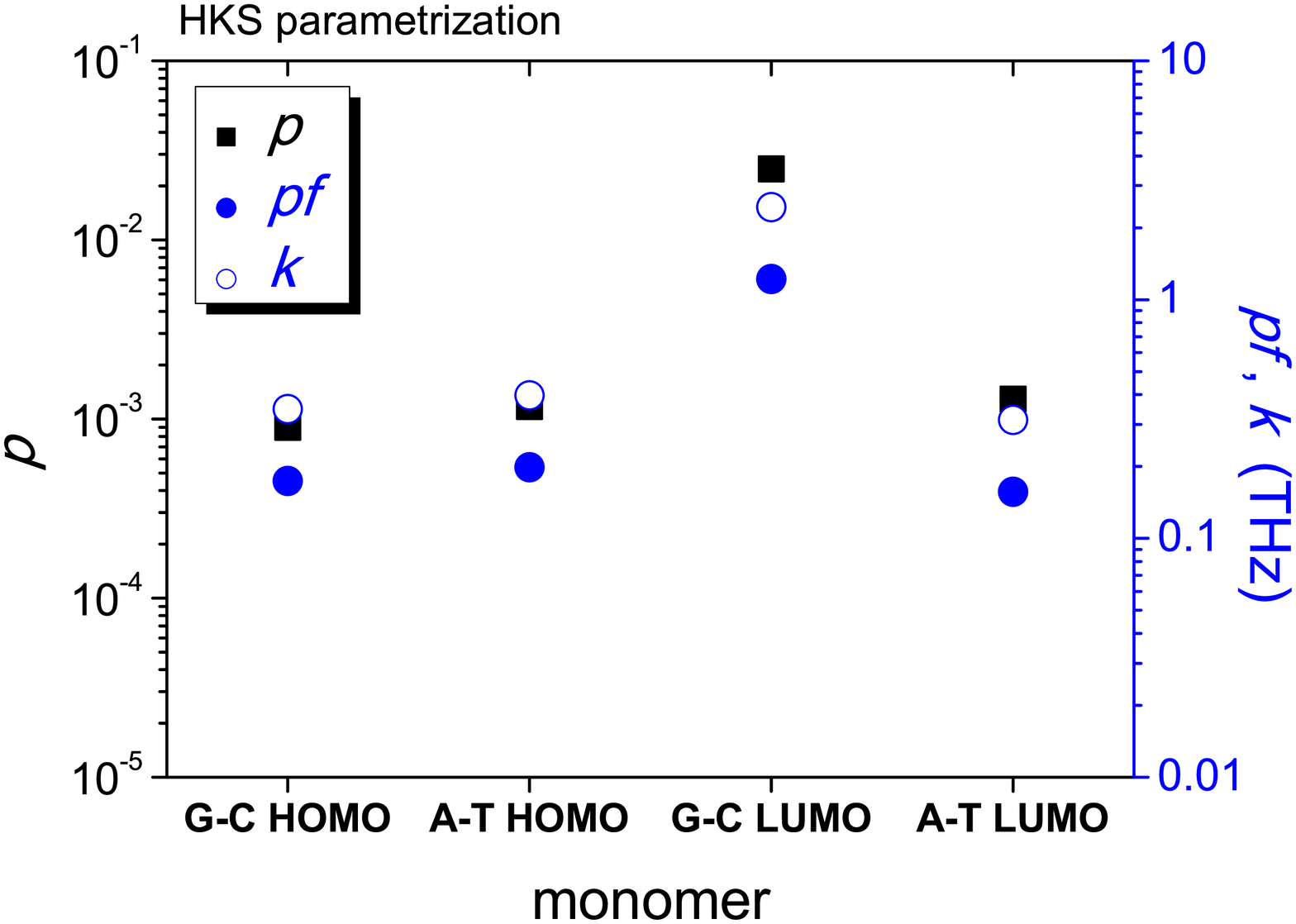}
\includegraphics[width=7.8cm]{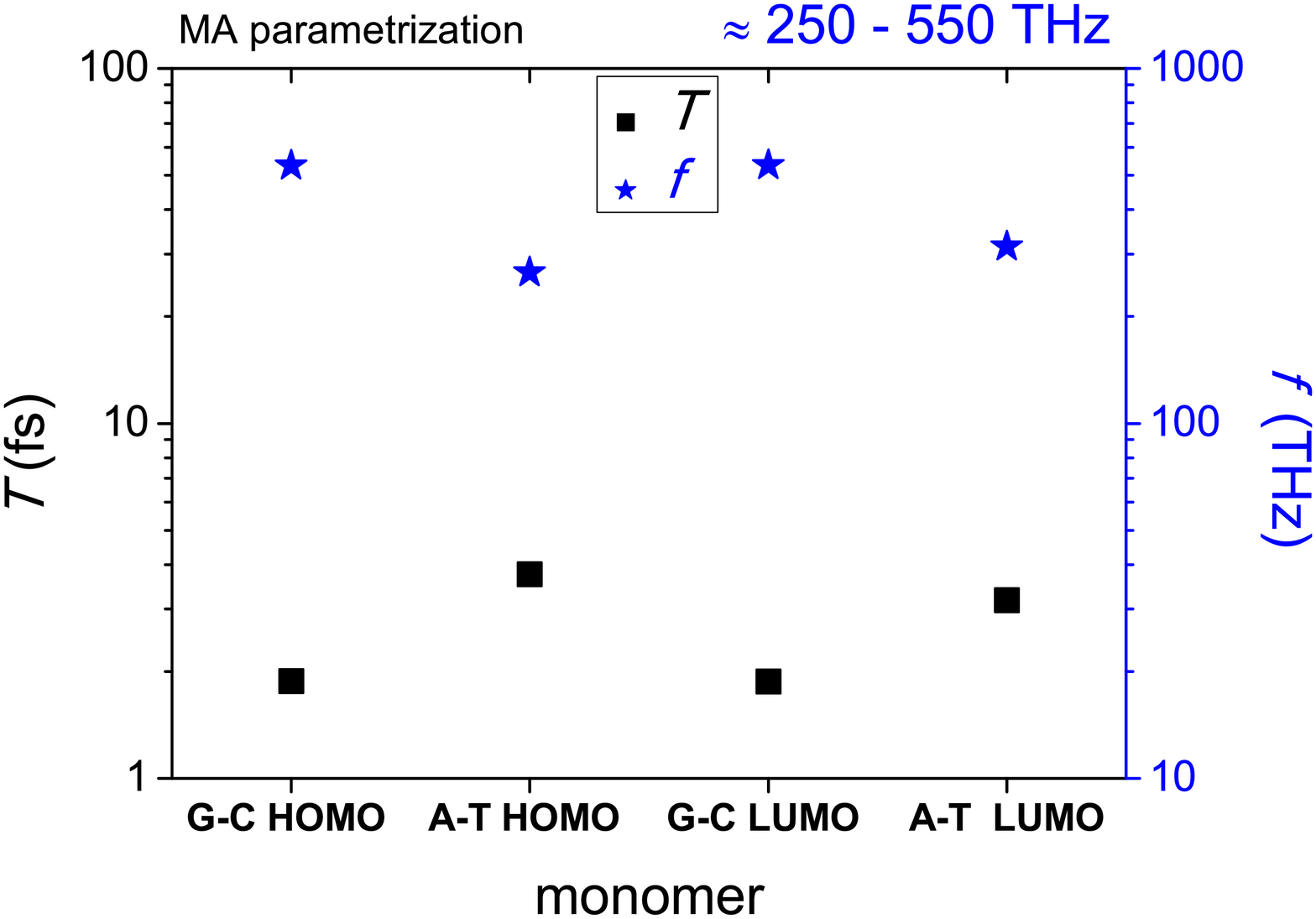}
\includegraphics[width=7.8cm]{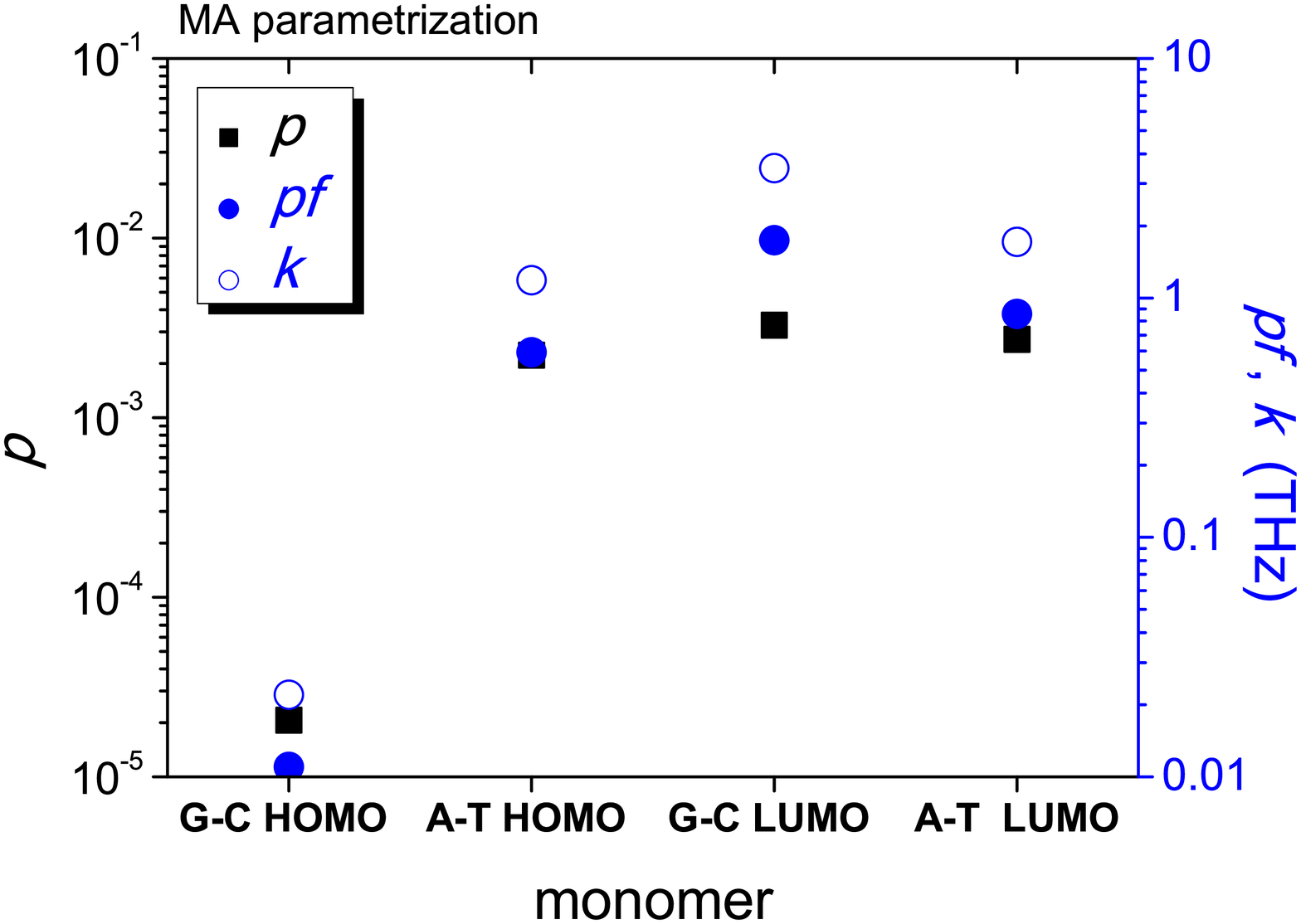}
\caption{Charge oscillations in A-T and G-C according to the TB single-base approach II:
period $T$, frequency $f$, maximum transfer percentage $p$, pure maximum transfer rate $pf$ and pure mean transfer rate $k$.
1st row: TB parameters from Ref.~\cite{HKS:2010-2011} (HKS parametrization).
2nd row: TB parameters from Ref.~\cite{MehrezAnantram:2005} (MA parametrization).
}
\label{fig:monomers}
\end{figure}

\begin{figure} [h!]
\centering
\includegraphics[width=15cm]{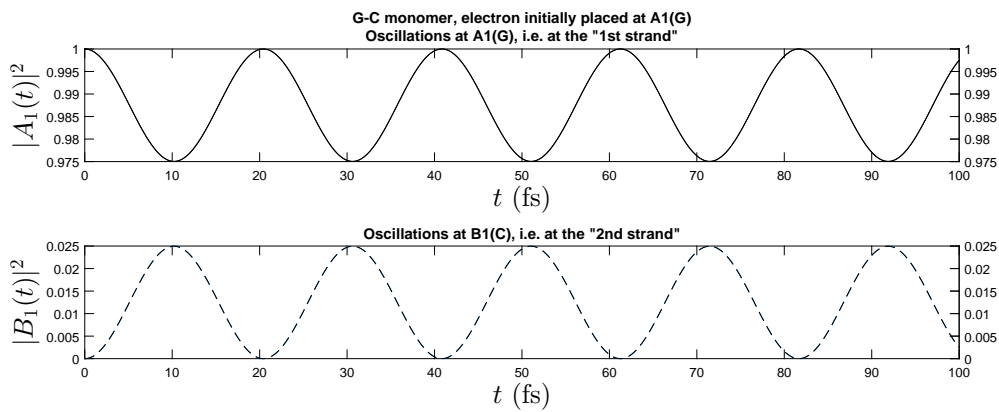}
\caption{Electron oscillations within G-C, according to TB single-base approach II and HKS parametrization.}
\label{fig:ElectronOscillationsG-C}
\end{figure}

TB I cannot be used for charge transfer in monomers, since it considers a monomer as a single site. Hence, we use TB II, supposing that initially we place the carrier at one of the bases. We can prove that an extra hole or electron oscillates between the bases  of the two possible monomers (G-C and A-T) with frequency (or period)
\begin{equation}\label{fandT}
f = \frac{1}{T} = \frac{\sqrt{(2t)^2 + \Delta^2}}{h},
\end{equation}
where $t$ is the hopping integral between the complementary bases and $\Delta$ is the energy gap between the on-site energies of the complementary bases.
Our results for A-T and G-C, both for holes and electrons, are shown in Fig.~\ref{fig:monomers}, with parameters from Ref.~\cite{HKS:2010-2011} (``HKS parametrization'') and from Ref.~\cite{MehrezAnantram:2005} (``MA parametrization'').
For HKS parametrization, $f \approx$  50-200 THz ($T \approx$ 5-20 fs),
for MA  parametrization, $f \approx$ 250-550 THz ( $T \approx$ 2-4 fs).
These ranges correspond to wavelength $\lambda \approx$ 545 nm - 6000 nm i.e. from visible to near-infrared and mid-infrared~\cite{ISO20473}.
We can prove that the maximum transfer percentage $p$
[e.g. $\max{(|B_{1}(t)|^2)}$ for initial conditions $A_{1}(0)=1, B_{1}(0)=0$ or vice versa], is given by
\begin{equation}\label{p}
p = \frac{(2t)^2}{(2t)^2 + \Delta^2}.
\end{equation}
We observe that the carrier is not very likely to be transferred between the monomer bases ($p$ is very small in all cases).
The pure maximum transfer rate defined as $pf$ is also here very small in all cases.
The pure mean transfer rate $k$ is also shown.
It can be analytically proven and numerically shown that here $k=2pf$.
$T$, $f$, $p$, $pf$ and $k$ do not depend on which base the carrier is initially placed at.
A snapshot of electron oscillations between G and C in G-C, according to the HKS parametrization, is given in Fig.~\ref{fig:ElectronOscillationsG-C}.

This is a two-level system of given stationary states (the two HOMOs or the two LUMOs) with a ``perturbation'' represented by the hopping integral, which impels an extra carrier to oscillate between these stationary states. Mathematically, the problem is equivalent to a two-level system (e.g. atom) under the influence of an electric field, which impels an electron to oscillate between the two eigenstates [semiclassical approach after Rotating Wave Approximation or the time-dependent problem with a Jaynes-Cummings Hamiltonian in a full quantum mechanical approach].
These problems are well known in the context of quantum optics~\cite{QOL:2016}.
The same applies to the dimer problem within TB I discussed briefly below.
Increasing the number of monomers which make up the DNA segment, either with TB I or TB II, we have to solve generalizations of the above mentioned problem.
In other words, to determine the spatiotemporal evolution of electrons or holes along a $N$ base-pair DNA segment, we have to solve a system of (I) $N$ or (II) $2N$ coupled differential equations. The relevant matrices $A$ are given in Appendix B.
For $N$ base pairs, the system has $N$ states within TB I, or $2N$ states within TB II. Thus, in this work, for TB I we  examine systems composed of two (dimers) or three (trimers) states, while for TB II we examine systems composed of two (monomers), four (dimers) or six (trimers) states.

\section{Dimers}          
\label{sec:ResultsDimers} 
\begin{figure} [h!]
\centering
\includegraphics[width=7.5cm]{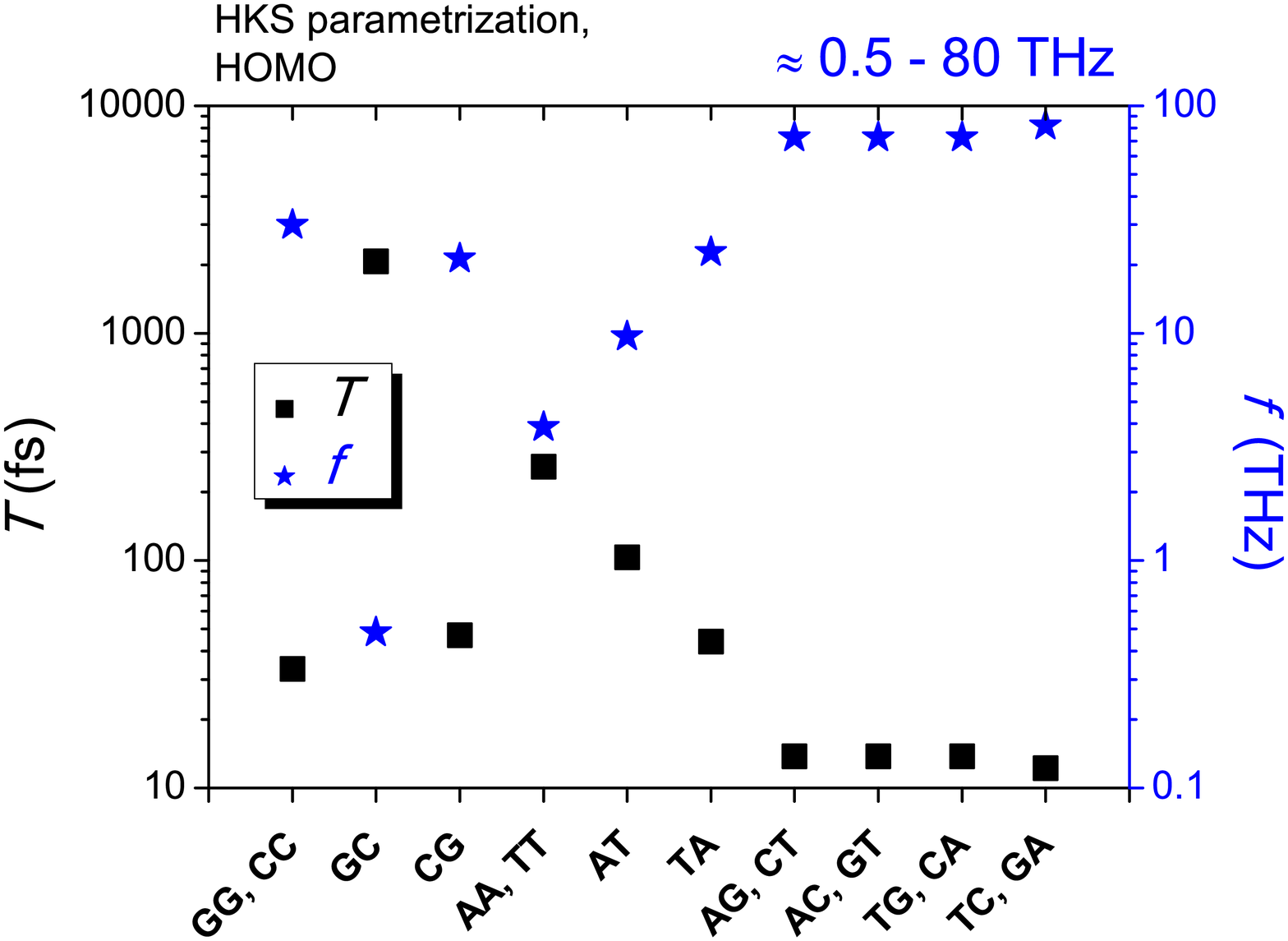}
\includegraphics[width=7.5cm]{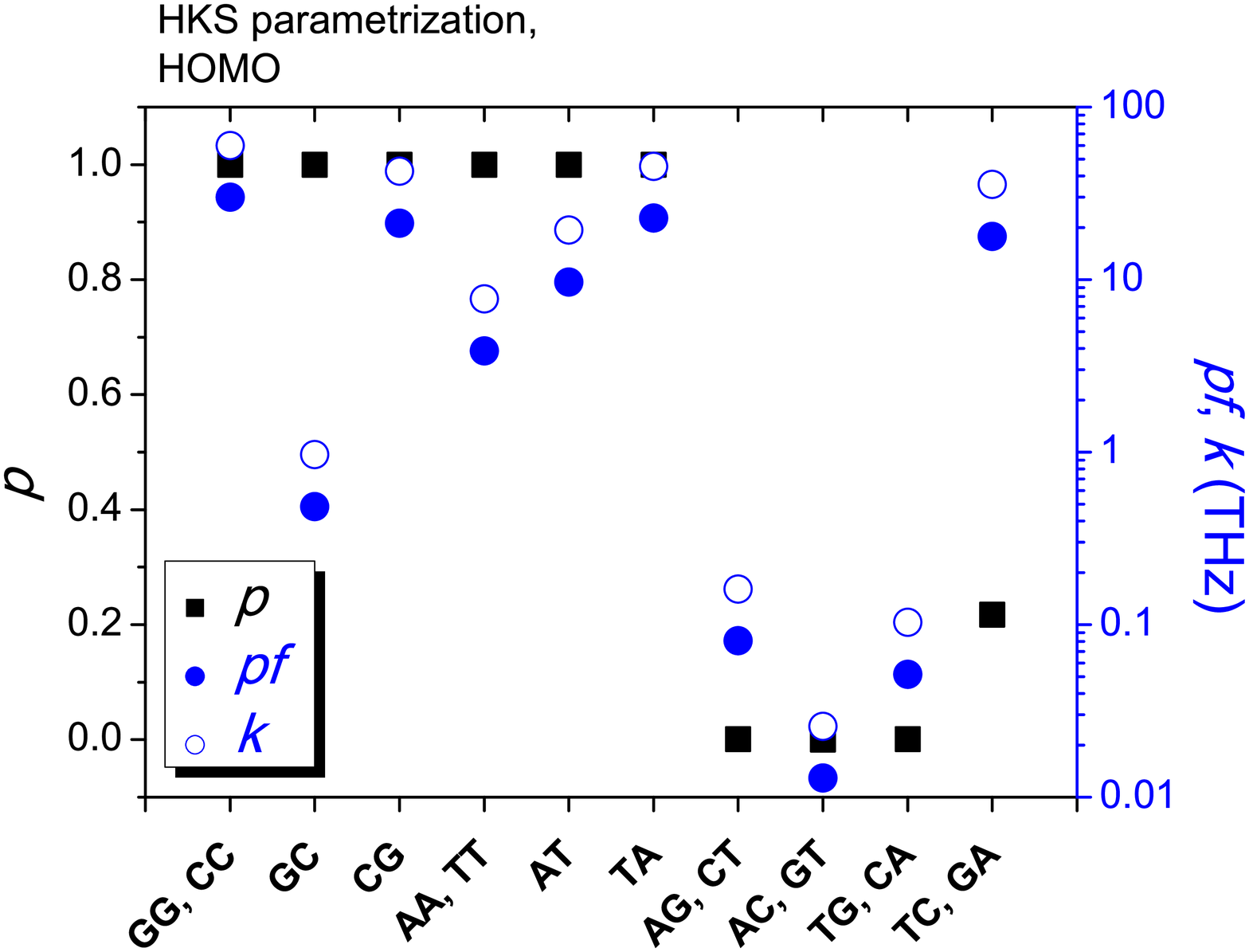}
\includegraphics[width=7.5cm]{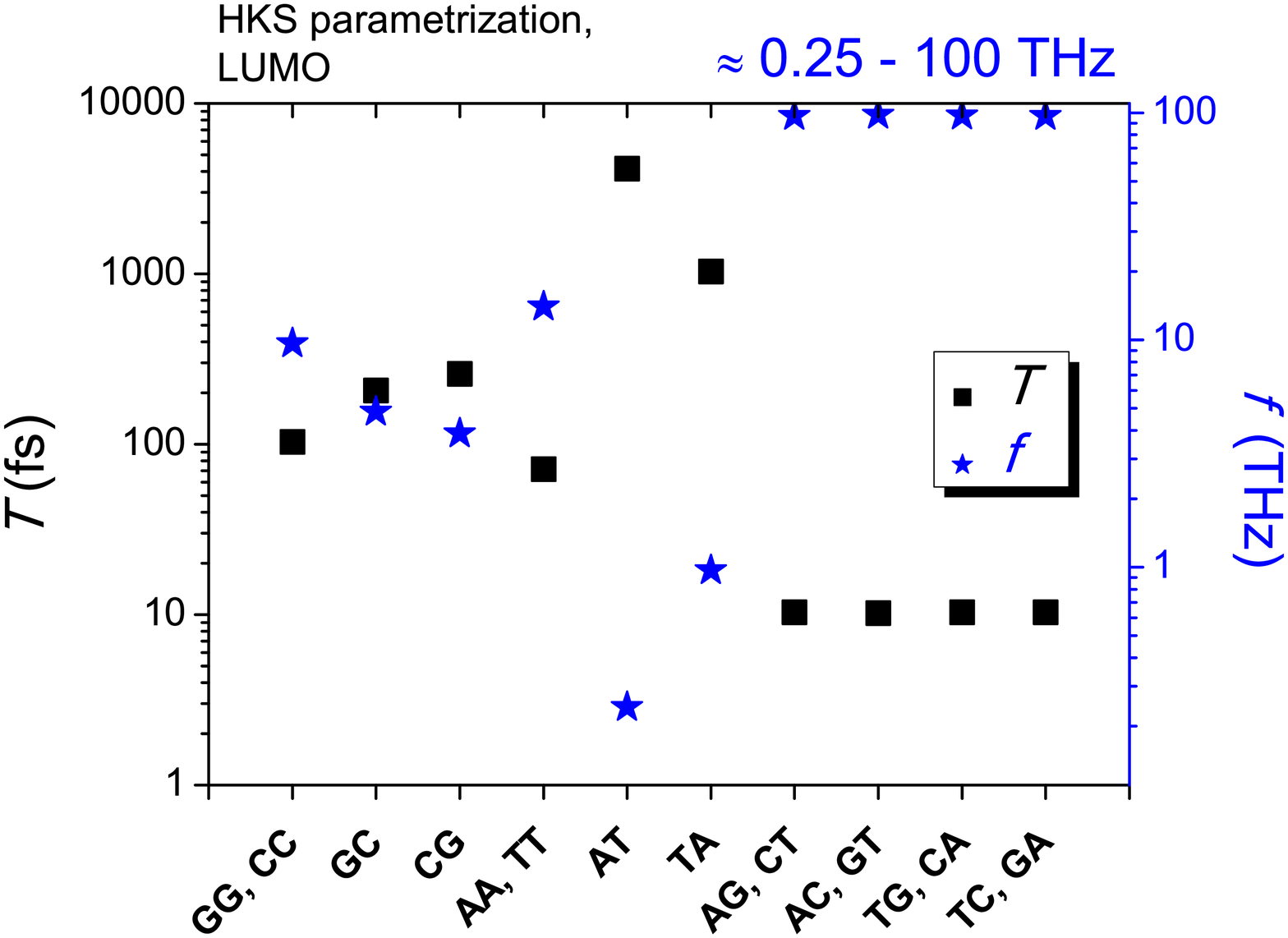}
\includegraphics[width=7.5cm]{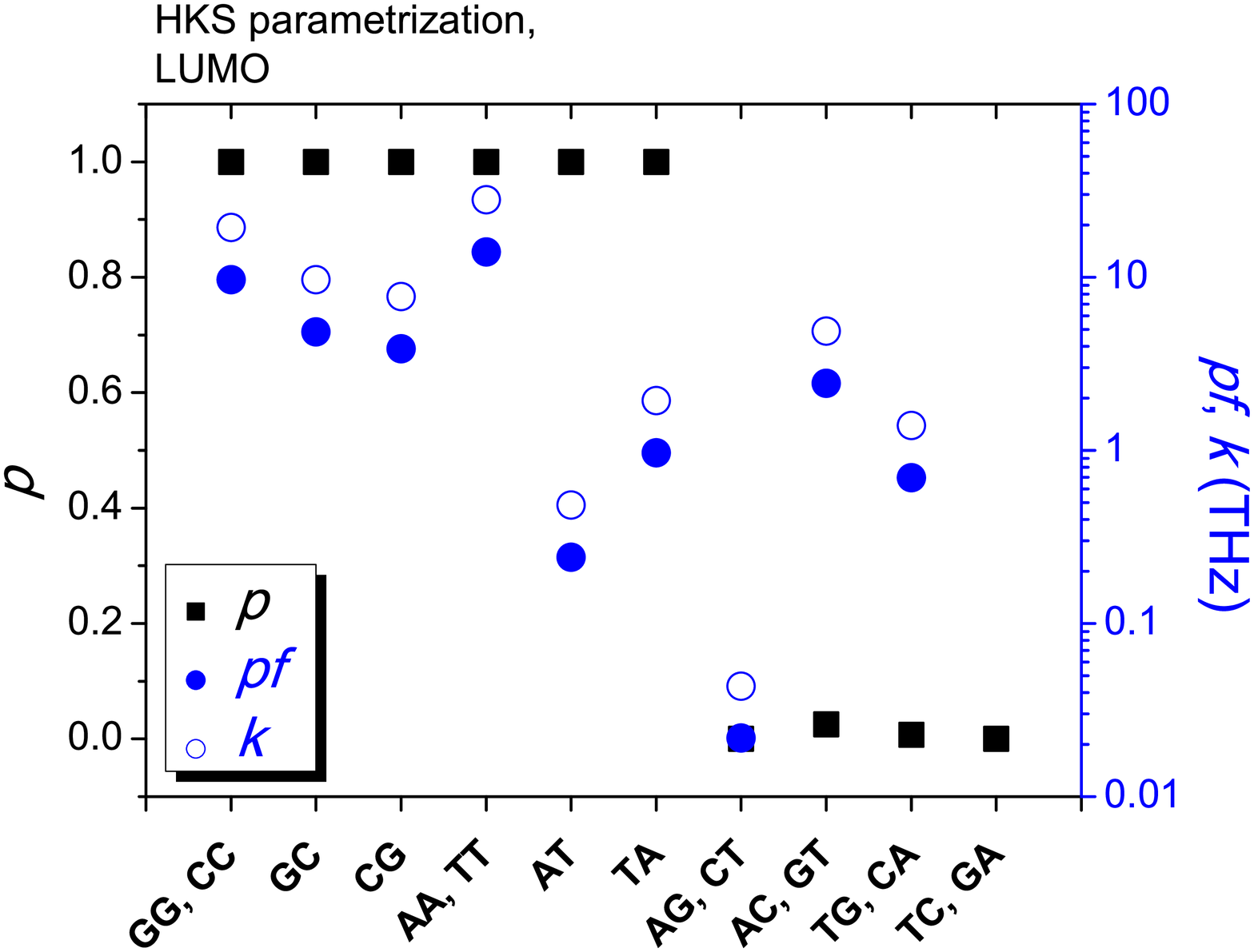}
\caption{Charge oscillations in all possible dimers according to the TB base-pair approach I and the HKS parametrization~\cite{HKS:2010-2011}:
period $T$, frequency $f$, maximum transfer percentage $p$, pure maximum transfer rate $pf$ and pure mean transfer rate $k$. It can be analytically proven and numerically shown that here $k=2pf$.}
\label{fig:dimersHKS}
\end{figure}

The possible dimers result from permutations with repetition of two out of the four bases taking additionally into account that
at the base located in helix $\sigma=1$ corresponds always its complementary base in helix $\sigma=2$ .
The number of possible permutations with repetition is $PR(4,2) = 4^2 = 16$. However, six of them are equivalent to other six i.e. GG$\equiv$CC, AA$\equiv$TT, AG$\equiv$CT, AC$\equiv$GT, TG$\equiv$CA, TC$\equiv$GA. Hence, the possible dimers are 10.
Recently, with TB I, we proved that the carrier movement in all dimers is strictly periodic~\cite{Simserides:2014,LKGS:2014,Lambropoulos:2014}.
The frequencies (or periods) are given by Eq.~\ref{fandT}, where, now $t$ is the hopping integral between the base pairs and $\Delta$ is the energy gap between the on-site energies of the base pairs.
Using the TB parameters of Ref.~\cite{Simserides:2014}, we found~\cite{Simserides:2014,LKGS:2014,Lambropoulos:2014} $f \approx$ 0.25-100 THz,  i.e. $T \approx$ 10-4000 fs, i.e. wavelength $\lambda \approx$ 3-1200 $\mu$m, in other words mainly in the MIR and the FIR range~\cite{ISO20473}.
We found that the maximum transfer percentage $p=1$ for dimers made of identical monomers, but $p<1$ for dimers made of different monomers and that the values of $f, T, p, pf, k$ do not depend on which of the two monomers the carrier is initially placed at (cf. also Eq.~\ref{periodsorfrequencies}).
The HKS parametrization~\cite{HKS:2010-2011} results in the same frequency range, although,  the predicted frequencies vary slightly due to the different values of the TB parameters (a summarizing graph is given in Fig.~\ref{fig:dimersHKS}).

Let us now compare TB approaches I and II relatively to the frequency content. Using TB II, one cannot strictly determine periodicity in the carrier movement between the four bases. Hence, $f, T, p, pf$ cannot be defined, but Fourier analysis shows similar frequency content in the THz domain, cf. Appendix C, where we depict the Fourier spectra given by Eq.~\ref{FourierSpectra}, within TB II and HKS parametrization~\cite{HKS:2010-2011}.

Let us start with GG, a dimer made of identical monomers with purines on purines (Fourier spectra in Appendix C, Fig.~\ref{fig:FourierGG}).
If we initially place the hole at A1(G) or A2(G), we obtain the main Fourier amplitude at $f \approx$ 30 THz.
If we initially place the hole at B1 (C) or B2 (C), we obtain the main Fourier amplitude at $f \approx$ 32 THz.
The rest of the frequencies show up with almost negligible amplitudes.
These results are in accordance with TB I and HKS parametrization~\cite{HKS:2010-2011} where for the GG dimer we obtain $f \approx$ 30 THz.
With TB I and the parametrization of Refs.~\cite{Simserides:2014,LKGS:2014} we had obtained $f \approx$ 48 THz~\cite{Simserides:2014,LKGS:2014}.
The amplitudes at the main frequencies are $\approx$ 0.5, in accordance with Eq.~(\ref{FourierSpectra}), expressing the fact that for GG the mean probability to find the hole at a base is,  approximately, almost exclusively, equally divided between the base the carrier was initially placed at and the other base of the same strand, cf. also Fig.~\ref{fig:DimersIIHOMO} below.

Let us now continue with GC, a dimer made of identical monomers with crosswise purines (Fourier spectra in Appendix C, Fig.~\ref{fig:FourierGC}).
If we initially place the hole at A1(G) or B2(G), we obtain the main Fourier amplitude at $f \approx$ 0.3 THz.
If we initially place the hole at B1 (C) or A2 (C), we obtain the main Fourier amplitude peak at $f \approx$ 1.6 THz.
The rest of the frequencies show up with almost negligible amplitudes.
These results are in accordance with TB I and HKS parametrization~\cite{HKS:2010-2011} where for the GC dimer we obtain $f \approx$ 0.5 THz.
With TB I and the parametrization of Refs.~\cite{Simserides:2014,LKGS:2014} we had obtained $f \approx$ 4.8 THz~\cite{Simserides:2014,LKGS:2014}.
The amplitudes at the main frequencies are $\approx$ 0.5, in accordance with Eq.~\ref{FourierSpectra}, expressing the fact that for GC the mean probability to find the hole at a base is, approximately, almost exclusively, equally divided between the base the carrier was initially placed at and the diagonally located base at the opposite strand, cf. also Fig.~\ref{fig:DimersIIHOMO} below.

Let us finish with CT, a dimer made of different monomers (Fourier spectra in Appendix C, Fig.~\ref{fig:FourierCT}).
The main Fourier amplitude is at $f \approx$ 70.75 THz, if the hole is initially placed at C or T, otherwise hole transfer is negligibly small.
Within TB I and the HKS parametrization~\cite{HKS:2010-2011}, for the CT dimer we obtain $f \approx$ 72.5 THz.
With TB I and the parametrization of Refs.~\cite{Simserides:2014,LKGS:2014} we had obtained $f \approx$ 74 THz~\cite{Simserides:2014,LKGS:2014}.
The amplitudes at the main frequencies are $\approx$ 0.25, when the hole is initially placed at C or T and tiny when the hole is initially placed at G or A, in accordance with Eq.~\ref{FourierSpectra}, expressing the fact that for CT the mean probability to find the hole at a base is approximately 0.75 at the base the carrier was initially placed at and approximately 0.25 at the other base of the same strand when these bases are C and T, but, hole transfer is negligibly small when the hole is initially placed at G or A, cf. also Fig.~\ref{fig:DimersIIHOMO} below.

\begin{figure} [h!]
\centering
\includegraphics[width=15cm]{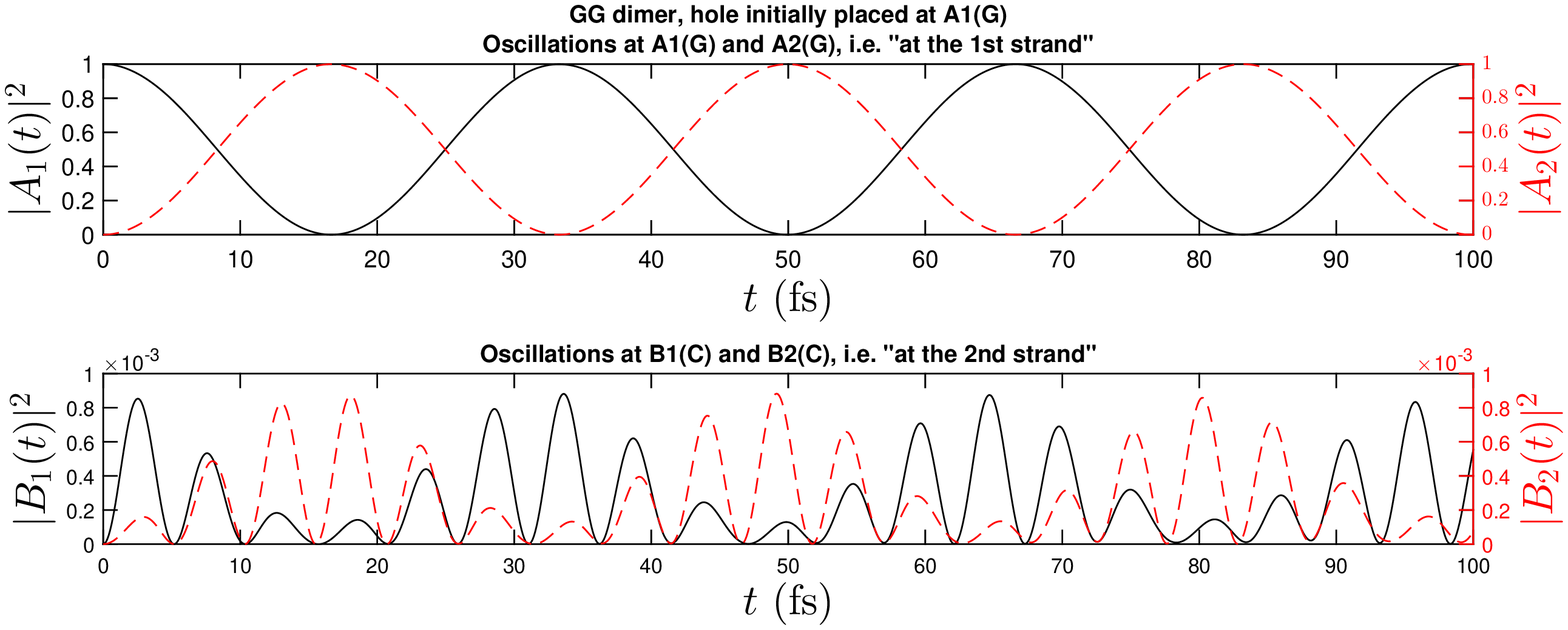}
\includegraphics[width=15cm]{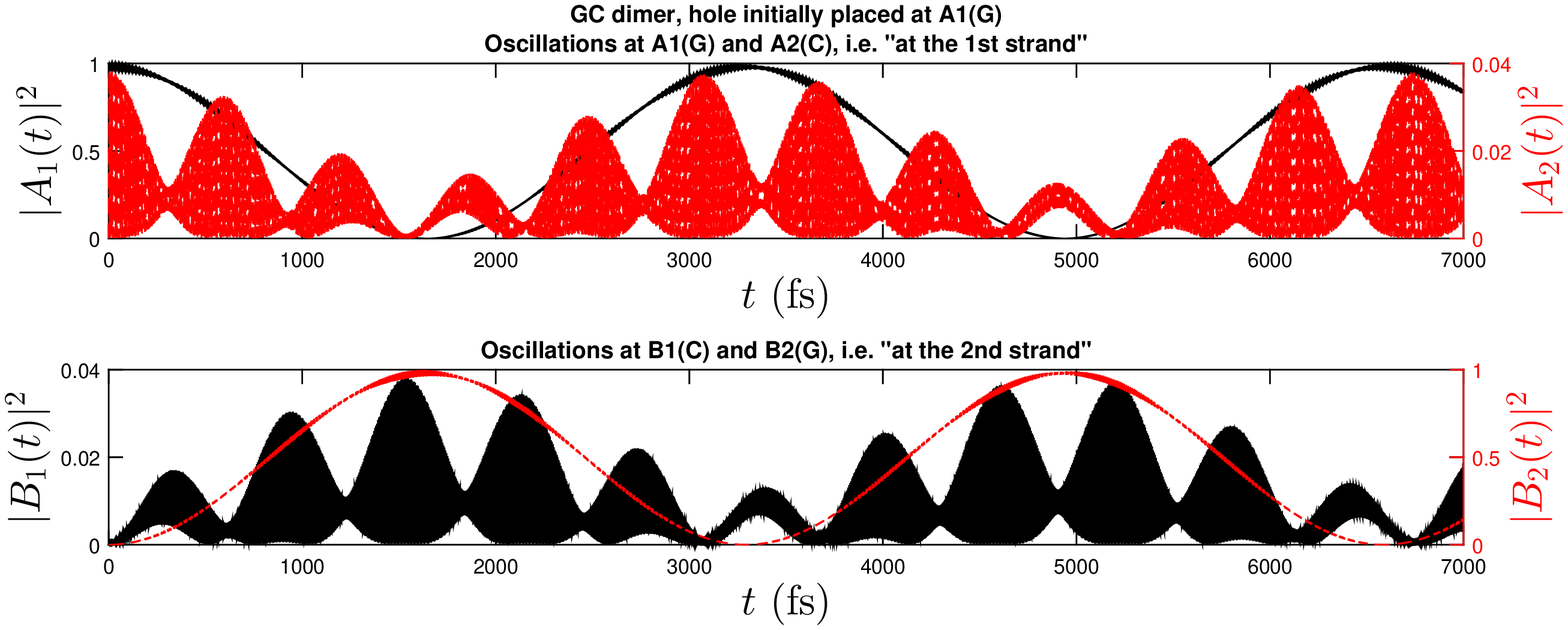}
\includegraphics[width=15cm]{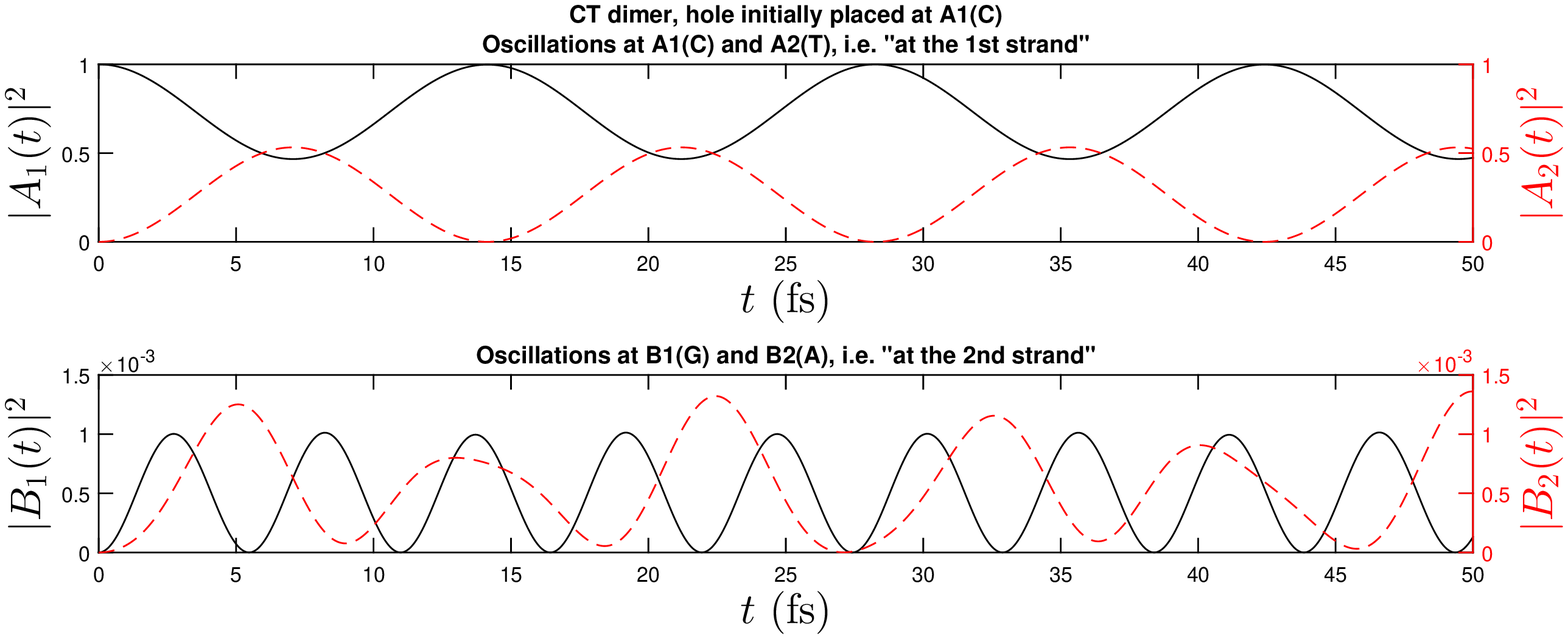}
\caption{ From top to bottom: Hole oscillations within the GG dimer, the GC dimer and the CT dimer, according to the TB single-base approach II and the HKS parametrization. Continuous black (dashed red) lines correspond to the 1st (2nd) base pair.}
\label{fig:HoleOscillationsDimers}
\end{figure}

Examples of snapshots of hole oscillations in GG, GC and CT dimers, according to the HKS parametrization and TB II, are given in Fig.~\ref{fig:HoleOscillationsDimers}.

Hence, generally, both TB approaches, independently of the specific parametrization, predict oscillations in the same THz range. The little differences arise from the specific values used for the on-site energies and the hopping parameters in each parametrization.

\begin{figure} [h!]
\centering
\includegraphics[width=7.5cm]{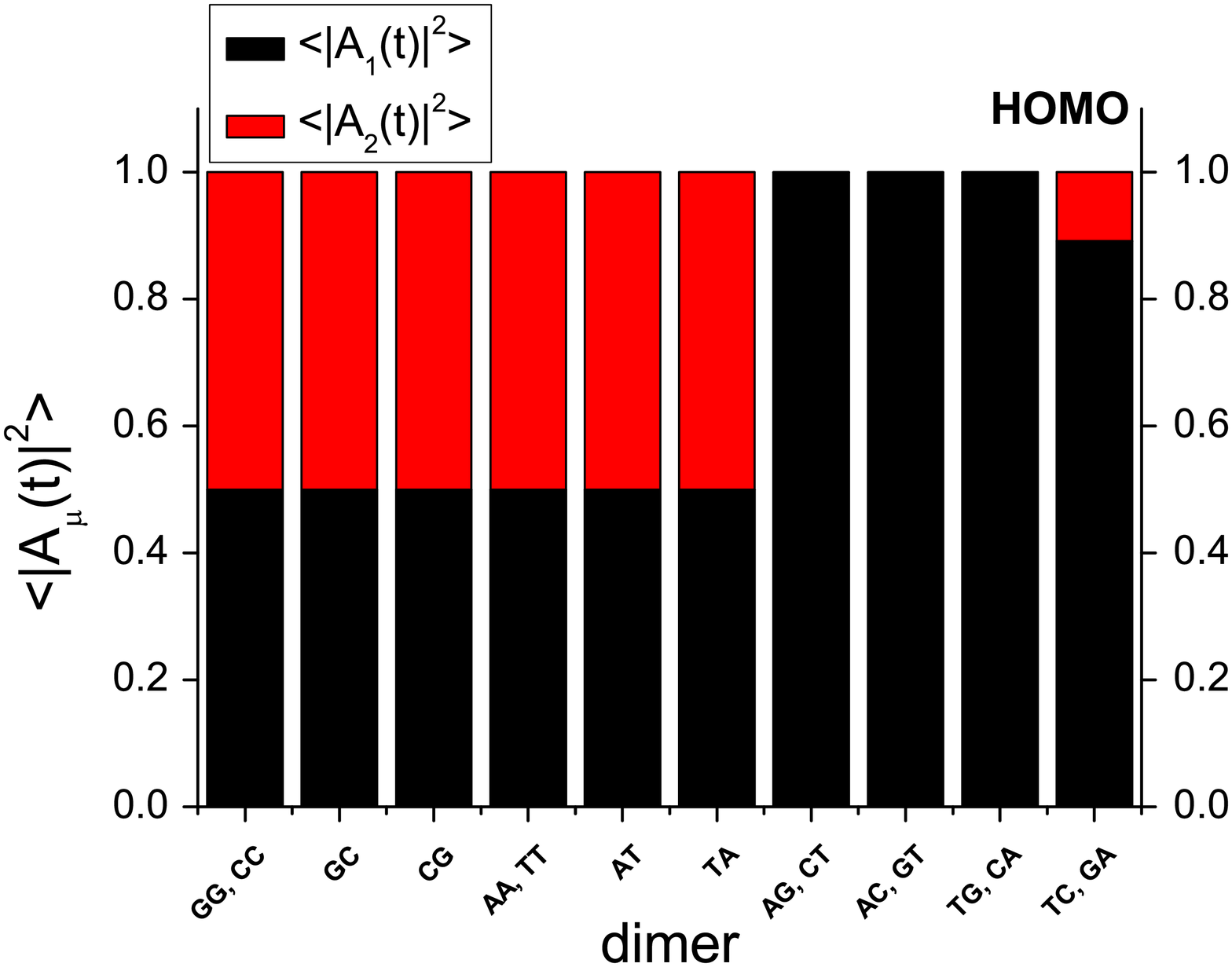}
\includegraphics[width=7.5cm]{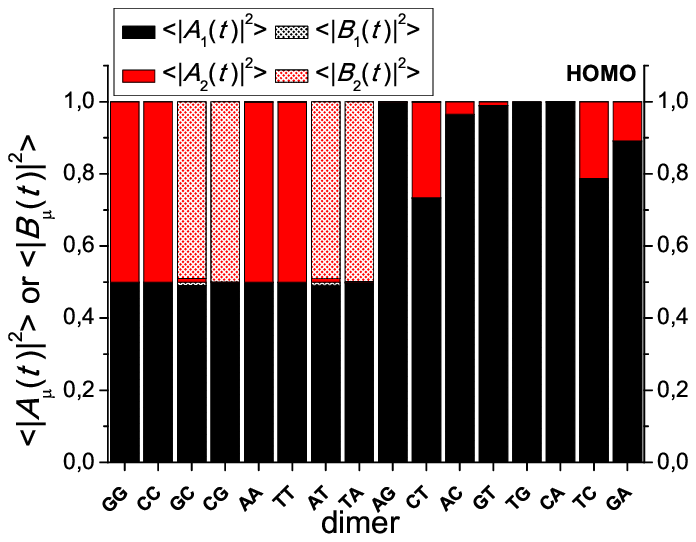}
\includegraphics[width=7.5cm]{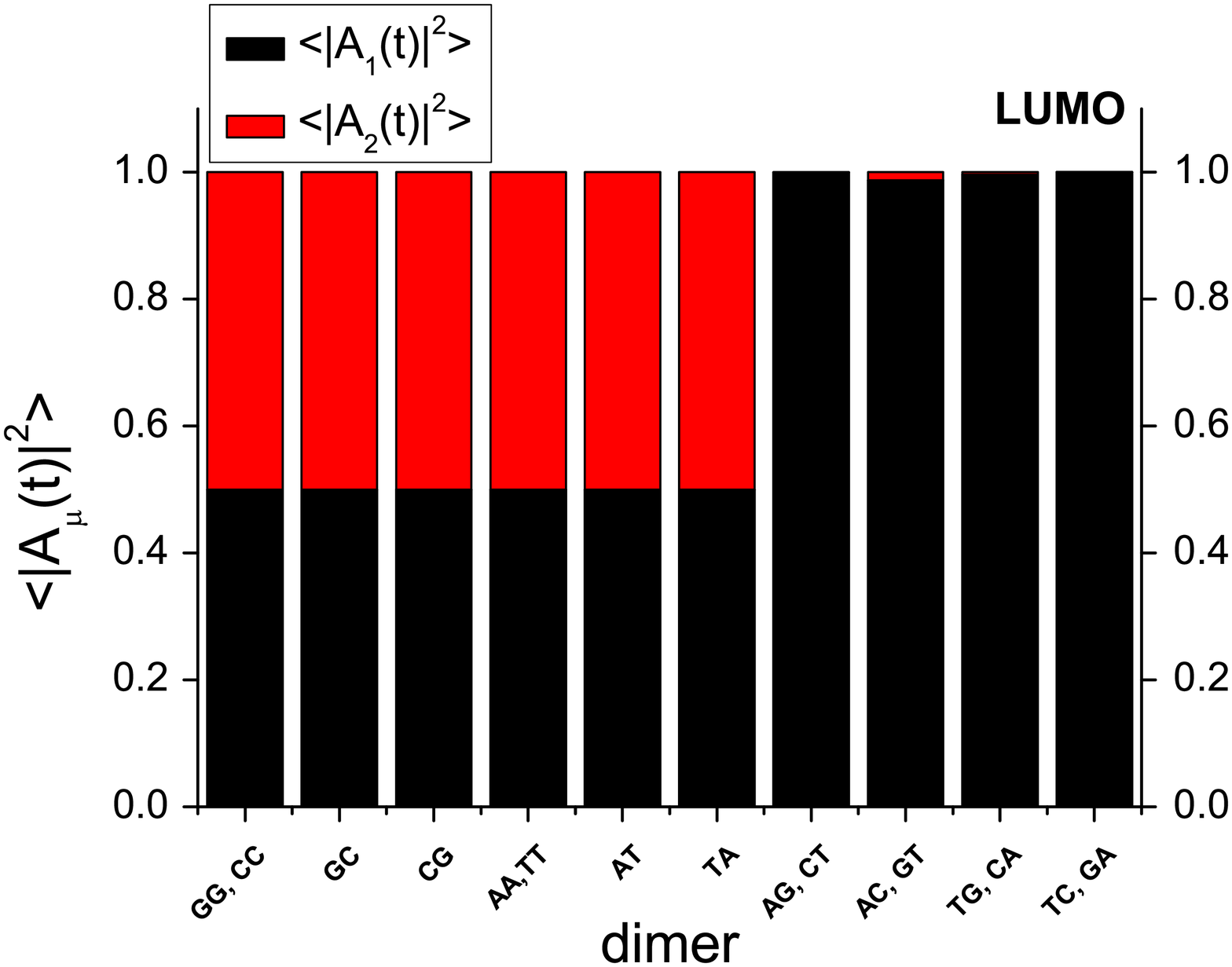}
\includegraphics[width=7.5cm]{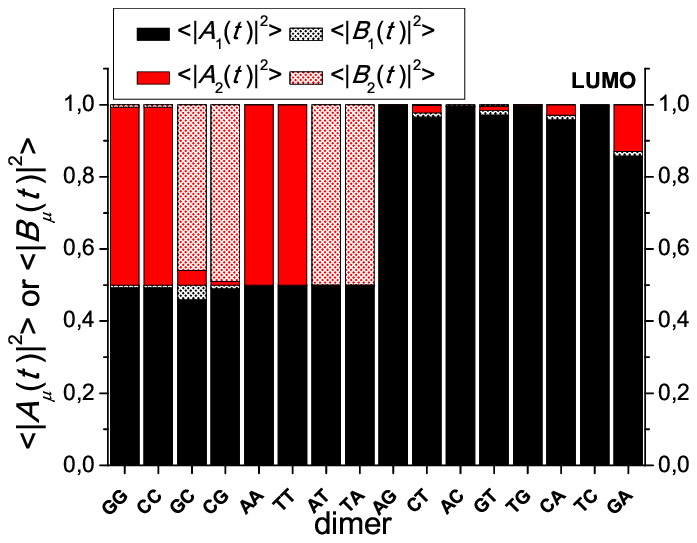}
\caption{The mean probabilities to find an extra carrier [hole (1st row) or electron (2nd row)] at each site of a DNA dimer,
as determined with
(I)  the base-pair  TB approach (left  column) and
(II) the single-base TB approach (right column).
For  TB I, the carrier is initially placed at the 1st monomer, while,
for TB II, it is initially placed at the base of the 1st monomer that belongs to the 1st strand.
We use the HKS parametrization~\cite{HKS:2010-2011}.}
\label{fig:comparison}
\end{figure}

TB approaches I and II allow us to determine the mean probability to find the carrier at a site [base pair for I, base for II].
A comparison between the mean probabilities obtained with TB approaches I and II is shown in Fig.~\ref{fig:comparison}, using the HKS parametrization~\cite{HKS:2010-2011}.
Comparing the two TB approaches e.g. looking at Fig.~\ref{fig:comparison}, but also focusing on TB II e.g. looking at Figs.~\ref{fig:DimersIIHOMO}-\ref{fig:DimersIILUMO} below, we reach the following conclusions:
(a) Carrier transfer is large in dimers made of identical monomers: Finally, the probability is equally shared between the two monomers which make up the dimer.
(b) For dimers made of identical monomers, if purines are crosswise to purines, the carrier changes strand (from strand 1 to strand 2 or vice versa), while if purines are on the same strand, the carrier is transferred through the strand it was initially placed at.
(c) For dimers made of different monomers, the carrier is transferred (albeit in small percentage) mainly through the strand it was initially placed at. Hence, carrier transfer is very small in dimers made of different monomers. The carrier basically remains in the base it was initially placed at, while a small percentage passes to the other base of the same strand.
A careful inspection in Figs.~\ref{fig:FourierGG},~\ref{fig:FourierGC} and \ref{fig:FourierCT}  shows that the Fourier analysis confirms conclusions (a),(b),(c).
For TB I, in Refs.~\cite{Simserides:2014,LKGS:2014,LChMKTS:2015,Lambropoulos:2014}, we used a somehow different set of TB parameters, the results are similar.

\begin{figure} [h!]
\centering
\includegraphics[width=7.5cm]{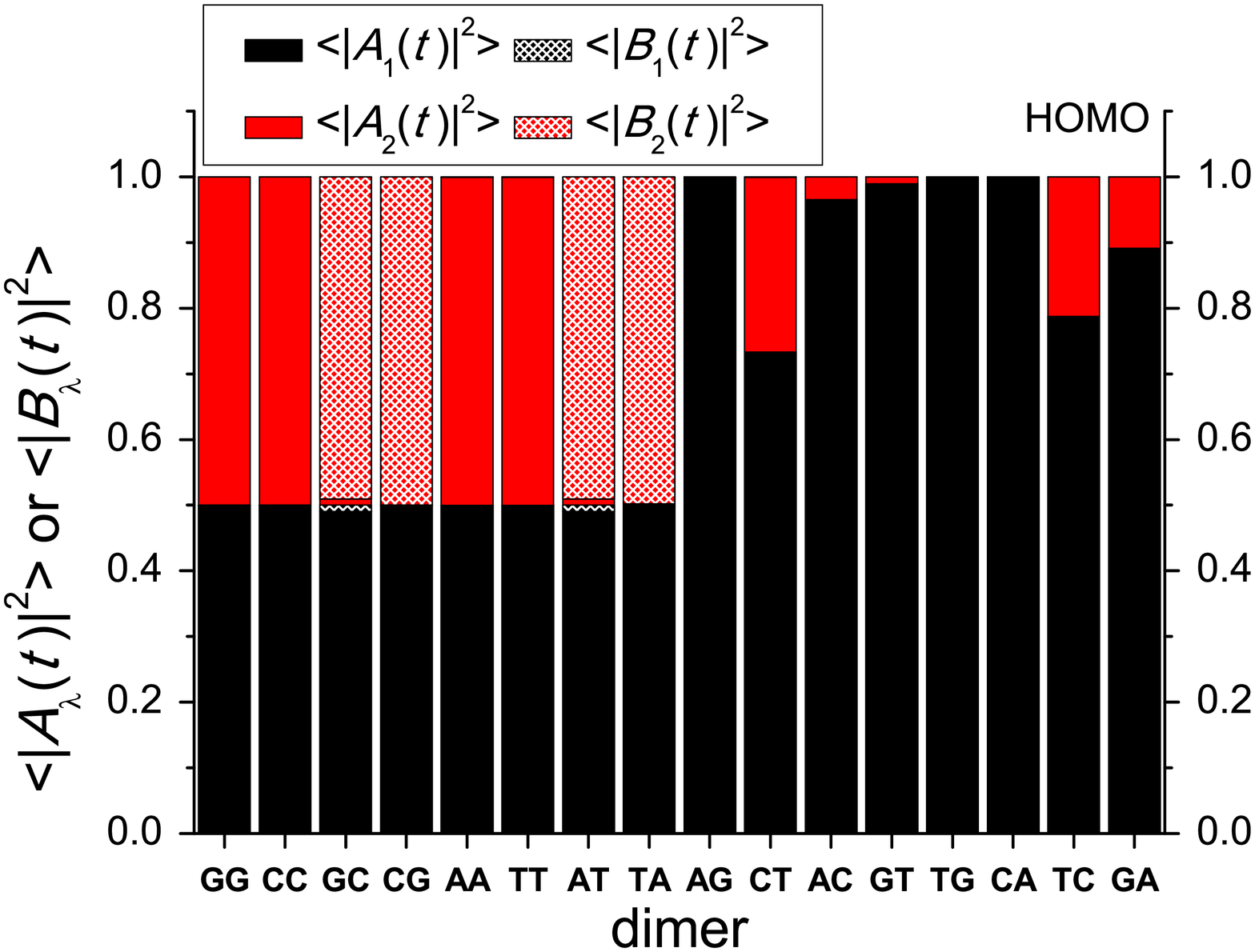}
\includegraphics[width=7.5cm]{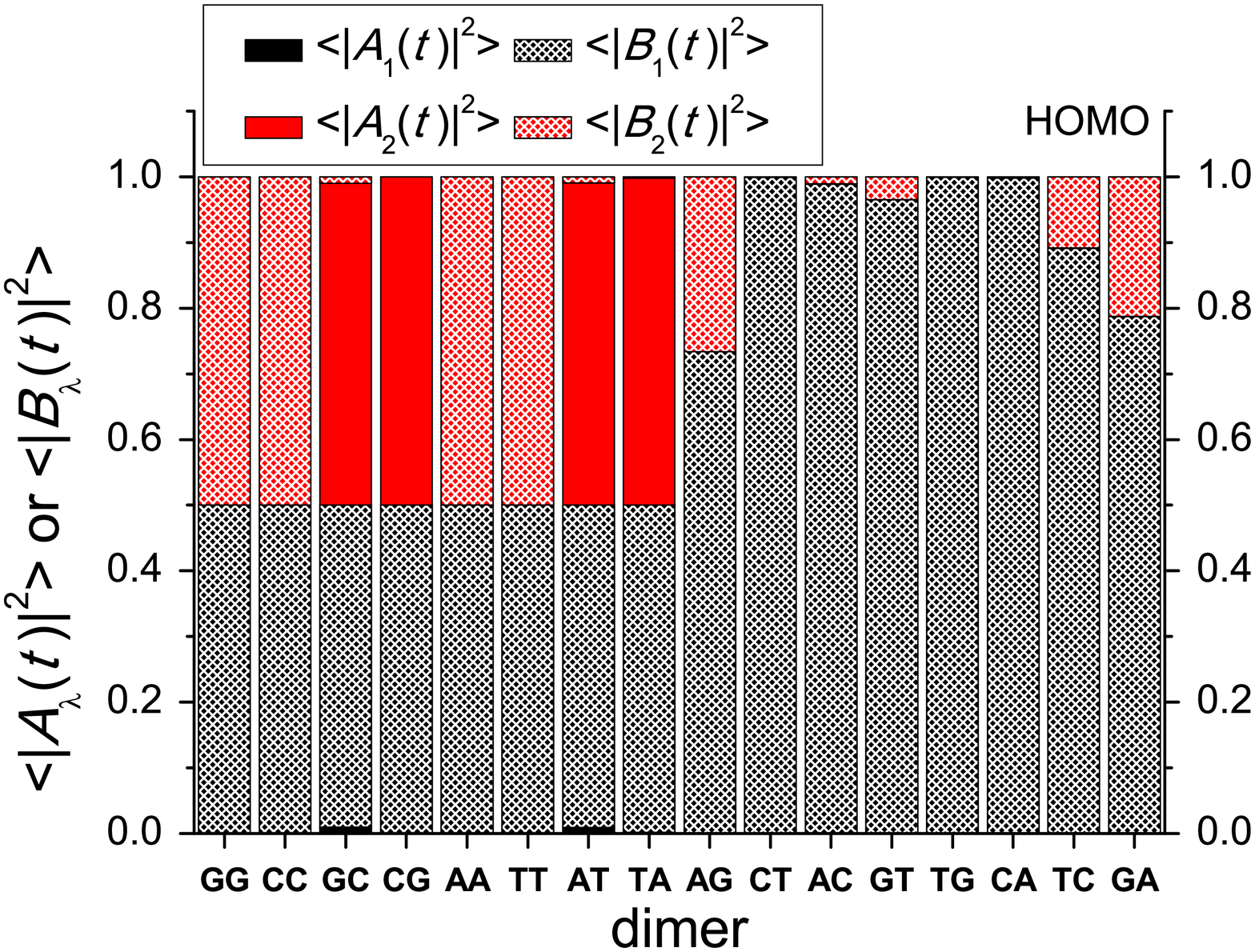}
\includegraphics[width=7.5cm]{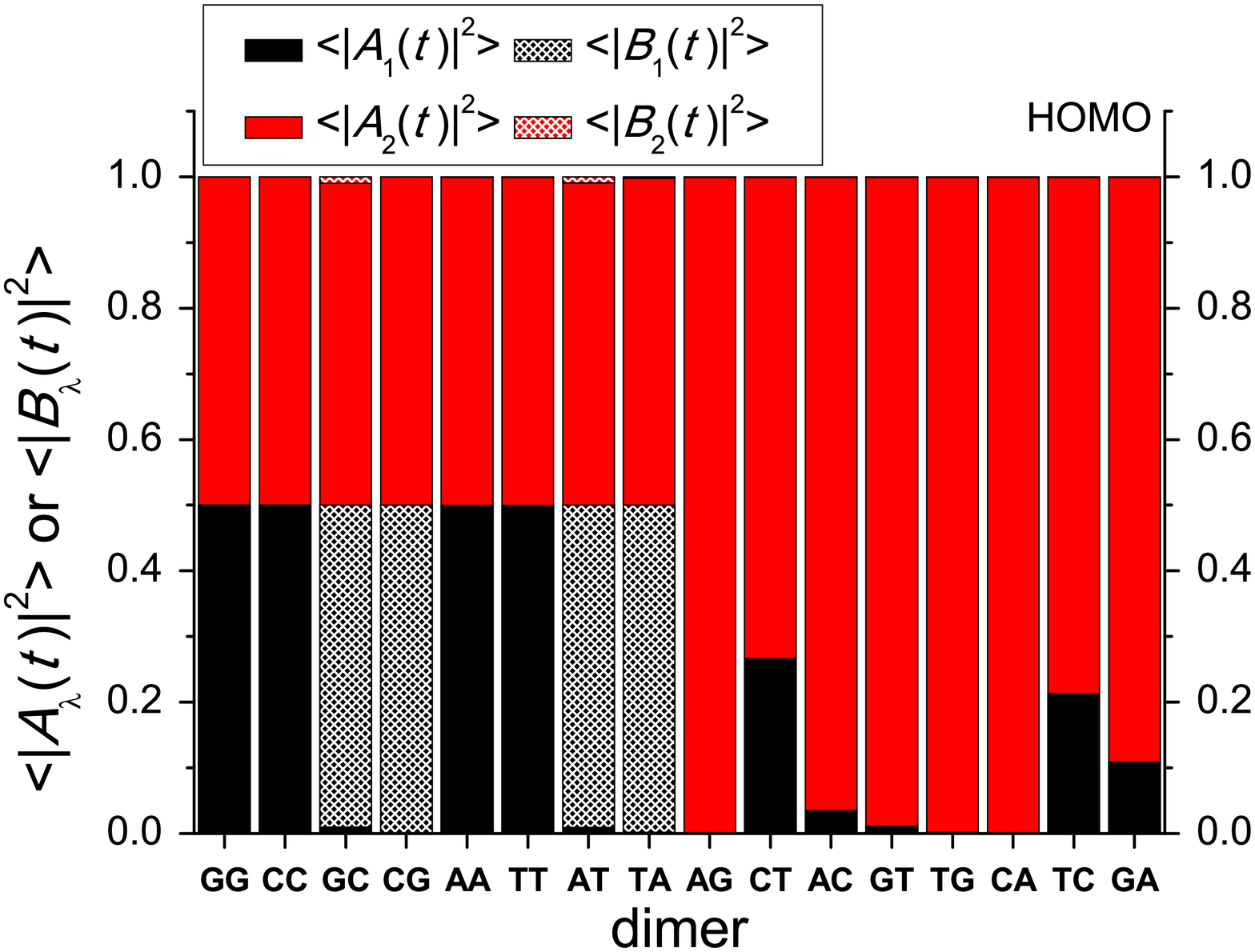}
\includegraphics[width=7.5cm]{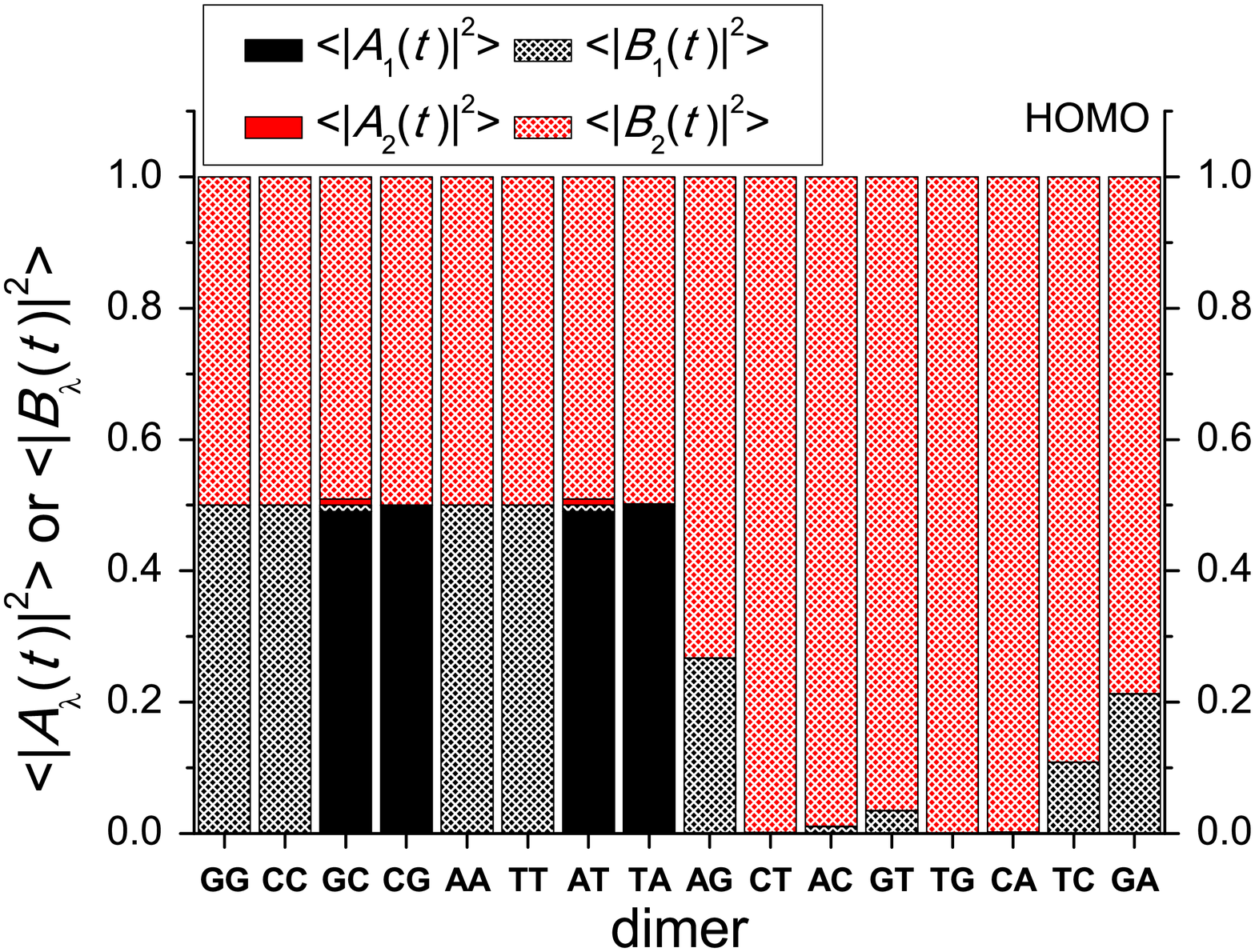}
\caption{The mean probabilities to find an extra hole at each base of a DNA dimer, as determined with TB approach II, the single-base approach.
The hole is initially placed at the 1st base (A1) (top left panel), the 2nd base (B1) (top right panel), the 3rd base (A2) (bottom left panel) or the 4th base (B2) (bottom right panel). We use the HKS parametrization~\cite{HKS:2010-2011}.}
\label{fig:DimersIIHOMO}
\end{figure}

\begin{figure} [h!]
\centering
\includegraphics[width=7.5cm]{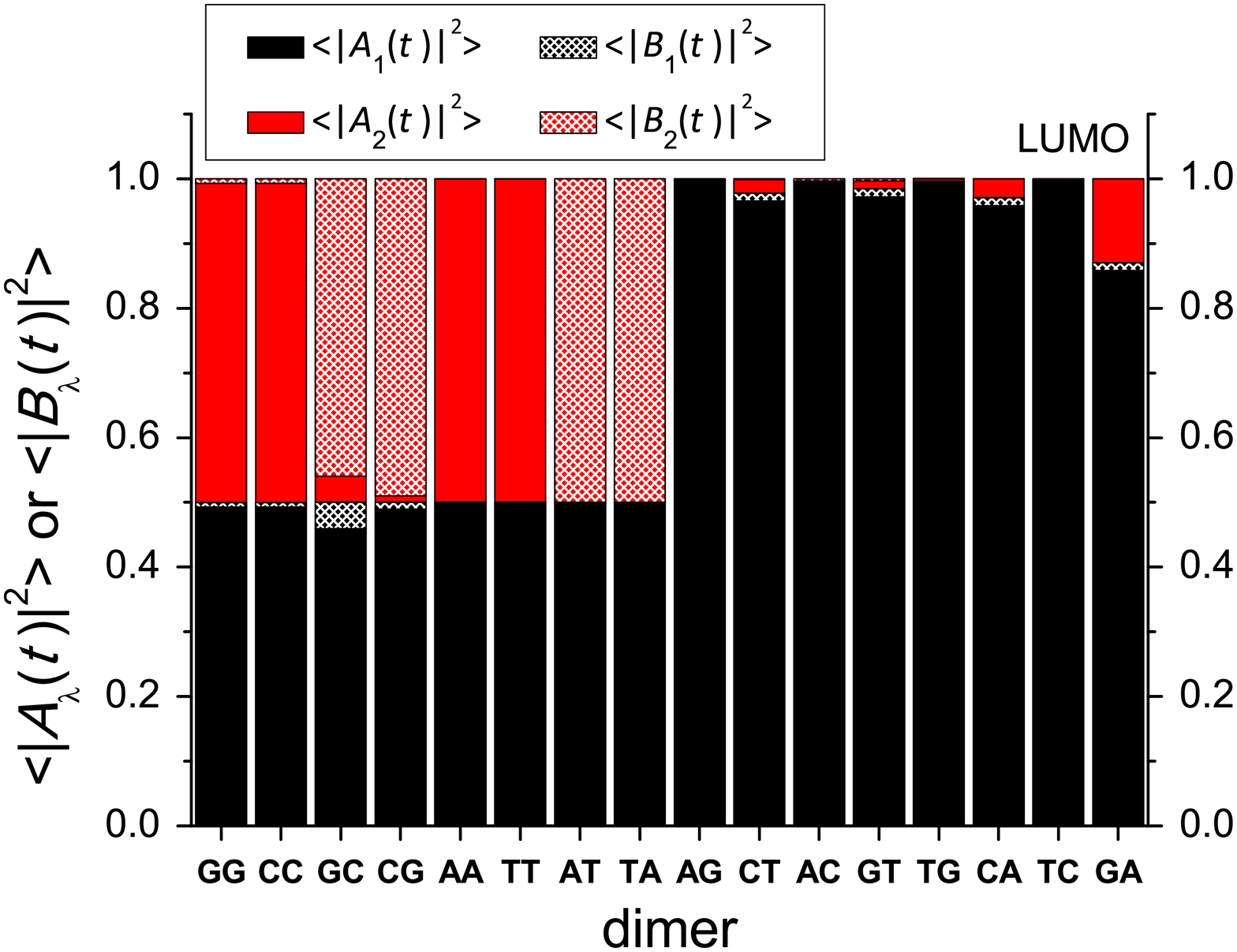}
\includegraphics[width=7.5cm]{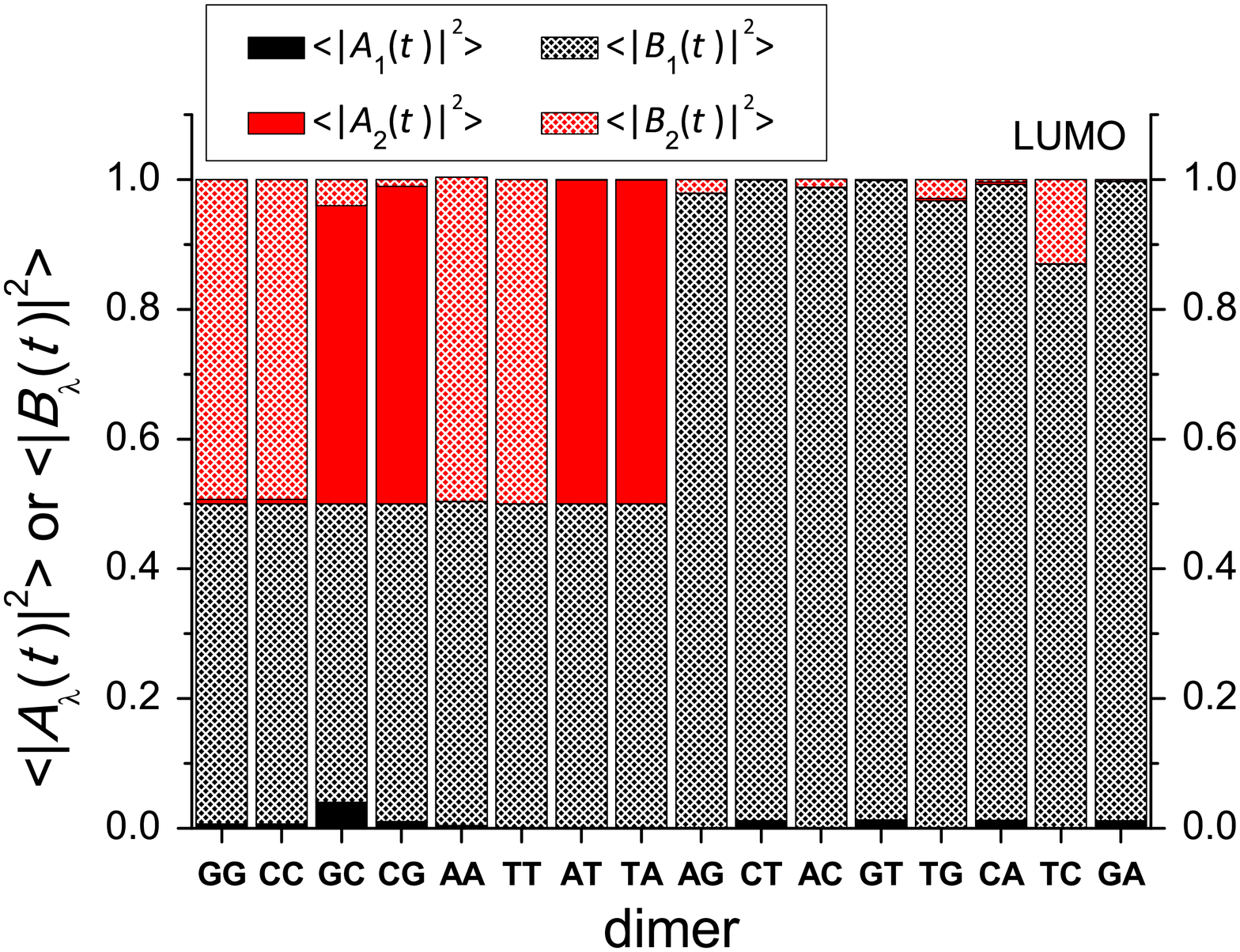}
\includegraphics[width=7.5cm]{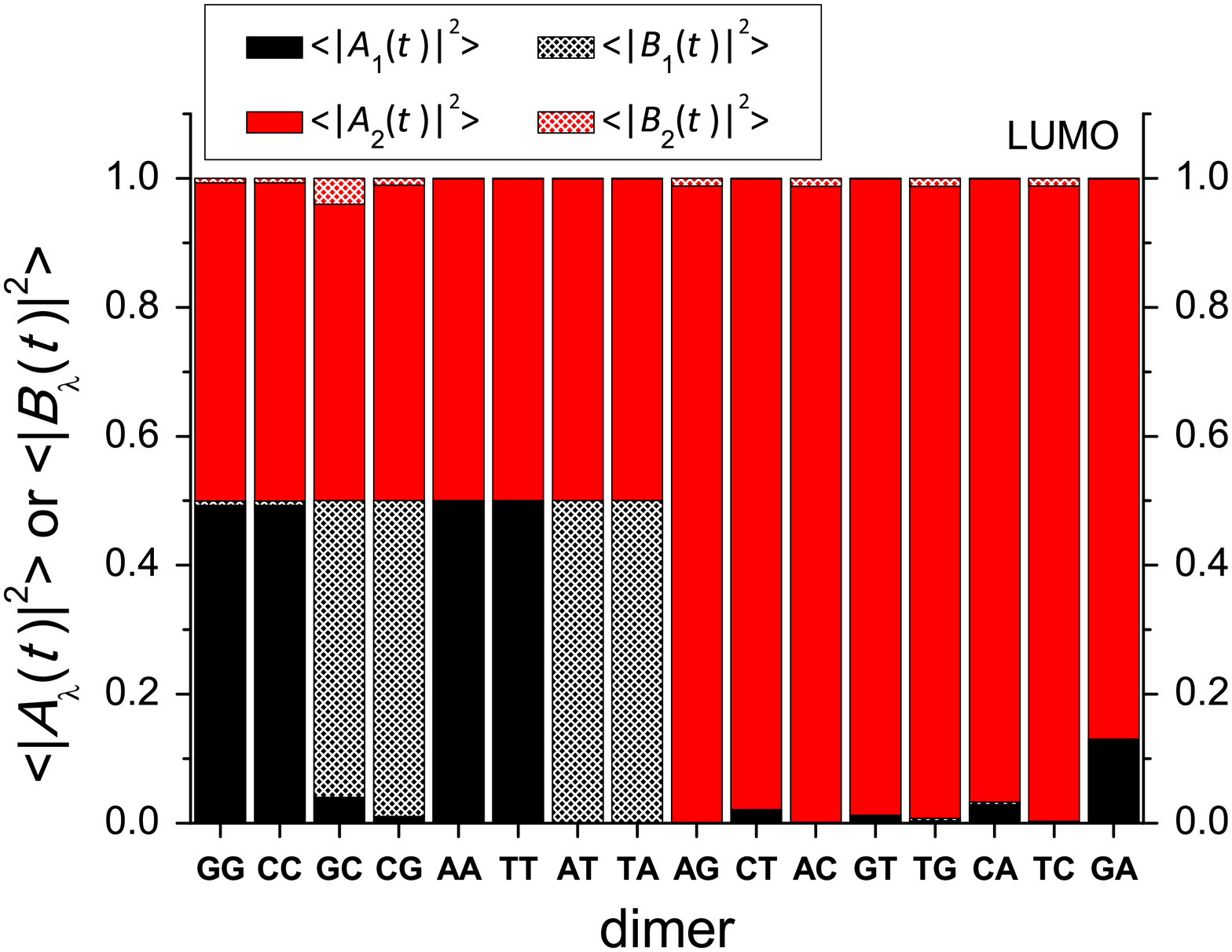}
\includegraphics[width=7.5cm]{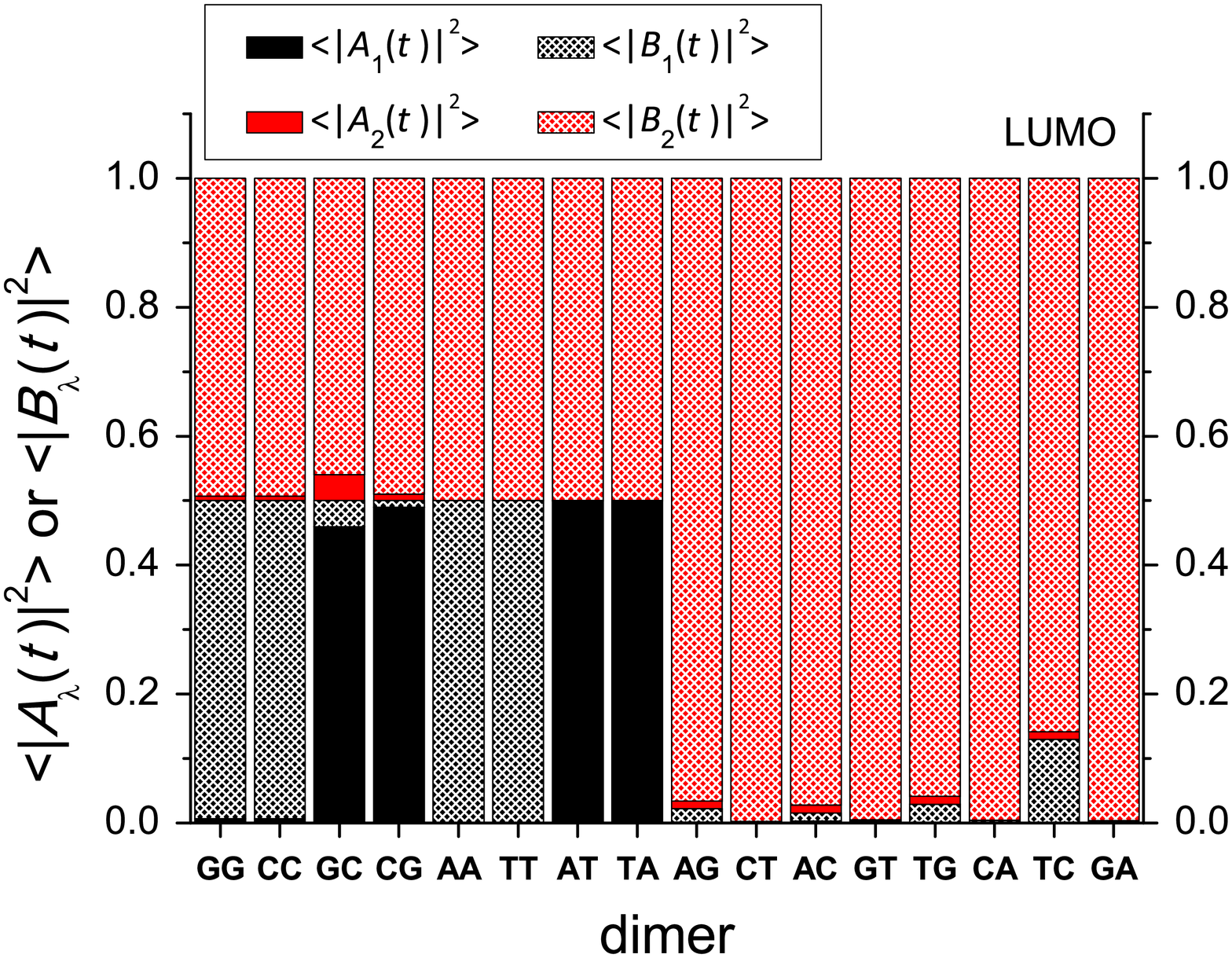}
\caption{The mean probabilities to find an extra electron at each base of a DNA dimer, as determined with TB approach II, the single-base approach.
The electron is initially placed at the 1st base (A1) (top left panel), the 2nd base (B1) (top right panel), the 3rd base (A2) (bottom left panel) or the 4th base (B2) (bottom right panel). We use the HKS parametrization~\cite{HKS:2010-2011}.}
\label{fig:DimersIILUMO}
\end{figure}

\begin{figure} [h!]
\centering
\includegraphics[width=7.5cm]{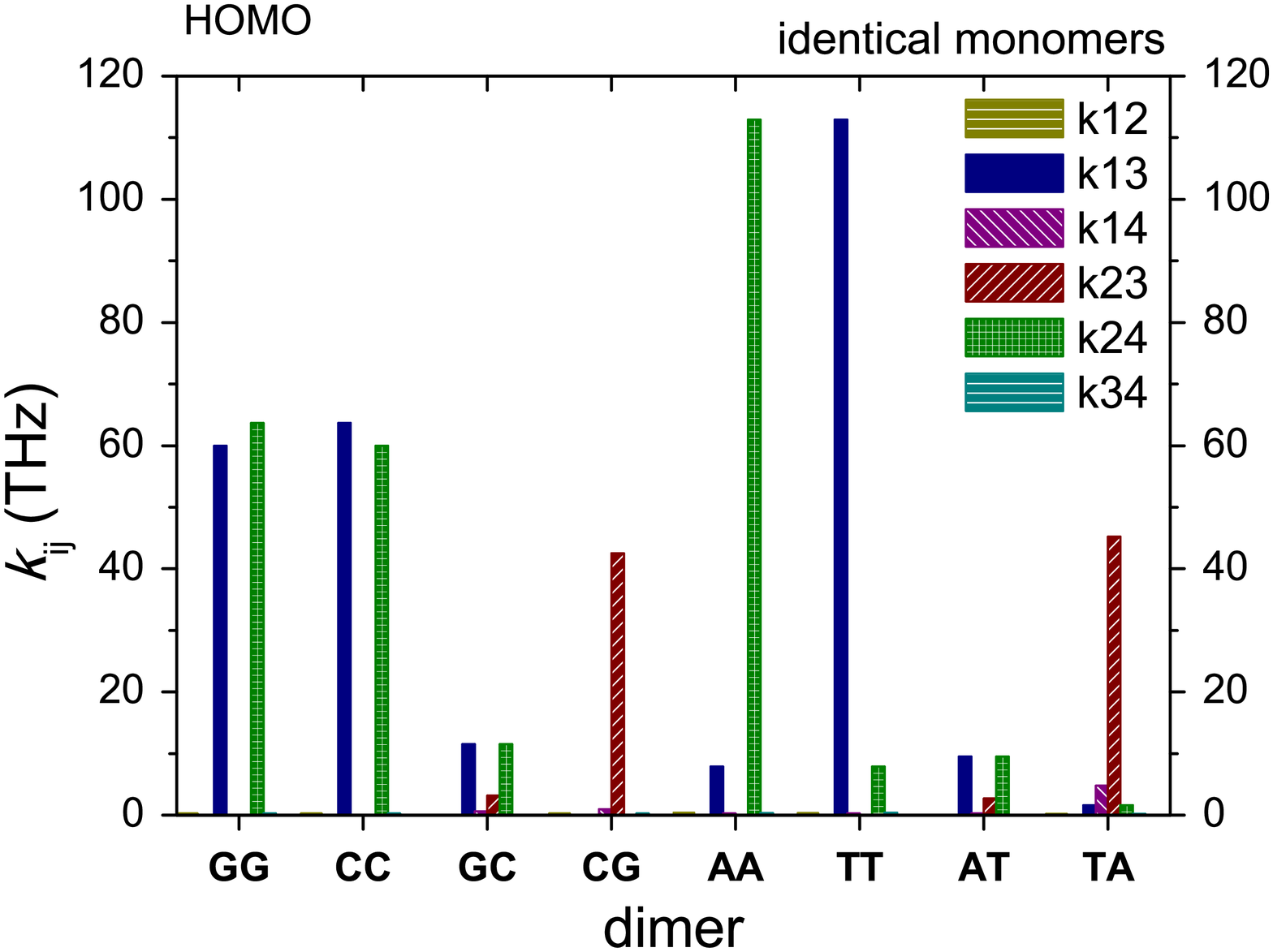}
\includegraphics[width=7.5cm]{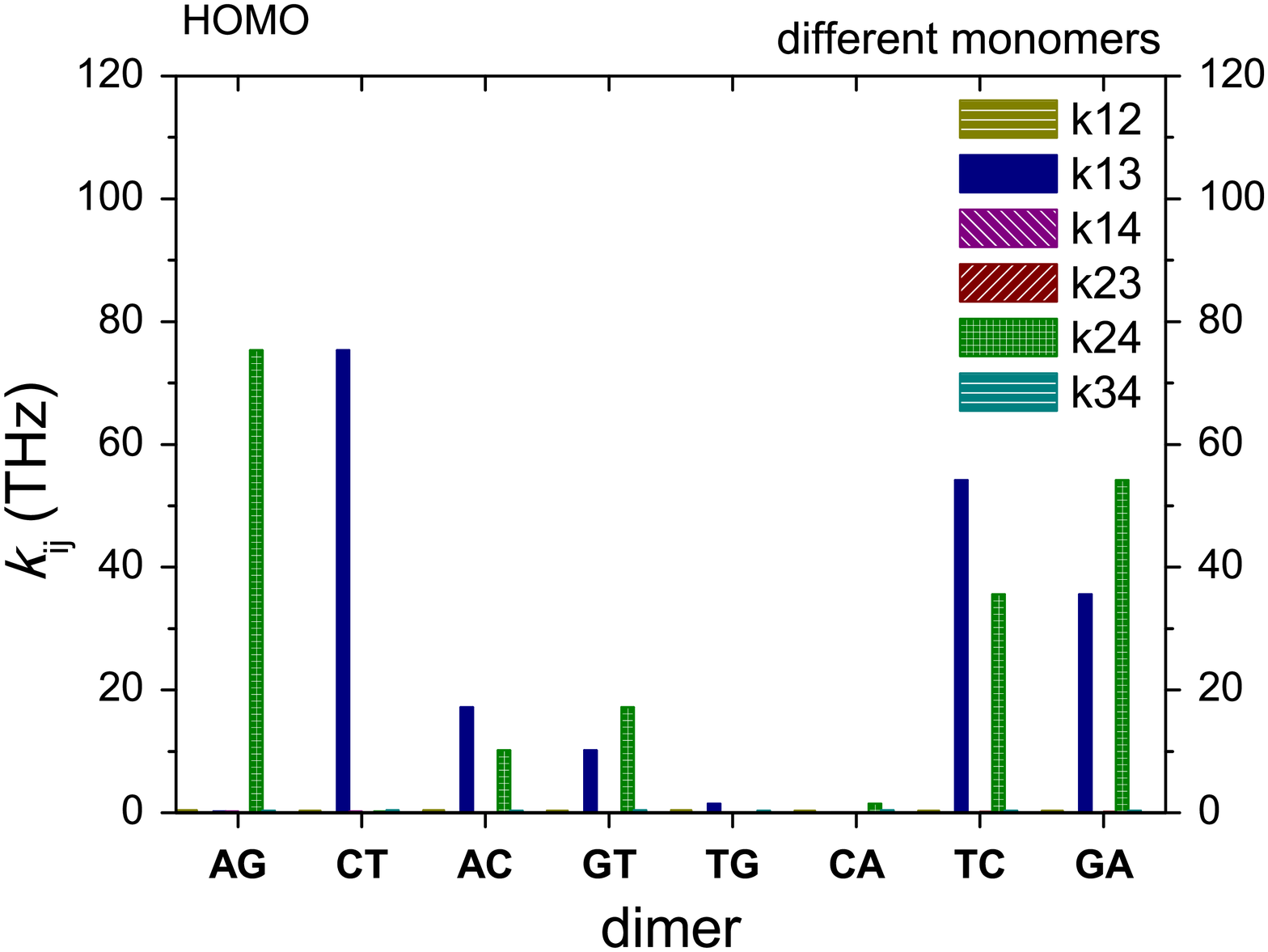}
\includegraphics[width=7.5cm]{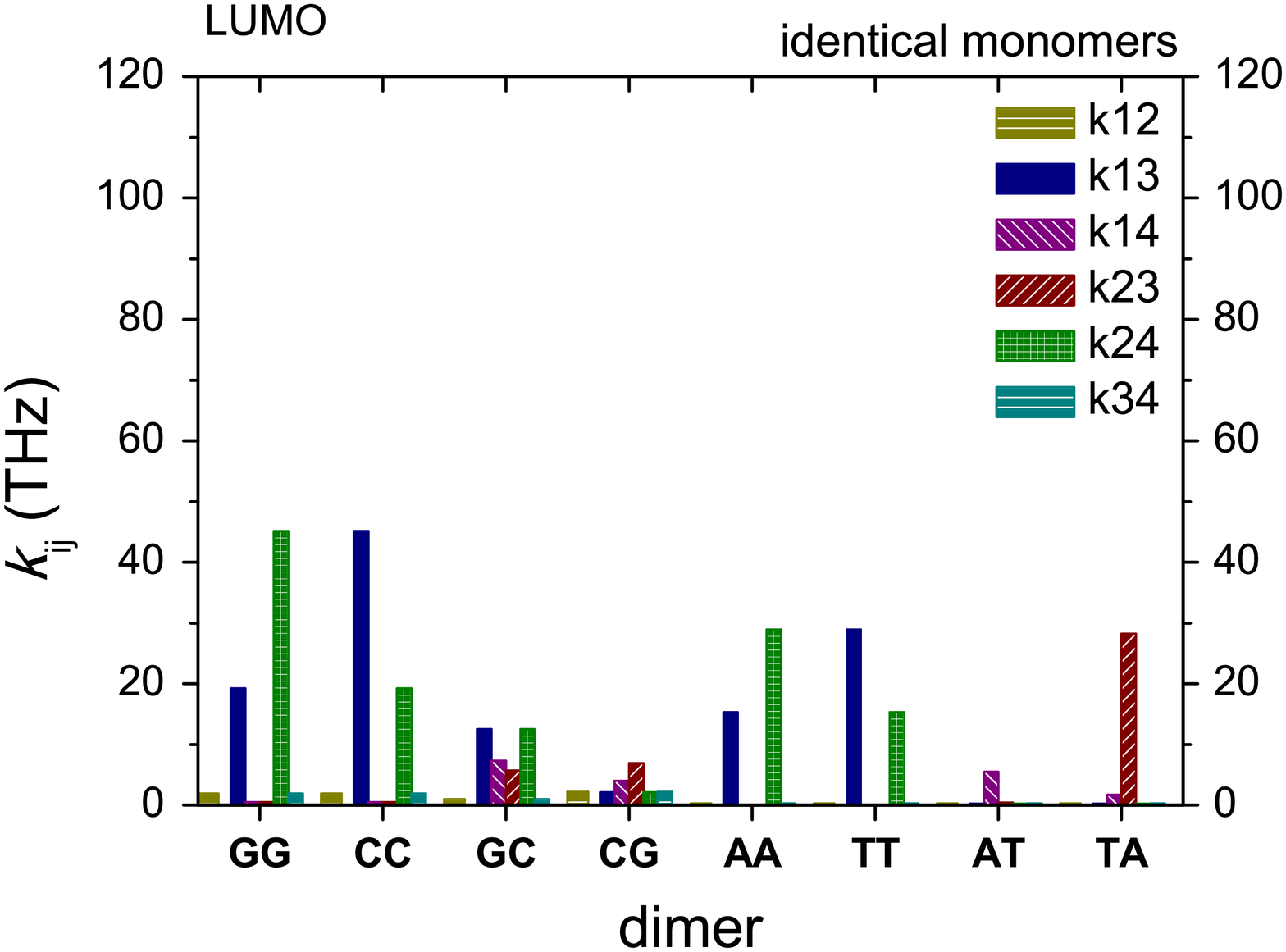}
\includegraphics[width=7.5cm]{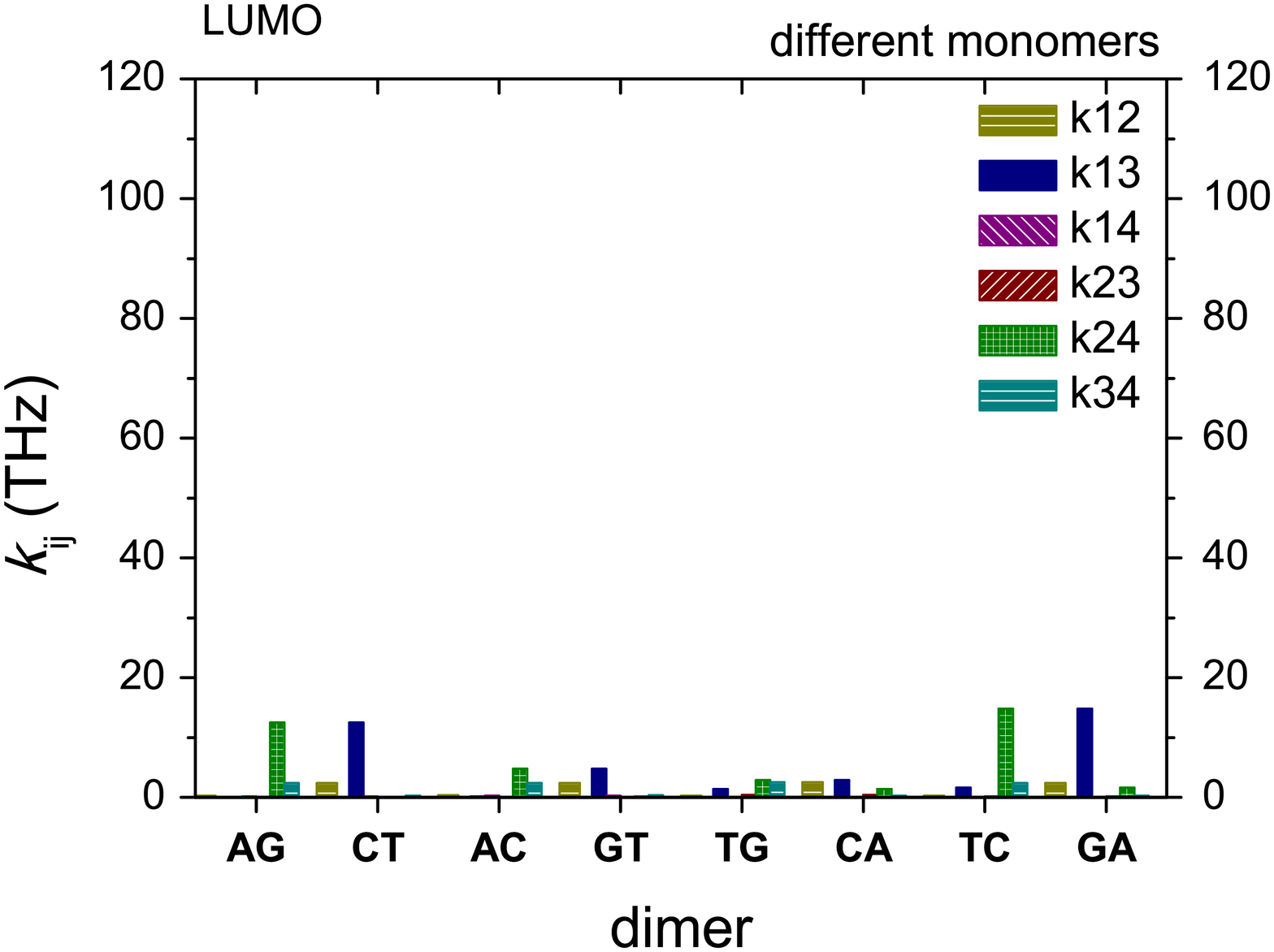}
\caption{Mean transfer rates $k_{ij}$ between bases $i$ and $j$ either for HOMO (hole transfer, 1st row) or for LUMO (electron transfer, 2nd row), for all dimers,
within TB approach II and the HKS parametrization~\cite{HKS:2010-2011}.
The 1st column corresponds to dimers made of identical monomers and
the 2nd row to dimers made of different monomers.}
\label{fig:DimersHOMOLUMOkijall}
\end{figure}

Let us now turn our discussion to carrier mean transfer rates $k_{ij}$, within TB II and the HKS parametrization~\cite{HKS:2010-2011}, cf. Fig.~\ref{fig:DimersHOMOLUMOkijall}. The specific values of $k_{ij}$ depend, of course, on the TB hopping parameters used~\cite{Kaklamanis:2015}. Let us call $k_{ij}$ the mean transfer rates of a dimer YX and $k^{\textrm{\scriptsize{compl}}}_{ij}$ the mean transfer rates of the equivalent dimer $\textrm{X}_{\textrm{\scriptsize{compl}}}\textrm{Y}_{\textrm{\scriptsize{compl}}}$. The following two properties hold:
\begin{equation}
k_{ij}=k_{ji},
\label{kij=kji}
\end{equation}
\begin{equation}
k_{ij}=k^{\textrm{\scriptsize{compl}}}_{(5-i)(5-j)}.
\label{k,k' relation for equivalent dimers}
\end{equation}
For dimers made of identical monomers with purine on purine (GG$\equiv$CC, AA$\equiv$TT),   hole transfer is almost entirely of intrastrand character i.e. it is along the $5'$-$3'$ or $3'$-$5'$ directions. Moreover, since $k_{13}$ and $k_{24}$ satisfy Eq.~\ref{k,k' relation for equivalent dimers}, in Fig.~\ref{fig:DimersHOMOLUMOkijall} we observe a symmetry in the alternation of colors for the couples of equivalent dimers GG$\equiv$CC and AA$\equiv$TT.
For dimers made of identical monomers with crosswise purines (CG, GC, TA, AT), there is significant diagonal hole transfer, and furthermore, for CG and TA the stronger hole transfer is along the $3'$-$3'$ direction. For dimers made of different monomers (AG$\equiv$CT, AC$\equiv$GT, TG$\equiv$CA, TC$\equiv$GA), hole transfer is almost exclusively of intrastrand character i.e. along the $5'$-$3'$ or $3'$-$5'$ directions; since $k_{13}$ and $k_{24}$ satisfy Eq. \ref{k,k' relation for equivalent dimers}, we observe the same symmetry in the alternation of colors. For the couple TG$\equiv$CA, although $k_{13}$ and $k_{24}$ are the biggest among all other $k_{ij}$, they are very small.
Electron transfer in dimers made of identical monomers with purine on purine (GG$\equiv$CC, AA$\equiv$TT) is qualitatively similar to hole transfer in such dimers. For dimers made of identical monomers with crosswise purines (GC, CG, AT, TA) electron transfer is slightly different than hole transfer in such dimers in the sense that diagonal channels are important but are not, quantitatively, identically important.
Electron transfer in dimers made of different monomers has a significant intrastrand character, but there is also intra-base-pair character in some cases. For the same reasons described above, we observe symmetry in color alternation for $k_{ij}$ of equivalent dimers.

\clearpage

\section{Trimers}\label{sec:ResultsTrimers} 
\begin{figure} [h!]
\centering
\includegraphics[width=7.5cm]{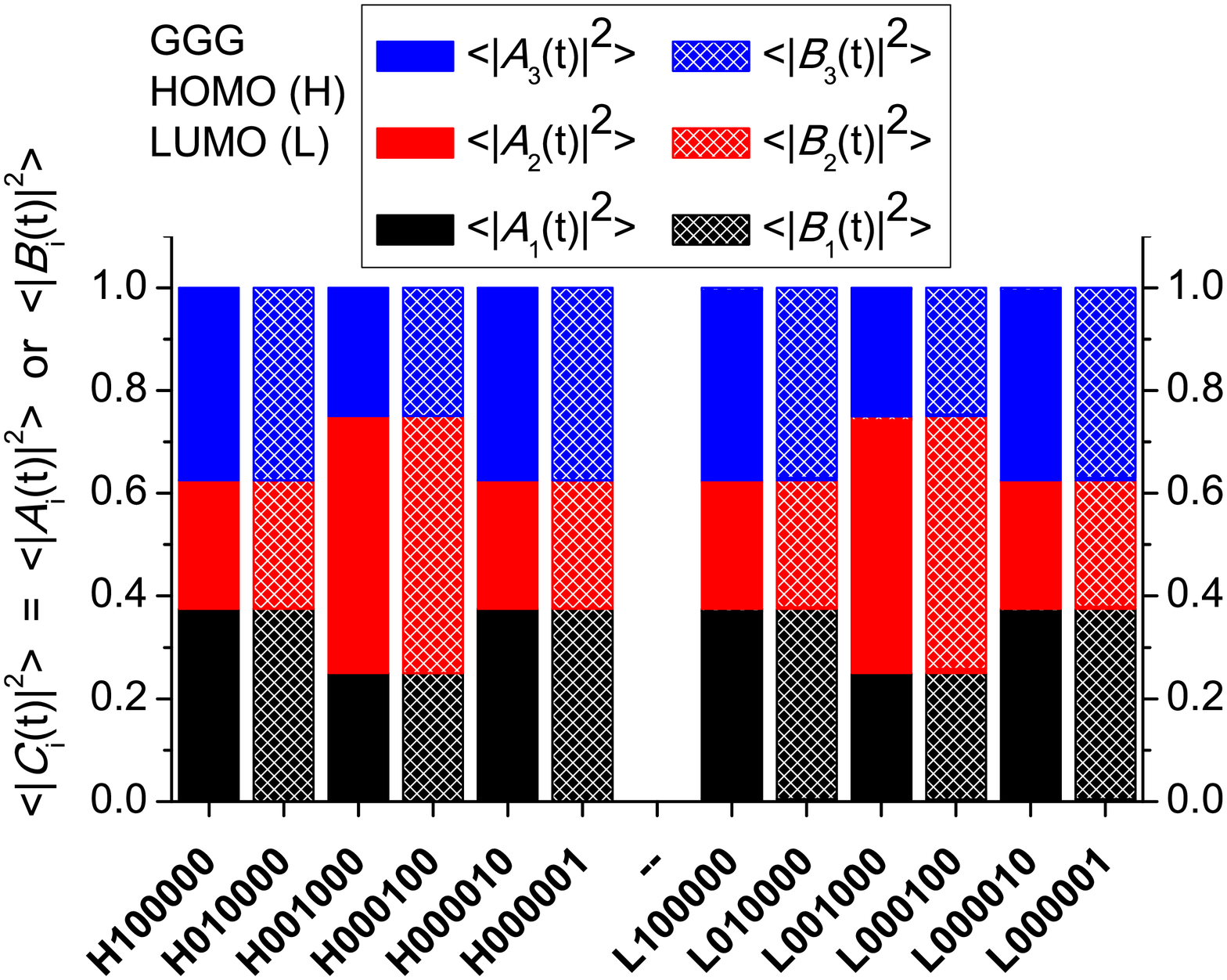}
\includegraphics[width=7.5cm]{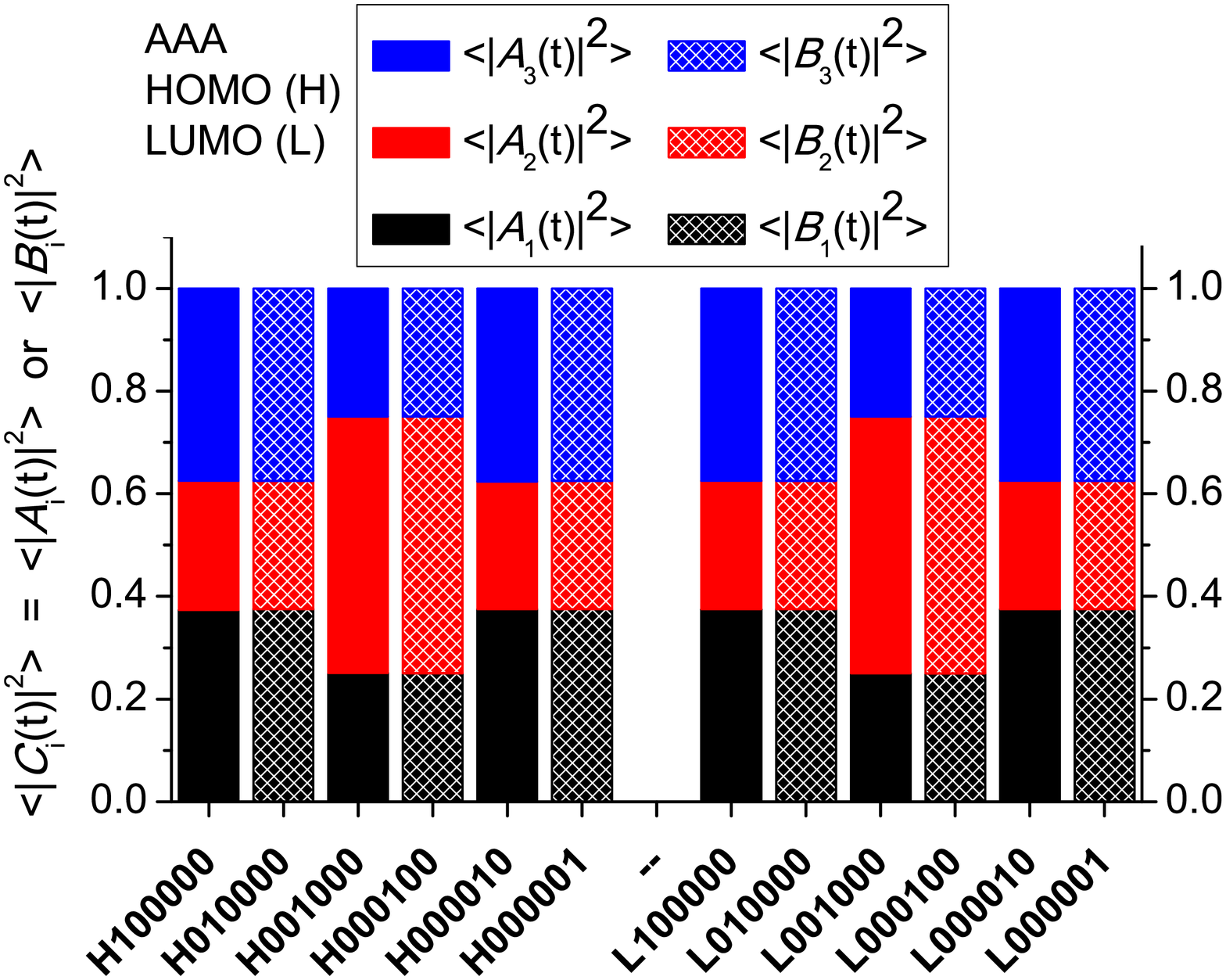}
\includegraphics[width=15cm]{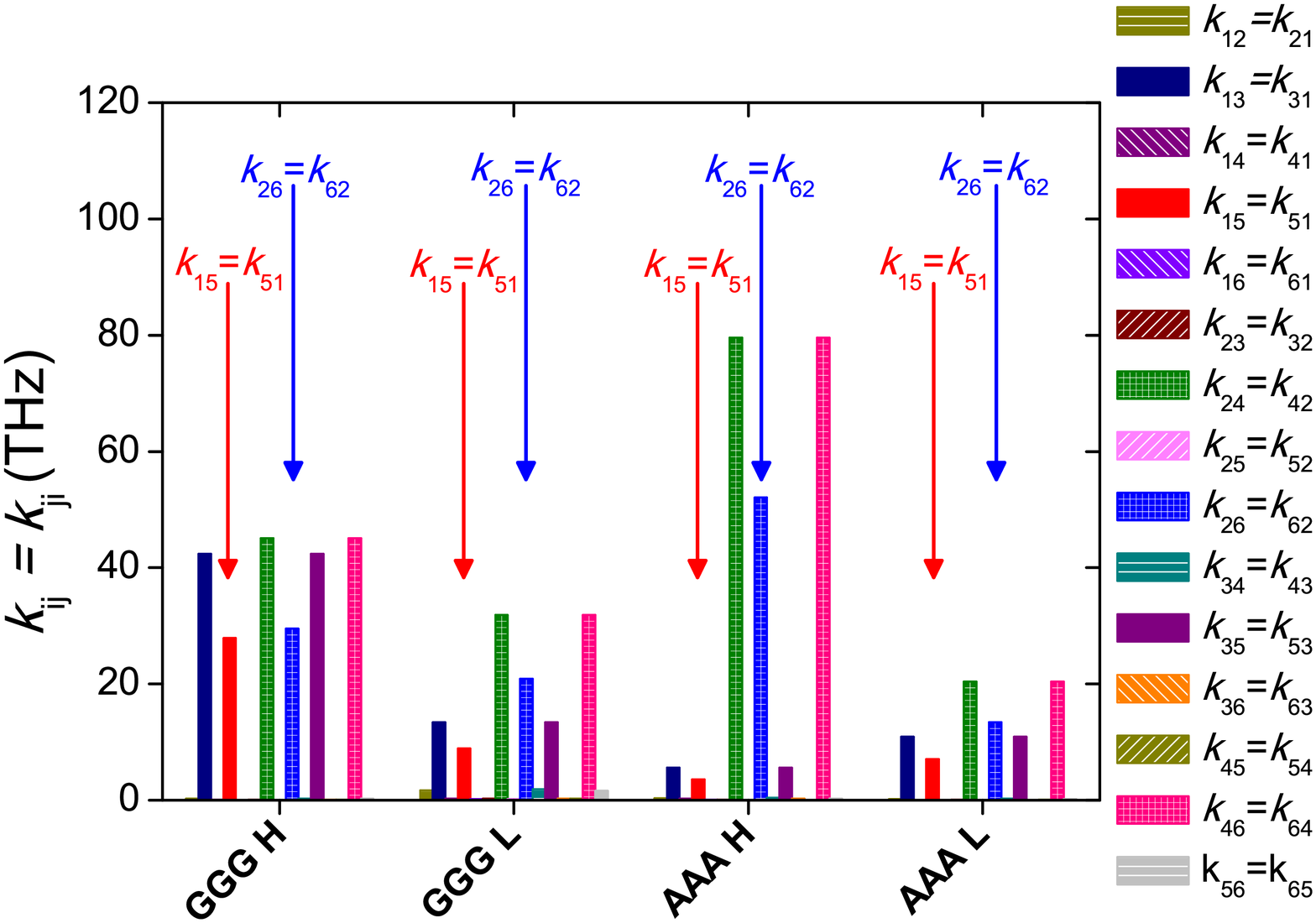}
\caption{GGG and AAA trimers within TB approach II and HKS parametrization~\cite{HKS:2010-2011}.
\textit{Upper panels:}
Mean probabilities to find the carrier at each base after having placed it initially at a particular base.
100000 means that the carrier was initially placed at base A1, 010000 at base B1 etc
(e.g. for GGG the bases are A1(G), B1(C), A2(G), B2(C), A3(G), B3(C)).
The carrier movement is almost exclusively of intrastrand character.
The probabilities to find the carrier at each base pair are either $\approx$ 0.375, 0.25, 0.375 or $\approx$ 0.25, 0.5, 0.25 depending on the initial placement of the carrier, which agrees remarkably with the rules established in Ref.~\cite{LChMKTS:2015} for TB I.
\textit{Lower panel:} The mean transfer rates $k_{ij}$ from base $i$ to base $j$.
The arrows indicate intrastrand transfer from the strand start to the strand end.}
\label{fig:TrimersGGGAAA}
\end{figure}

Let us now compare TB I and TB II in trimers made of identical monomers. Using TB I we proved~\cite{Simserides:2014,LKGS:2014} that in trimers made of identical monomers an extra carrier oscillates periodically with
\begin{equation}\label{fandTtrimers}
f = \frac{1}{T} = \frac{\sqrt{t^2 + t'^2}}{h},
\end{equation}
where $t, t'$ are the hopping integrals between the base pairs (when all purines are on the same strand, $t = t'$). With the parametrization of Refs.~\cite{Simserides:2014,LKGS:2014}, we found $f \approx$ 0.5-33 THz ($T \approx$ 30-2000 fs)~\cite{LKGS:2014}; with the parametrization of Ref.~\cite{HKS:2010-2011}, we find $f \approx$ 0.5-21 THz ($T \approx$ 48-2000 fs).
In other words, for trimers made of identical monomers,
the frequency range is narrower than for dimers.
For 0 times crosswise purines, the maximum transfer percentage $p = 1$, while for 1 or 2 times crosswise purines $p < 1$  \cite{Simserides:2014,LKGS:2014}.
TB II in trimers, generally, does not allow one to strictly determine periodicity;
$T, f, p$, and $pf$ cannot be defined.
However, Fourier analysis shows similar frequency content.
(Specific examples are given in Appendix C, where we present Fourier analysis of hole oscillations in two trimers made of identical monomers:
GGG in Fig.~\ref{fig:FourierGGG} and
GCG in Fig.~\ref{fig:FourierGCGandCACandCTC}.)
For hole transfer in GGG, within TB I and the parametrization of Refs.~\cite{Simserides:2014,LKGS:2014}, we found $f \approx $ 34.2 THz;
now with the HKS parametrization~\cite{HKS:2010-2011}, we find $f \approx $ 21.2 THz.
This is in remarkable agreement with the frequencies obtained by Fourier Transform,
within TB II and HKS parametrization~\cite{HKS:2010-2011} shown in Fig.~\ref{fig:FourierGGG}.
Specifically, e.g. for initial placement of the hole at base A1(G), the main frequencies are around 21.2 THz (a double peak) and 42.4 THz (a single peak), while, e.g. for initial placement of the hole at base B1(C), the main frequencies are  around 22.5 THz (a double peak) and 45 THz (a single peak).
In other words, for hole transfer in GGG, within the HKS parametrization, the period predicted by TB I agrees with the approximate period predicted by TB II.
The mean probabilities to find an extra carrier at a base and the mean transfer rates in GGG and AAA, within TB II and HKS parametrization~\cite{HKS:2010-2011},
are shown in Fig.~\ref{fig:TrimersGGGAAA}.
We observe that the carrier movement is almost exclusively of intrastrand character,
a fact also evident from the Fourier analysis in Fig.~\ref{fig:FourierGGG}.
Remarkably, with TB II, the probabilities to find the carrier at each base pair are either $\approx$ 0.375, 0.25, 0.375 or $\approx$ 0.25, 0.5, 0.25 depending on the initial placement of the carrier, in agreement with the rules established in Ref.~\cite{LChMKTS:2015} for TB I.
The mean transfer rates $k_{ij}$ confirm the intrastrand character of charge transfer in GGG and AAA.
For hole transfer in GCG, within the HKS parametrization~\cite{HKS:2010-2011},
TB I and TB II give similar results, indicating rather weak transfer.
For example,
placing the hole initially at the first base pair for TB I or placing the hole initially at the first base for TB II, the probability to find the hole at the first base pair is $\approx$ 0.9990 for TB I and 0.9848 for TB II and at the last base pair $\approx$ 0.0008 for TB I and 0.0006 for TB II. This is  mirrored in the very small Fourier amplitudes for GCG (Appendix C, Fig.~\ref{fig:FourierGCGandCACandCTC}).

Within TB I, for trimers made of different monomers, carrier movement may be non-periodic~\cite{Simserides:2014,LKGS:2014}. Within TB I, generally, increasing the number of monomers above three, the system becomes more complex and periodicity is lost~\cite{LKGS:2014}; even in the simplest cases, e.g. tetramers made of identical monomers with all the purines  on the same strand, there is no periodicity~\cite{Lambropoulos:2014}. Within TB II, as mentioned before, in trimers, generally, periodicity cannot be strictly determined. As a last point, we restrict ourselves in giving (Appendix C, Fig.~\ref{fig:FourierGCGandCACandCTC}) two examples of the frequency content of hole oscillations in trimers made of different monomers, specifically in CAC and CTC.

\section{Conclusion}   
\label{sec:Conclusion} 
Using two TB approaches,
a wire model called here TB I, where a site coincides with a DNA base pair, and
an extended (i.e. including also diagonal hoppings) ladder model called here TB II,
where a site coincides with a DNA base,
we demonstrated that THz oscillations in DNA monomers, dimers and trimers exist.
We studied various aspects of the effect, e.g.
frequency content,
maximum transfer percentages and transfer rates between sites, and
mean probabilities to find the carrier at a particular site.
We also compared successfully the two TB approaches.
Naturally, TB II allows for greater detail.

For DNA monomers, i.e. for adenine-thymine and guanine-cytosine, with TB II,
we predicted electron or hole oscillations in the range
$f \approx$ 50-550 THz ($T \approx$ 2-20 fs), i.e. $\lambda \approx$ 545 nm - 6 $\mu$m, from visible to near- and mid infrared~\cite{ISO20473}. We found that the maximum transfer percentage $p$ and the pure maximum transfer rate $pf$ between the bases are very small.

For DNA dimers, with TB I, we predicted electron or hole oscillations in the range $f \approx$ 0.25-100 THz ($T \approx $ 10-4000 fs) i.e. $\lambda \approx$ 3-1200 $\mu$m, approximately in the mid- and far-infrared. For dimers made of identical monomers the maximum transfer percentage $p = 1$, but for dimers made of different monomers $p < 1$. With TB II, the carrier oscillations are not strictly periodic but the frequency content is similar to that predicted with TB I. For the mean probabilities to find the carrier at a particular site, the two approaches give coherent, complementary results.
TB II shows that for dimers made of identical monomers, when purines are crosswise to purines, interstrand carrier transfer dominates, i.e. we have significant diagonal transfer, justifying the inclusion of diagonal hoppings in our model, while if purines are on the same strand, intrastrand carrier transfer dominates. For dimers made of different monomers, we carrier transfer is mainly intrastrand character but the transfer percentage is small.

With TB I, for trimers made of identical monomers, the carrier oscillates periodically with $f \approx$ 0.5-33 THz ($T \approx$ 30-2000 fs) if we use the parametrization of Refs.~\cite{Simserides:2014,LKGS:2014}.
With the HKS parametrization~\cite{HKS:2010-2011}, $f \approx$ 0.5-21 THz.
For 0 times crosswise purines $p = 1$, for 1 or 2 times crosswise purines $p < 1$.
With TB II, the carrier oscillations are not strictly periodic but the frequency content is similar to that predicted with TB I. For the mean probabilities to find the carrier at a particular site, the two approaches give coherent, complementary results.

Finally, we would like to mention that increasing the number of monomers,
i.e. constructing an oligomer or a polymer, the frequency spectrum becomes more fragmented and moves towards lower frequencies.
A systematic study of longer segments is beyond the scope of the present manuscript.
It seems that a source or receiver of electromagnetic radiation
made of DNA monomers, dimers or trimers,
with frequencies from fractions of THz to just below PHz,
could be envisaged. \\

\noindent \textbf{* Related Work at} \\
\verb"http://users.uoa.gr/~csimseri/physics_of_nanostructures_and_biomaterials.html" \\

\noindent \textbf{Acknowledgements}
A. Morphis thanks the State Scholarships Foundation-IKY for a Ph.D. research scholarship
via ``IKY Fellowships of Excellence'', Hellenic Republic-Siemens Settlement Agreement.
M. Tassi thanks the State Scholarships Foundation-IKY for a post-doctoral research fellowship
via ``IKY Fellowships of Excellence'', Hellenic Republic-Siemens Settlement Agreement. \\

\clearpage

\appendix{\noindent \textbf{Appendix A}}\label{AA}
According to TB I [description at the base-pair level] the HOMO or LUMO wave function of a given DNA segment,
made of $N$ base pairs, $\Psi_{DNA}({\bf r},t)$, is considered as a linear combination of the base-pair wave functions $\Psi_{bp}^{\mu}({\bf r})$
with time-dependent coefficients
\begin{equation}\label{bplSchroedinger}
\Psi_{DNA}({\bf r},t) = \sum_{\mu=1}^N A_{\mu}(t) \; \Psi_{bp}^{\mu}({\bf r}).
\end{equation}
$|A_{\mu}(t)|^2$ gives the probability of finding the carrier (hole for HOMO, electron for LUMO) at base pair $\mu$.
Using the time-dependent Schr\"{o}dinger equation,
\begin{equation}\label{tdse}
i \hbar \frac{\partial \Psi_{DNA}({\bf r},t)}{\partial t} =H_{DNA}\Psi_{DNA}({\bf r},t),
\end{equation}
we find that
the time evolution of the coefficients $A_{\mu}(t)$ obeys the system of equations~\cite{HKS:2010-2011}
\begin{equation}\label{bpequations}
i \hbar \frac{dA_{\mu}} {dt} = E^{\mu} A_{\mu} + t^{\mu,\mu-1} A_{\mu-1} + t^{\mu,\mu+1} A_{\mu+1},
\end{equation}
where $E^{\mu}, \; \mu=1, 2, ...N$ are the HOMO or LUMO on-site energies of the base pairs, and $t^{\mu,\mu'}$ are the HOMO or LUMO hopping integrals between two nearest neighbouring base pairs $\mu$ and $\mu'$.

According to TB II [description at the single-base level] $\Psi_{DNA}({\bf r},t)$ is derived from the single-base wave functions, according to the expression
\begin{equation} \label{blSchroedinger}
\Psi_{DNA}({\bf r}) = \sum_{\mu=1}^{N}[A_{\mu}(t) \Psi_{b}^{\mu,1}({\bf r})  + B_{\mu}(t) \Psi_{b}^{\mu,2}(\bf r)]
\end{equation}
where $\Psi_{b}^{\mu,\sigma}, \sigma=1,2$ and $\mu=1,2,..N$, is the wave function of the base at the $\mu$-th base pair and in the $\sigma$-th strand.
$|A_{\mu}(t)|^2$, $|B_{\mu}(t)|^2$ give the probability to find the carrier at the base of strand 1 and 2, respectively, of the $\mu$-th base pair.
Again, using the time-dependent Schr\"{o}dinger equation i.e. Eq.~\ref{tdse}, we find that
in this case, the system of equations is~\cite{HKS:2010-2011}
\begin{eqnarray} \label{bequations-strand1}
i\hbar\frac{dA_{\mu}}{dt} \! = \! E^{\mu,1} \! A_{\mu} +
t^{\mu,1;\mu,2} \! B_\mu + t^{\mu,1;\mu-1,1} \! A_{\mu-1} + \\ \nonumber
t^{\mu,1;\mu+1,1} \! A_{\mu+1} + t^{\mu,1;\mu-1,2} \! B_{\mu-1}
+ t^{\mu,1;\mu+1,2} \! B_{\mu+1},
\end{eqnarray}
\begin{eqnarray} \label{bequations-strand2}
i\hbar\frac{dB_{\mu}}{dt} \! = \! E^{\mu,2} \! B_{\mu} +
t^{\mu,2;\mu,1} \! A_\mu + t^{\mu,2;\mu-1,2} \! B_{\mu-1} + \\  \nonumber
t^{\mu,2;\mu+1,2} \! B_{\mu+1} + t^{\mu,2;\mu-1,1} \! A_{\mu-1} +
t^{\mu,2;\mu+1,1} \! A_{\mu+1}.
\end{eqnarray}
$E^{\mu,\sigma}$ are the HOMO or LUMO on-site energies of the base at the $\mu$-th base pair and in the $\sigma$-th strand, and $t^{\mu,\sigma;\mu',\sigma'}$ are the HOMO or LUMO hopping parameters between neighbouring bases, i.e. between
(a) two successive bases in the same strand,
(b) complementary bases that define a base pair, and
(c) diagonally located bases of successive base pairs.

To determine the temporal and spatial evolution of electrons or holes along a $N$ base-pair DNA segment, we solve the system of (I) $N$ or (II) $2N$ coupled differential equations i.e. Eq.~\ref{bpequations} or Eqs.~\ref{bequations-strand1}-\ref{bequations-strand2}, respectively, with the eigenvalue method, which is explained below. TB II allows us to examine the system in higher detail than with TB I, but TB I uses a smaller number of parameters and it is readily implemented.

To solve Eq.~\ref{bpequations} for TB I or Eqs.~\ref{bequations-strand1}-\ref{bequations-strand2} for TB II, we define the vector matrix
\begin{equation}\label{x}
\hspace{-2cm}
\vec{x}(t) = \left[
\begin{array}{c}
A_1(t) \\
A_2(t) \\
\vdots \\
A_N(t)  \end{array} \right] \; \textrm{ for TB I \quad \textit{or} } \quad
\vec{x}(t) =
\left[
\begin{array}{c}
A_1(t) \\
B_1(t) \\
\vdots \\
A_N(t) \\
B_N(t) \end{array} \right] \; \textrm{ for TB II}
\end{equation}
and therefore Eq.~\ref{bpequations} for TB I or Eqs.~\ref{bequations-strand1}-\ref{bequations-strand2} for TB II become
\begin{equation}\label{xdotmathcalAx}
\dot{\vec{x}}(t) = \widetilde{\mathcal{A}} \vec{x}(t),
\end{equation}
where we can define
\begin{equation}\label{mathcalA}
\widetilde{\mathcal{A}} = - \frac{i}{\hbar} \textrm{A}.
\end{equation}
$\mathcal{A}$ is the TB hamiltonian matrix.
We solve Eq.~\ref{xdotmathcalAx} via the {\it eigenvalue method}, i.e., we look for solutions of the form
\begin{equation}
\vec{x}(t) = \vec{v} e^{\tilde{\lambda} t} \Rightarrow \dot{\vec{x}}(t) = \tilde{\lambda} \vec{v} e^{\tilde{\lambda} t}.
\end{equation}
Hence, Eq.~\ref{xdotmathcalAx} reads
\begin{equation}\label{eigenvalueproblem1}
\widetilde{\mathcal{A}} \vec{v} = \tilde{\lambda} \vec{v},
\end{equation}
or
\begin{equation}\label{eigenvalueproblem2}
\textrm{A} \vec{v} = \lambda \vec{v},
\end{equation}
\begin{equation}
\tilde{\lambda} = - \frac{i}{\hbar} \lambda.
\end{equation}
In other words, we have to solve an eigenvalue problem.
The order of the matrices $\widetilde{\mathcal{A}}$ or $\textrm{A}$ is $N$ for TB I or $2N$ for TB II.
The matrices $\textrm{A}$ for TB I and TB II are shown in Appendix B.
Already, we called $\mu$ the base-pair index i.e. $\mu= 1, 2, \dots, N$ and $\sigma$ the strand index i.e. $\sigma=1$ or $\sigma=2$.
Then, the base index $\beta(\mu,\sigma) = 2(\mu-1)+\sigma$ and $\beta = 1, 2, 3, 4, \dots 2N-1, 2N$. Schematically,
\begin{eqnarray}
\mu \qquad & \sigma \qquad & \beta \nonumber \\
 1  & 1      & 1 \nonumber \\
 1  & 2      & 2 \nonumber \\
 2  & 1      & 3 \nonumber \\
 2  & 2      & 4 \nonumber \\
 \vdots & \vdots & \vdots \nonumber \\
\label{indices}
\end{eqnarray}
Having checked that the normalized eigenvectors $\vec{v}_k$ corresponding to the eigenvalues $\lambda_k$ of Eq.~\ref{eigenvalueproblem2} are linearly independent, the solution of our problem is
\begin{equation}\label {xsolution}
\vec{x}(t) = \sum_{k=1}^{MD} c_k \vec{v}_k e^{-\frac{i}{\hbar} \lambda_k t}.
\end{equation}
where $MD$ is the \textrm{A} matrix dimension, i.e., $N$ for TB I or $2N$ for TB II.

If we initially place the carrier at the first base pair (TB I) or
if we initially place the carrier at the first base pair either at strand 1 or at strand 2 (TB II), then, the initial conditions would be
\begin{equation}\label{x0}
\hspace{-2.5cm}
\vec{x}(0) \! = \! \left[
\begin{array}{c}
A_1(0) \\
A_2(0) \\
\vdots \\
A_N(0)  \end{array} \right]
\! = \!
\left[
\begin{array}{c}
1 \\
0 \\
\vdots \\
0  \end{array} \right] \; \textrm{(TB I)} \quad \textit{or} \quad
\vec{x}(0) \! = \! \left[
\begin{array}{c}
A_1(0) \\
B_1(0) \\
\vdots \\
A_N(0) \\
B_N(0)  \end{array} \right]
\! = \!
\left[
\begin{array}{c}
1 \\
0 \\
\vdots \\
0 \\
0  \end{array} \right] \textrm{or}
\left[
\begin{array}{c}
0 \\
1 \\
\vdots \\
0 \\
0  \end{array} \right] \; \textrm{(TB II)},
\end{equation}
respectively. Other initial conditions could be similarly defined.
From the initial conditions we determine the coefficients $c_k$.
Using $i = 1, 2, \dots, MD$ as a generic index for either $\mu$ in TB I or $\beta$ in TB II, for initial placement of the extra carrier at a site $i$,
we can show analytically that $c_k = v_{ik}$,
where $v_{ik}$ is the $i$-th component of the eigenvector $\vec{v}_k$.
Generally, the Hamiltonians describing equivalent $N$-mers are related with a similarity transformation, hence, they have the same eigenvalues and their eigenvectors are connected by
$v_{ik}^\textrm{\scriptsize{$N$-mer}} =
v_{(MD-i+1)k}^\textrm{\scriptsize{equiv $N$-mer}}$.

In TB I, the mean over time probability to find the extra carrier
at base pair $\mu$ is $\langle |A_{\mu}(t)|^2 \rangle$, while,
in TB II, the mean over time probability to find the extra carrier
at the 1st strand base of base pair $\mu$ is $\langle |A_{\mu}(t)|^2 \rangle$ and
at the 2nd strand base of base pair $\mu$ is $\langle |B_{\mu}(t)|^2 \rangle$).
Using $C_{i}(t)$, $i = 1, 2, \dots, MD$, as a generic symbol for either $A_{\mu}(t)$ in TB I or $A_{\mu}(t)$ and $B_{\mu}(t)$ in TB II, we can show analytically that
the mean over time probability to find the carrier at a site $i$ is given by
\begin{equation}\label{meanprobability}
\langle |C_{i}(t)|^2 \rangle = \sum_{k=1}^{MD} c_k^2 v_{ik}^2.
\end{equation}

Furthermore, supposing that the eigenvalues are arranged in ascending order, i.e.
$\lambda_1 < \lambda_2 < \dots < \lambda_{MD}$,
the frequencies or periods involved in charge transfer are
\begin{equation}\label{periodsorfrequencies}
f_{kk'} = \frac{1}{T_{kk'}} = \frac{\lambda_k - \lambda_{k'}}{h}, \quad \forall k > k'.
\end{equation}
We can analytically show that the one-sided Fourier spectrum for the probability $|C_{i}(t)|^2$ is given by
\begin{equation}\label{FourierSpectra}
|\mathcal{F}_i(f)| = \sum_{k=1}^{MD} c_k^2 v_{ik}^2 \delta(f)+ 2 \sum_{k'=1}^{MD} \sum_{k > k'}^{MD} |c_k c_{k'} v_{ik} v_{ik'}| \delta(f - f_{kk'}).
\end{equation}

Within TB I, an estimation of the transfer rate can be obtained~\cite{Simserides:2014} as follows:
Let's suppose that initially we place the carrier at the first monomer. Then, $|A_{1}(0)|^2 = 1$, while all other $|A_{j}(0)|^2 = 0$, $j=2, \dots, N$. Hence, for a polymer consisting of $N$ monomers, a \textit{pure} mean transfer rate can be defined as
\begin{equation}\label{meantransferrateN(I)}
k = \frac{\langle |A_{N}(t)|^2 \rangle}{{t_{N}}_{mean}},
\end{equation}
where ${t_{N}}_{mean}$ is the first time $|A_{N}(t)|^2$ becomes equal to $\langle |A_{N}(t)|^2 \rangle$ i.e.
``the mean transfer time''.
An analogous definition can be given within TB II. Suppose that initially we place the carrier at a given base, then, e.g.
$|A_1(0)|^2 = 1$, while all other $|A_{j}(0)|^2 = 0$, $j=2, \dots, N$ and all $|B_{j}(0)|^2 = 0$, $j=1, \dots, N$.
Then, if we are interested in $\langle |A_N(t)|^2 \rangle$ or $\langle |B_N(t)|^2 \rangle$, we can define
\begin{equation}\label{meantransferrateN(II)}
k = \frac{\langle |A_{N}(t)|^2 \rangle}{{t_{N}}_{mean}} \quad \textrm{or} \quad \frac{\langle |B_{N}(t)|^2 \rangle}{{t_{N}}_{mean}}
\end{equation}
where ``the mean transfer time'' ${t_{N}}_{mean}$ is
the first time $|A_{N}(t)|^2$ becomes equal to $\langle |A_{N}(t)|^2 \rangle$ or
the first time $|B_{N}(t)|^2$ becomes equal to $\langle |B_{N}(t)|^2 \rangle$, respectively.
It is possible to give a more general definition either for TB I or TB II.
From an initial base pair in TB I or base in TB II, $i$, to a final base pair in TB I or base in TB II, $j$,
\begin{equation}\label{meantransferrateN}
k_{ij} = \frac{\langle |C_{j}(t)|^2 \rangle}{{t_{j}}_{mean}},
\end{equation}
where $|C_i(0)|^2 = 1$ and all other $|C_{\ell \neq i}(0)|^2 = 0$. \\

\appendix{\noindent \textbf{Appendix B}}\label{BB}
For TB (I), the matrix $\textrm{A}$ is a symmetric tridiagonal matrix, i.e.,
\begin{equation}\label{Abp}
\textrm{A} = \left[
\begin{array}{ccccccc}
E^{bp(1)}   & t^{bp(1;2)} & 0           & \cdots &     0          &      0        & 0             \\
t^{bp(2;1)} & E^{bp(2)}   & t^{bp(2;3)} & \cdots &     0          &      0        & 0             \\
\vdots      & \vdots      & \vdots      & \vdots & \vdots         & \vdots        & \vdots        \\
0           & 0           & 0           & \cdots & t^{bp(N-1;N-2)}& E^{bp(N-1)}   & t^{bp(N-1;N)} \\
0           & 0           & 0           & \cdots &              0 & t^{bp(N;N-1)} & E^{bp(N)}   \end{array} \right].
\end{equation}
For TB II, the matrix $\textrm{A}$ has a different form, i.e.,
\begin{equation}\label{Abdimers}
\textrm{A} = \left[
\begin{array}{cccc}
E^{b(1)}   & t^{b(1,2)} & t^{b(1,3)} & t^{b(1,4)} \\
t^{b(2,1)} & E^{b(2)}   & t^{b(2,3)} & t^{b(2,4)} \\
t^{b(3,1)} & t^{b(3,2)} & E^{b(3)}   & t^{b(3,4)} \\
t^{b(4,1)} & t^{b(4,2)} & t^{b(4,3)} & E^{b(4)} \end{array} \right] \textrm{ (for dimers)}
\end{equation}
\begin{equation}\label{Abtrimers}
\textrm{A} = \left[
\begin{array}{cccccc}
E^{b(1)}   & t^{b(1,2)} & t^{b(1,3)} & t^{b(1,4)} & 0          & 0          \\
t^{b(2,1)} & E^{b(2)}   & t^{b(2,3)} & t^{b(2,4)} & 0          & 0          \\
t^{b(3,1)} & t^{b(3,2)} & E^{b(3)}   & t^{b(3,4)} & t^{b(3,5)} & t^{b(3,6)} \\
t^{b(4,1)} & t^{b(4,2)} & t^{b(4,3)} & E^{b(4)}   & t^{b(4,5)} & t^{b(4,6)} \\
0          & 0          & t^{b(5,3)} & t^{b(5,4)} & E^{b(5)}   & t^{b(5,6)} \\
0          & 0          & t^{b(6,3)} & t^{b(6,4)} & t^{b(6,5)} & E^{b(6)}
\end{array} \right] \textrm{ (for trimers)}
\end{equation}
\begin{equation}\label{Abtetramers}
\textrm{A} = \left[
\begin{array}{cccccccc}
E^{b(1)}   & t^{b(1,2)} & t^{b(1,3)} & t^{b(1,4)} & 0          & 0          & 0          & 0          \\
t^{b(2,1)} & E^{b(2)}   & t^{b(2,3)} & t^{b(2,4)} & 0          & 0          & 0          & 0          \\
t^{b(3,1)} & t^{b(3,2)} & E^{b(3)}   & t^{b(3,4)} & t^{b(3,5)} & t^{b(3,6)} & 0          & 0          \\
t^{b(4,1)} & t^{b(4,2)} & t^{b(4,3)} & E^{b(4)}   & t^{b(4,5)} & t^{b(4,6)} & 0          & 0          \\
0          & 0          & t^{b(5,3)} & t^{b(5,4)} & E^{b(5)}   & t^{b(5,6)} & t^{b(5,7)} & t^{b(5,8)} \\
0          & 0          & t^{b(6,3)} & t^{b(6,4)} & t^{b(6,5)} & E^{b(6)}   & t^{b(6,7)} & t^{b(6,8)} \\
0          & 0          & 0          & 0          & t^{b(7,5)} & t^{b(7,6)} & E^{b(7)}   & t^{b(7,8)} \\
0          & 0          & 0          & 0          & t^{b(8,5)} & t^{b(8,6)} & t^{b(8,7)} & E^{b(8)}
\end{array} \right] \textrm{ (for tetramers)}
\end{equation}
and so on. \\

\appendix{\noindent \textbf{Appendix C}}\label{CC}

We present the Fourier Analysis Figures mentioned in the main text.

\begin{figure} [h!]
\centering
\includegraphics[width=7.5cm]{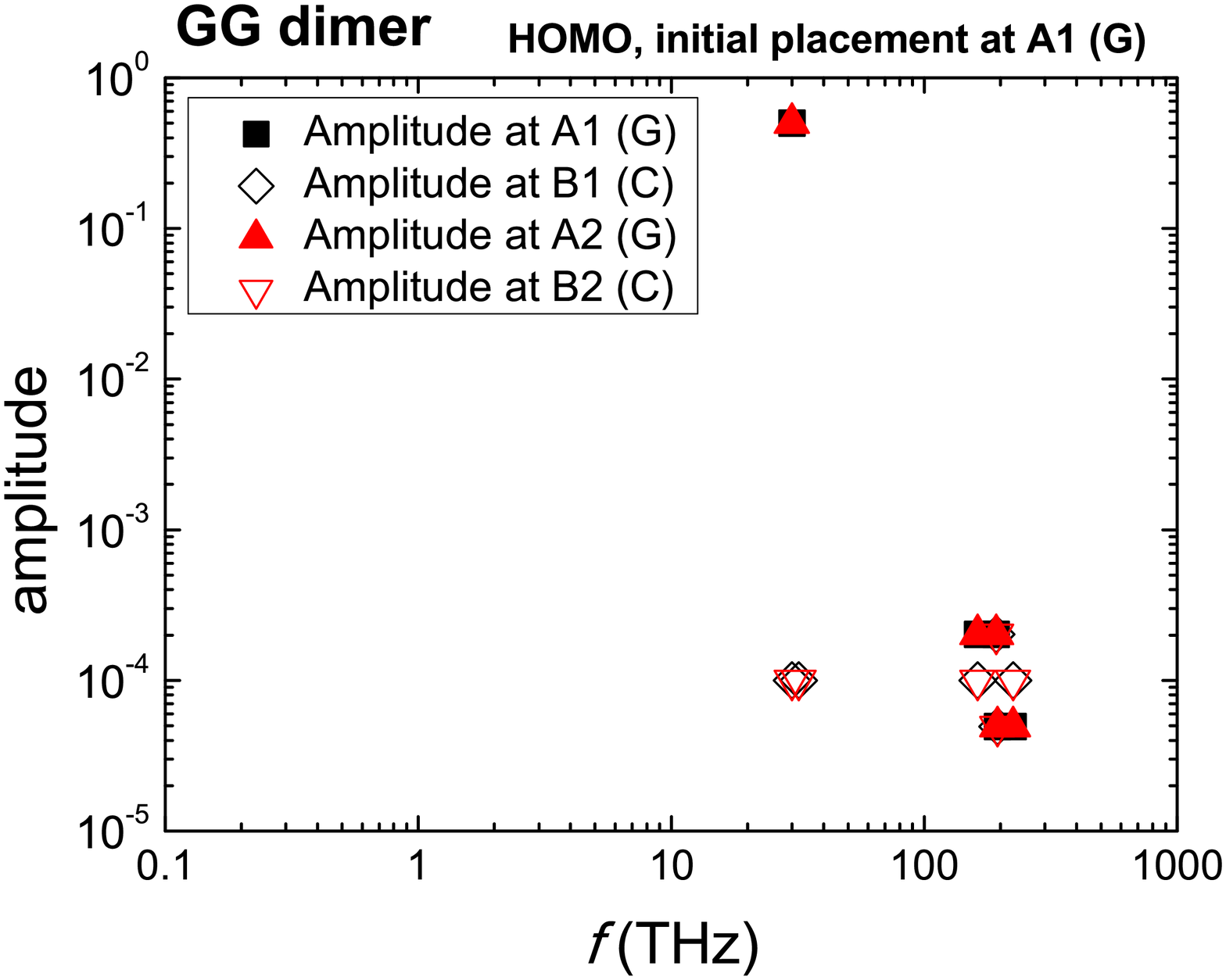}
\includegraphics[width=7.5cm]{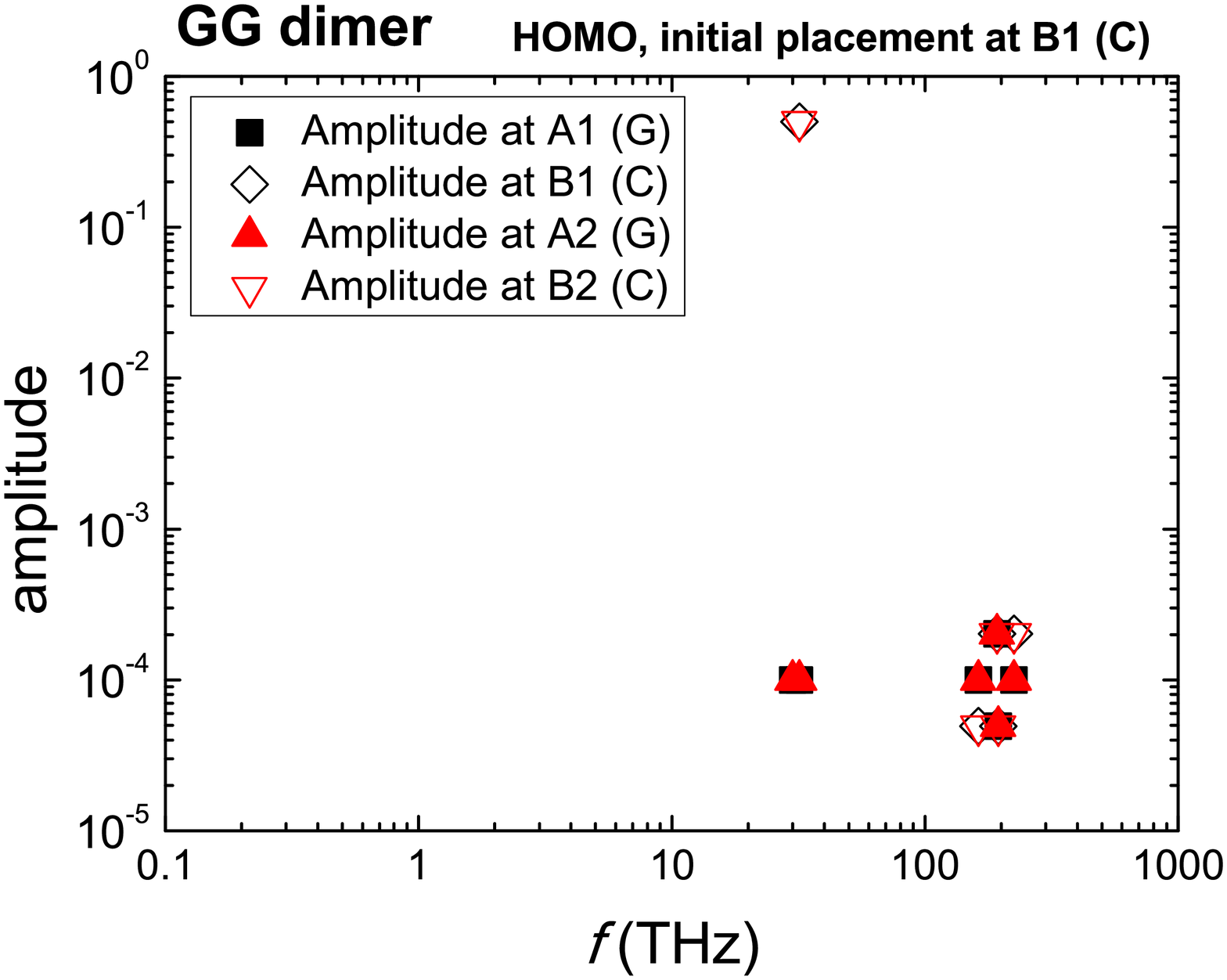}
\includegraphics[width=7.5cm]{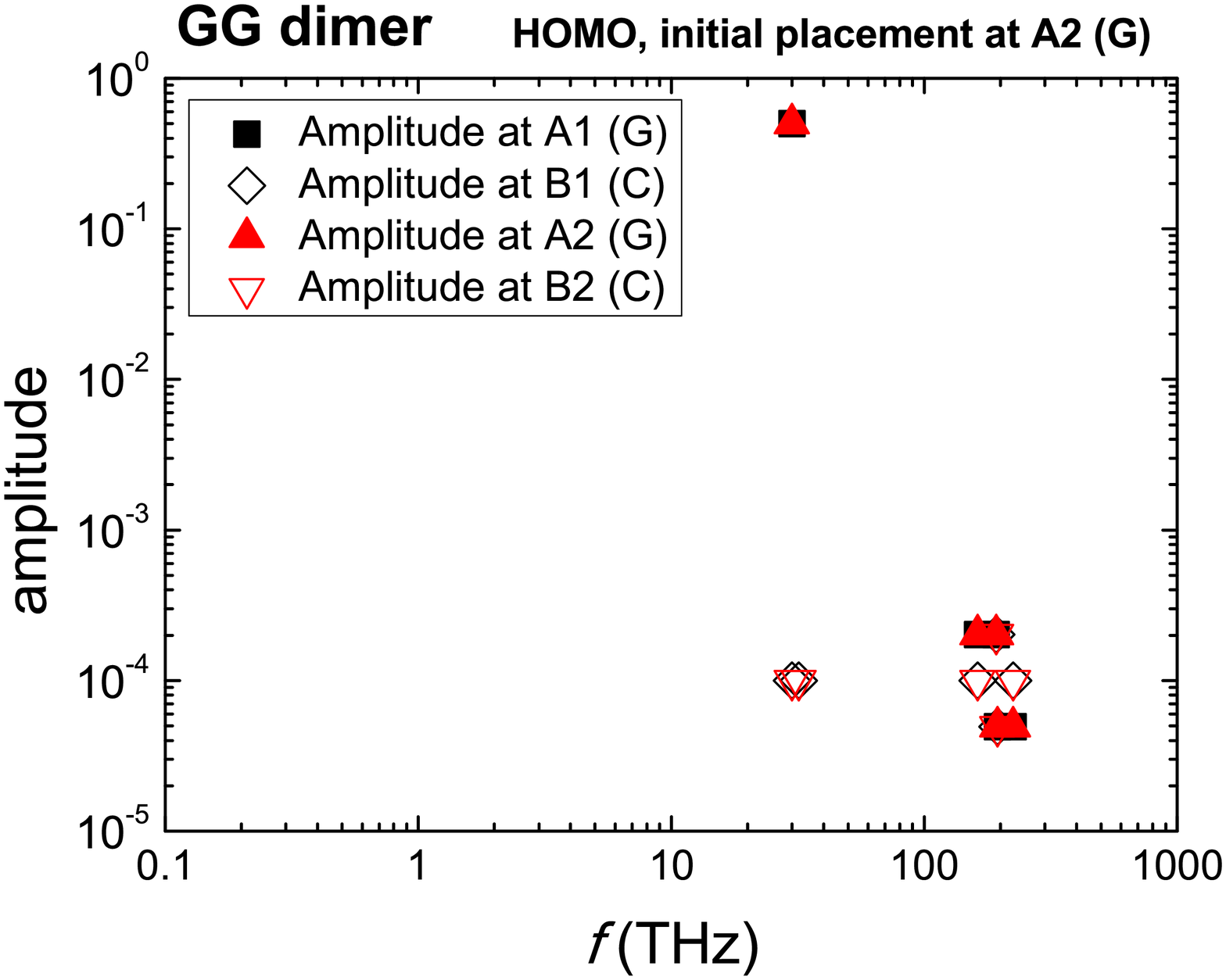}
\includegraphics[width=7.5cm]{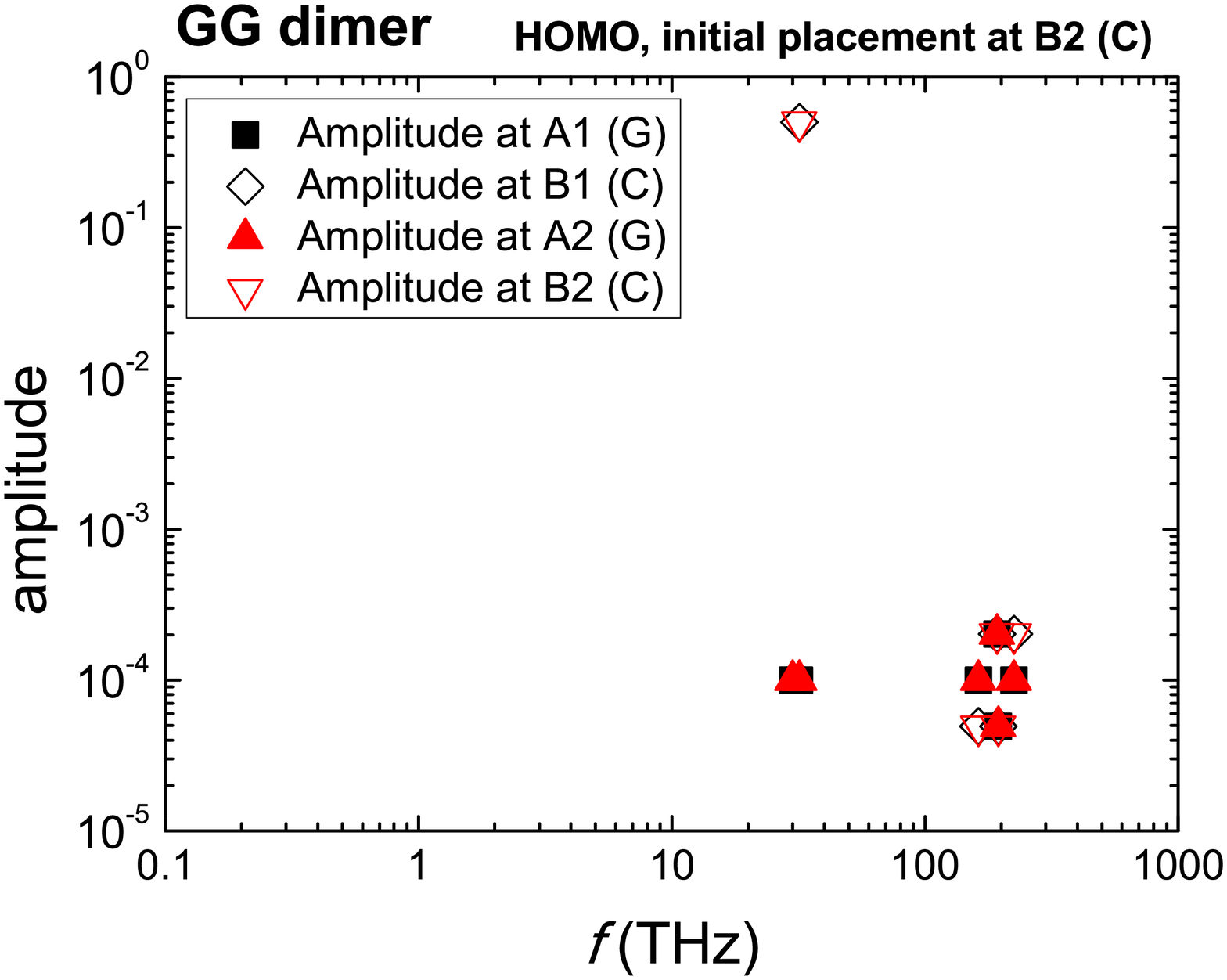}
\caption{Fourier analysis, within TB approach II and HKS parametrization~\cite{HKS:2010-2011}, of the GG dimer. A hole is placed initially at a base and we depict the frequency spectrum at all bases, A1(G), B1(C), A2(G), B2(C).}
\label{fig:FourierGG}
\end{figure}

\begin{figure} [h!]
\centering
\includegraphics[width=7.5cm]{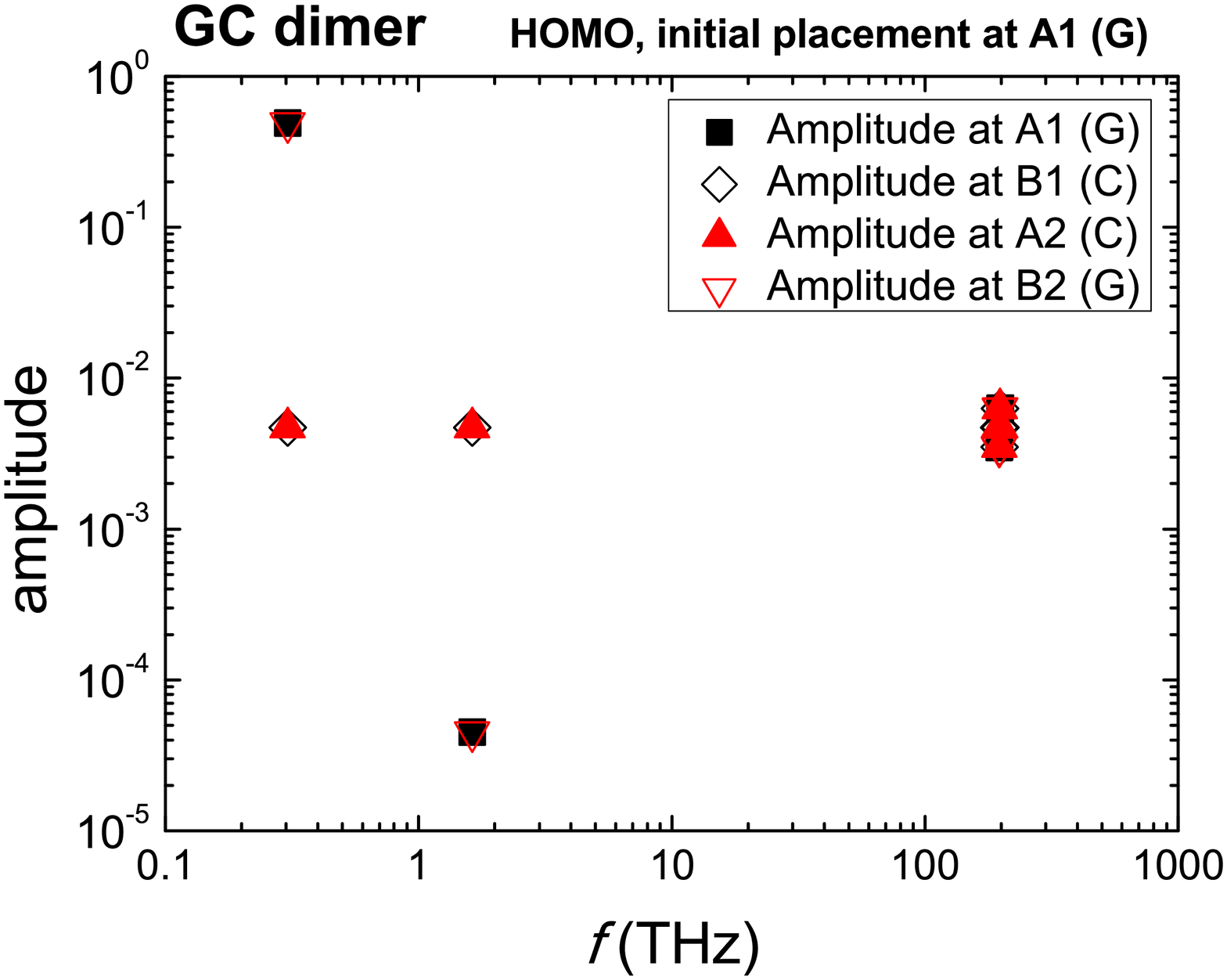}
\includegraphics[width=7.5cm]{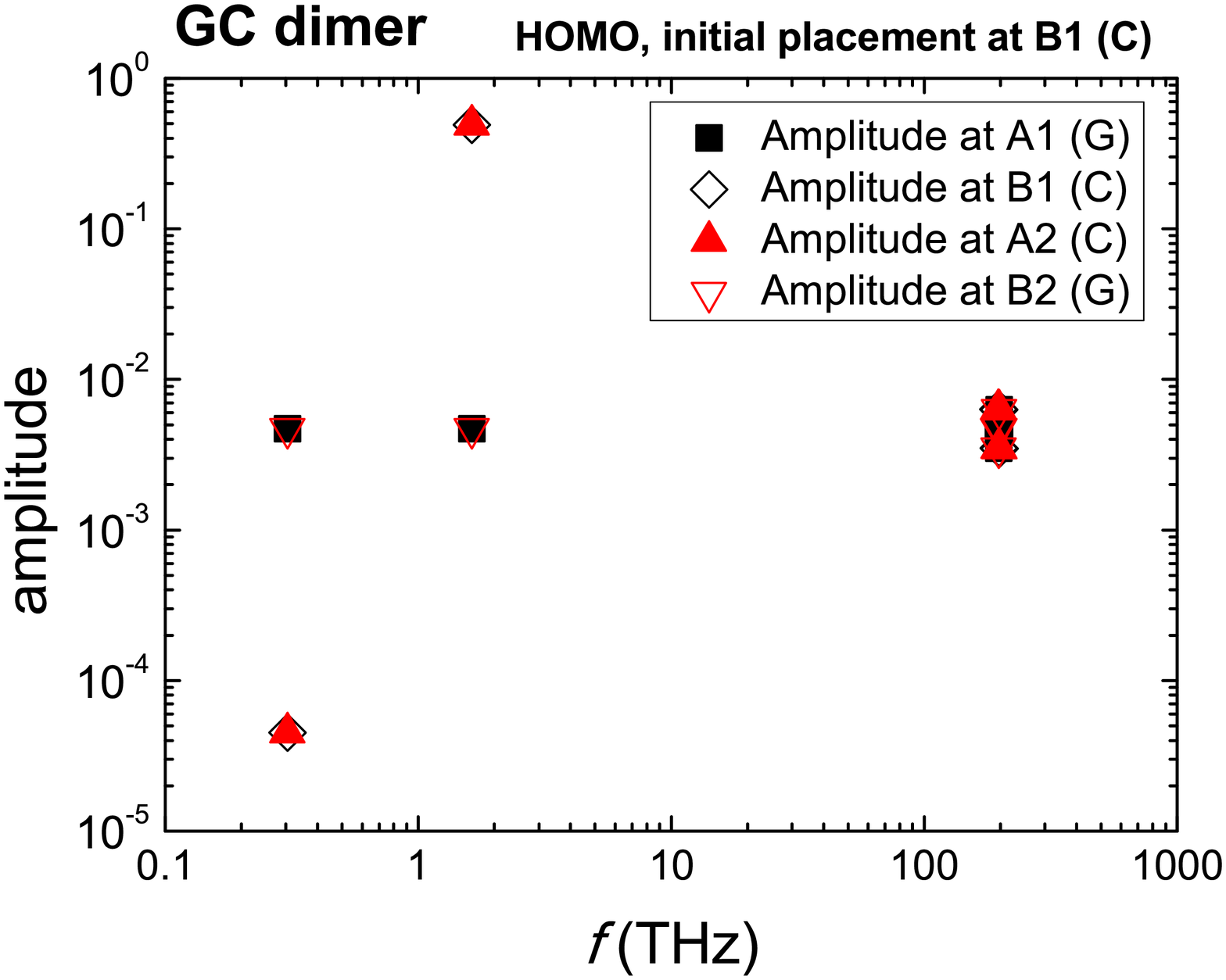}
\includegraphics[width=7.5cm]{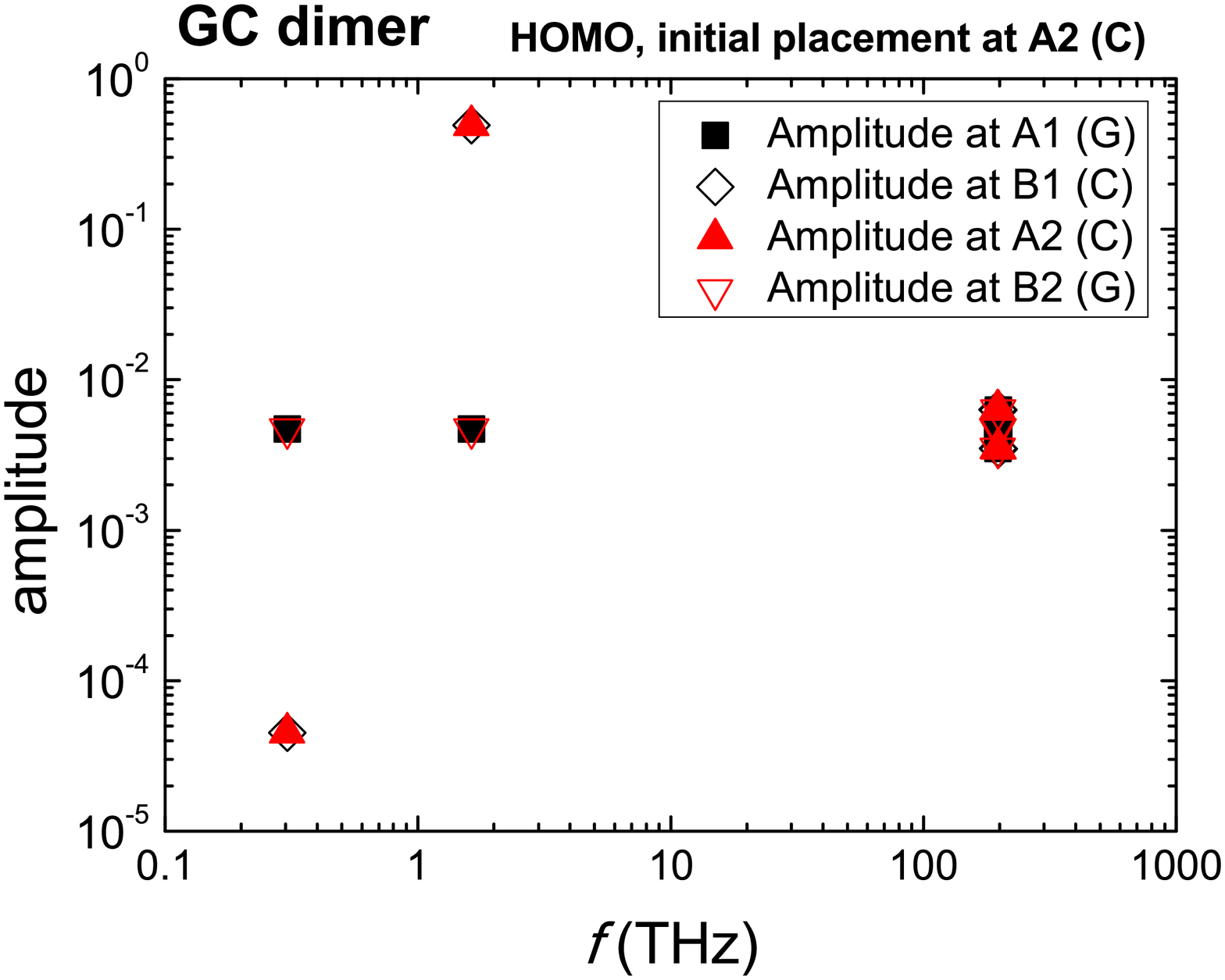}
\includegraphics[width=7.5cm]{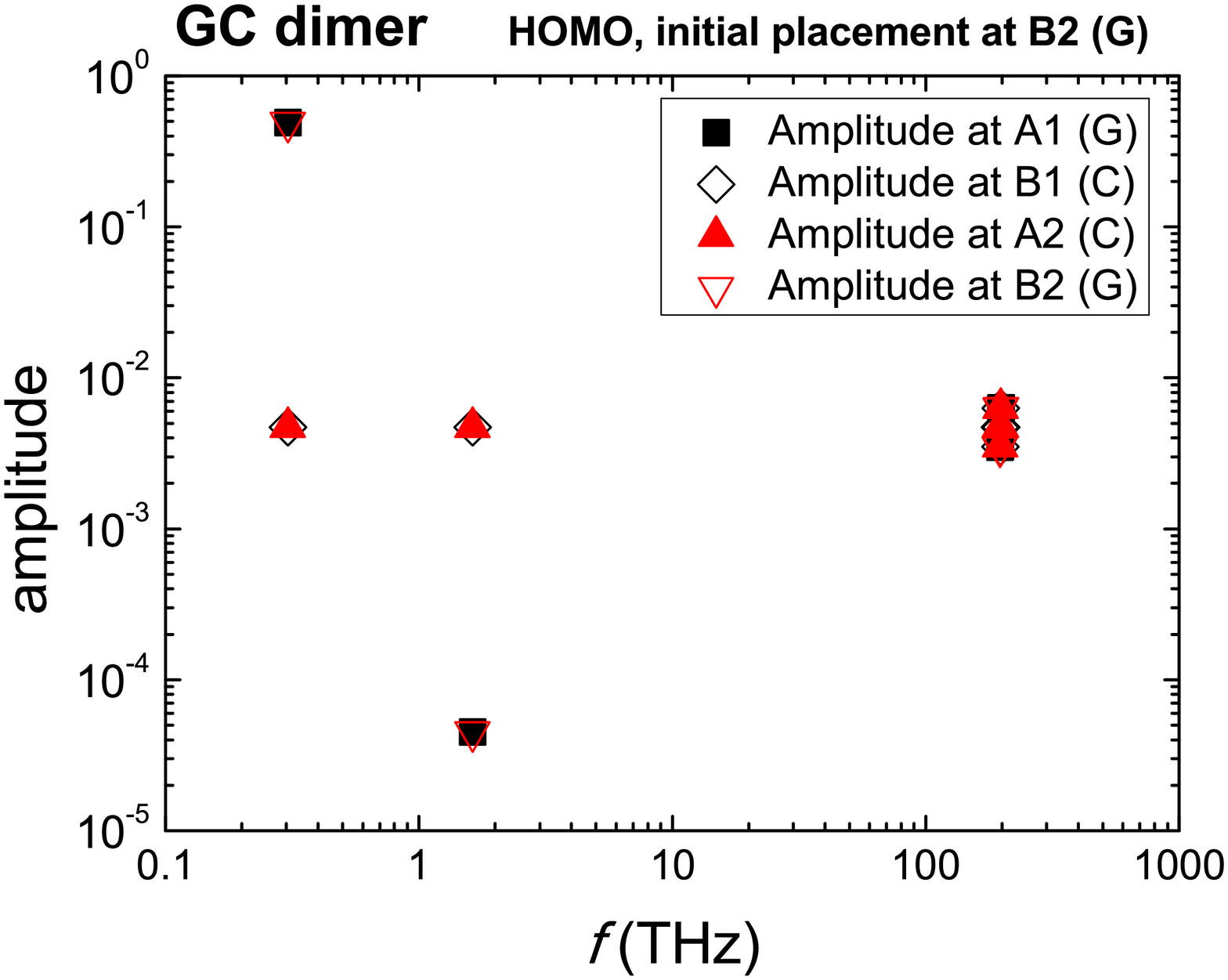}
\caption{Fourier Analysis, within TB approach II and HKS parametrization~\cite{HKS:2010-2011}, of the GC dimer. A hole is placed initially at a base and we depict the frequency spectrum at all bases, A1(G), B1(C), A2(C), B2(G).}
\label{fig:FourierGC}
\end{figure}

\begin{figure} [h!]
\centering
\includegraphics[width=7.5cm]{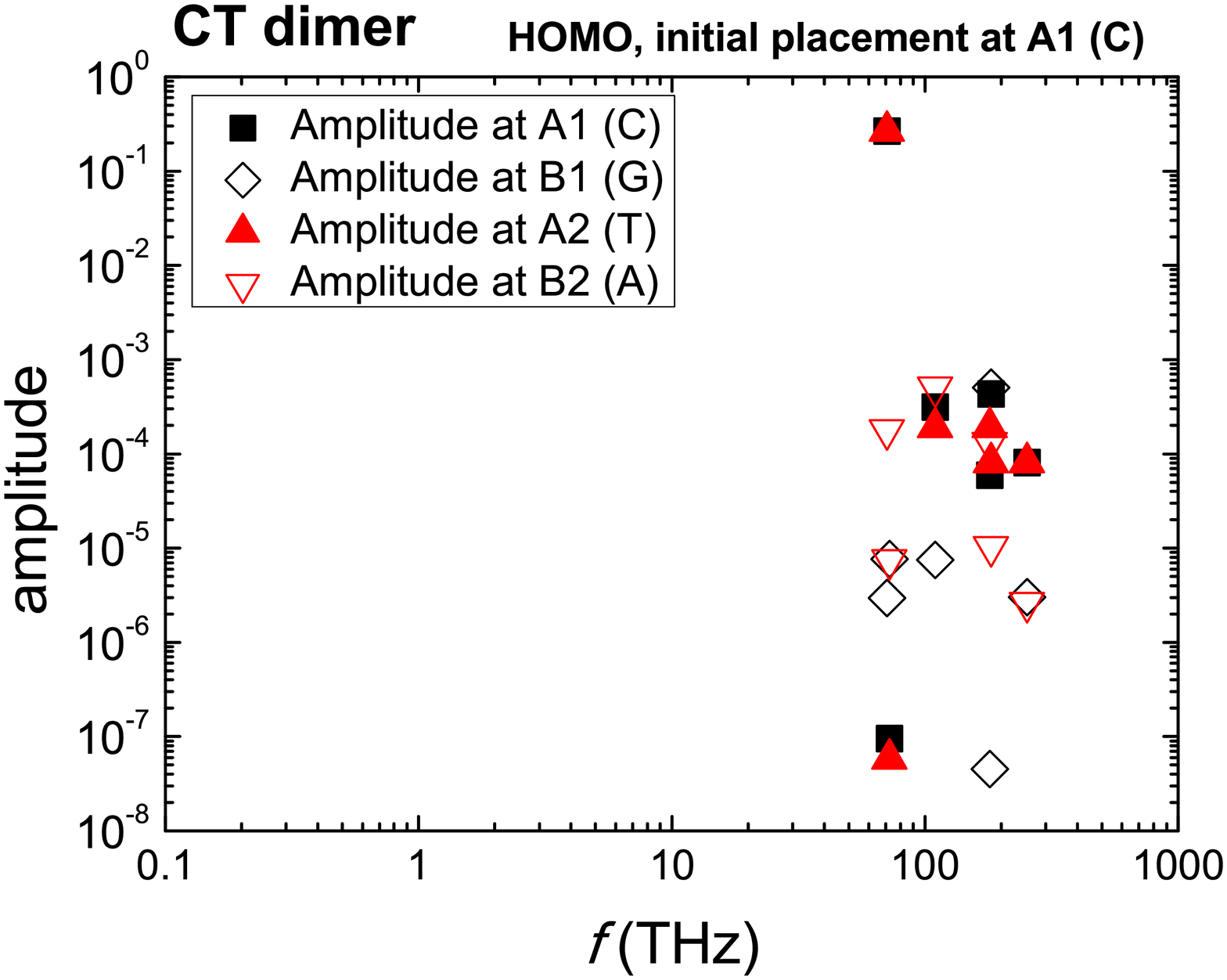}
\includegraphics[width=7.5cm]{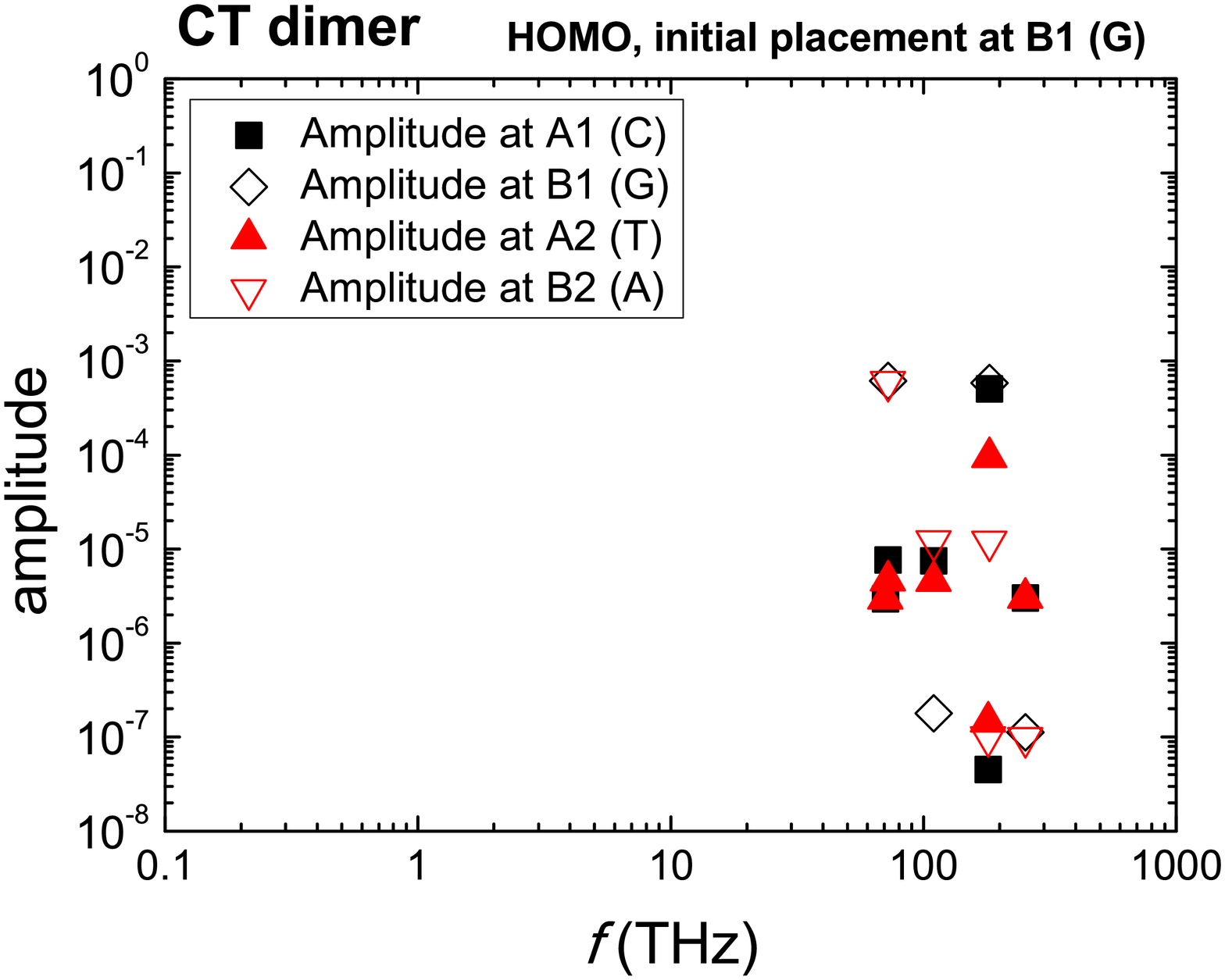}
\includegraphics[width=7.5cm]{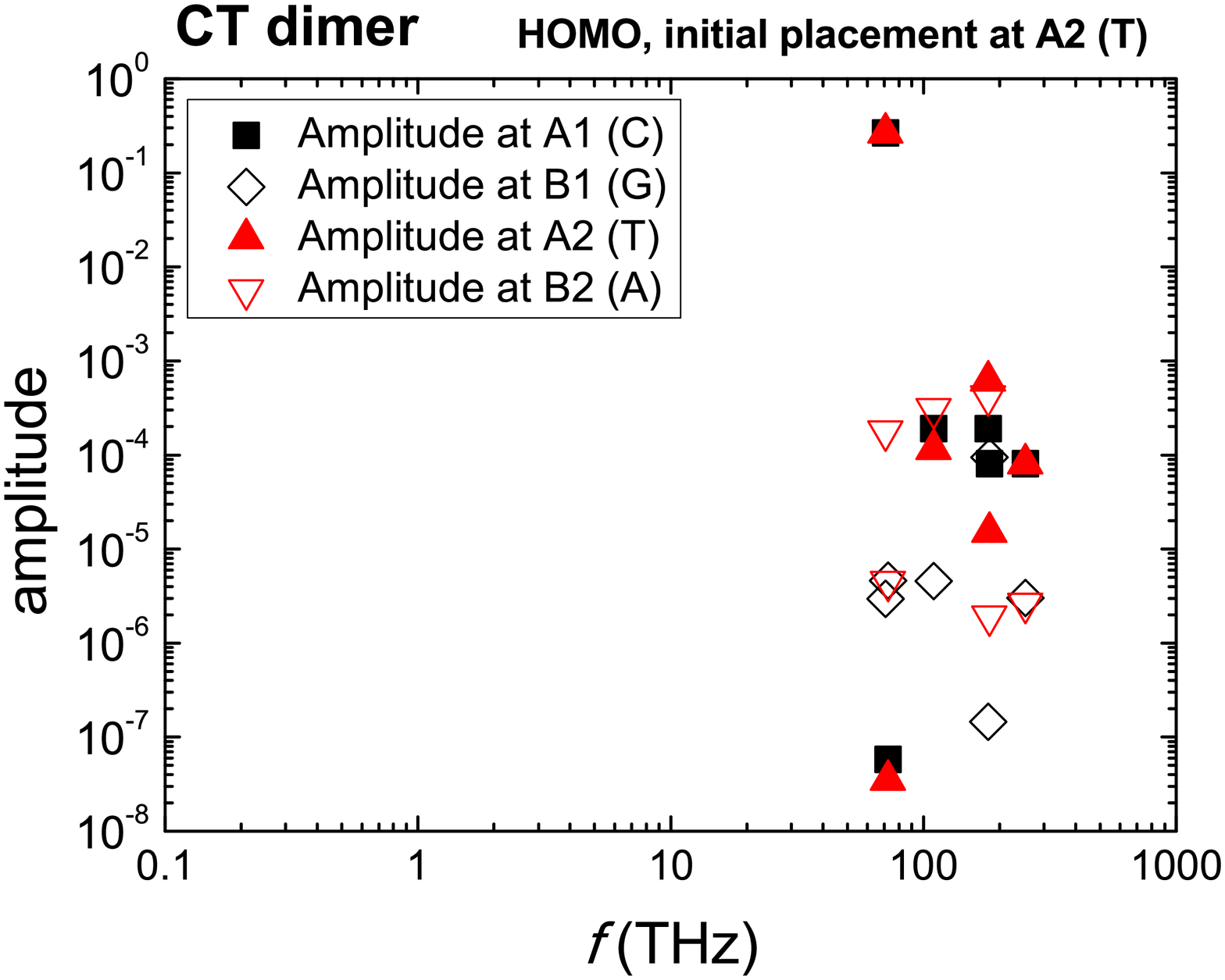}
\includegraphics[width=7.5cm]{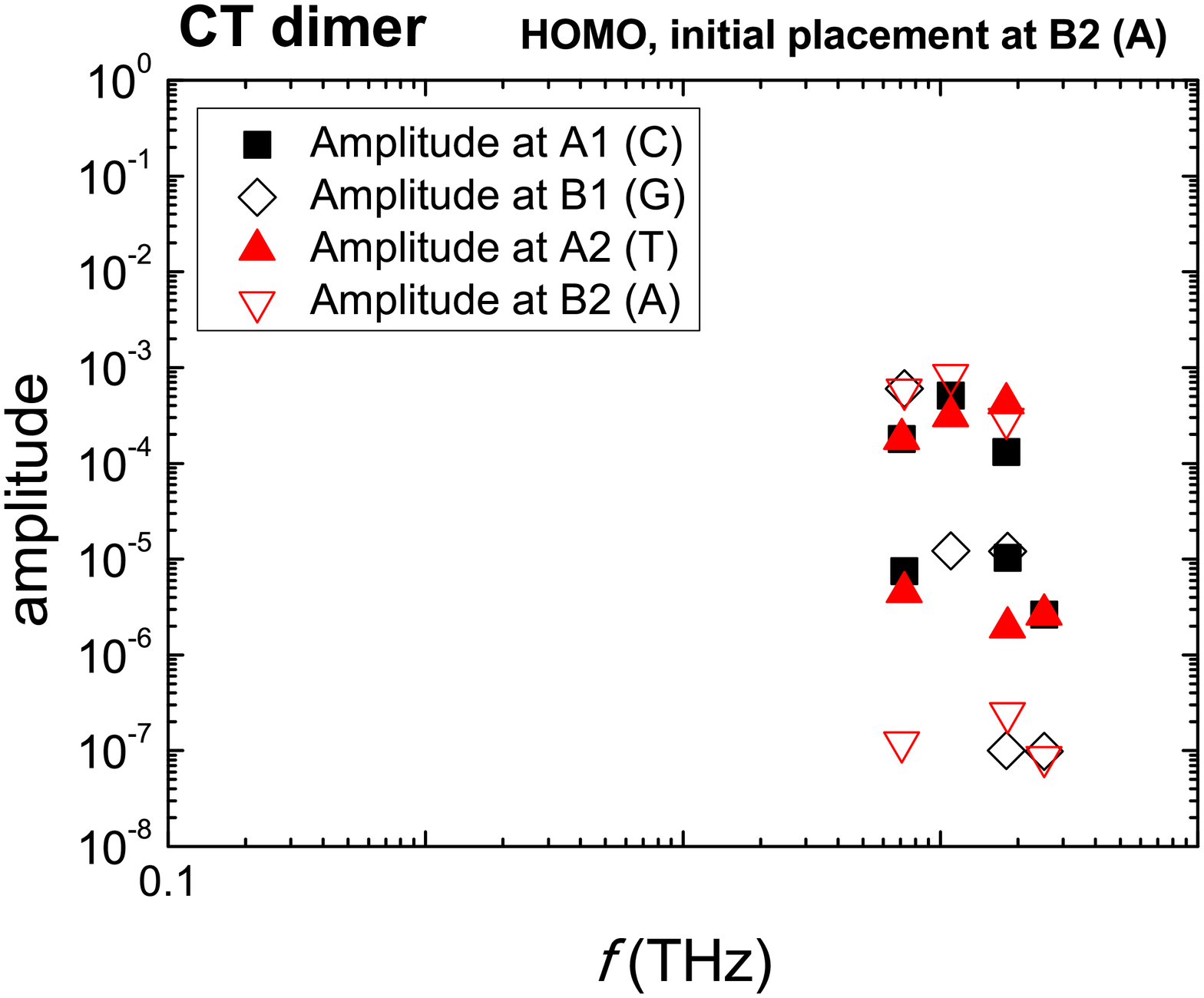}
\caption{Fourier Analysis, within TB approach II and HKS parametrization~\cite{HKS:2010-2011}, of the CT dimer. A hole is placed initially at a base and we depict the frequency spectrum at all bases, A1(C), B1(G), A2(T), B2(A).}
\label{fig:FourierCT}
\end{figure}

\begin{figure} [h!]
\centering
\includegraphics[width=7.5cm]{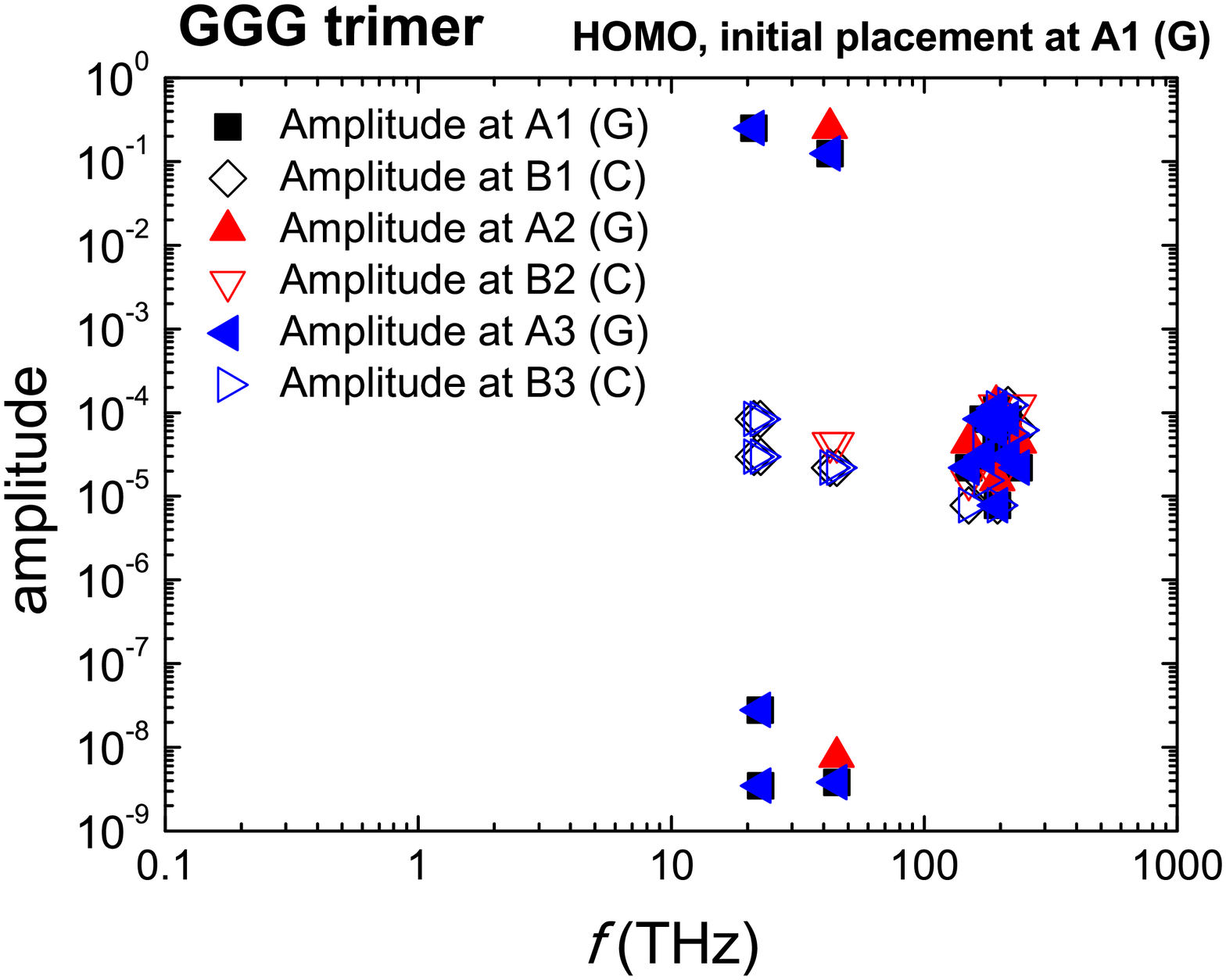}
\includegraphics[width=7.5cm]{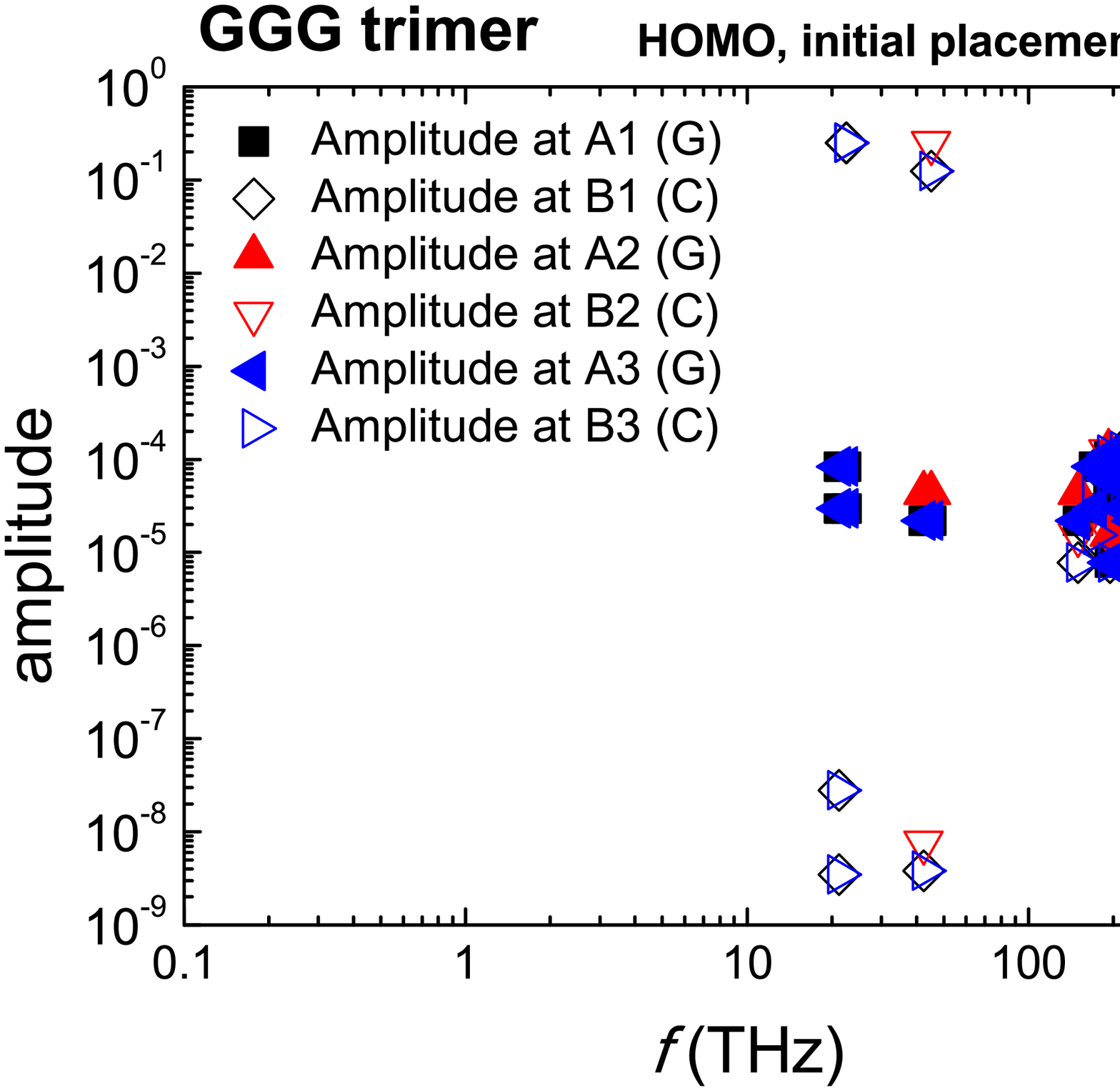}
\includegraphics[width=7.5cm]{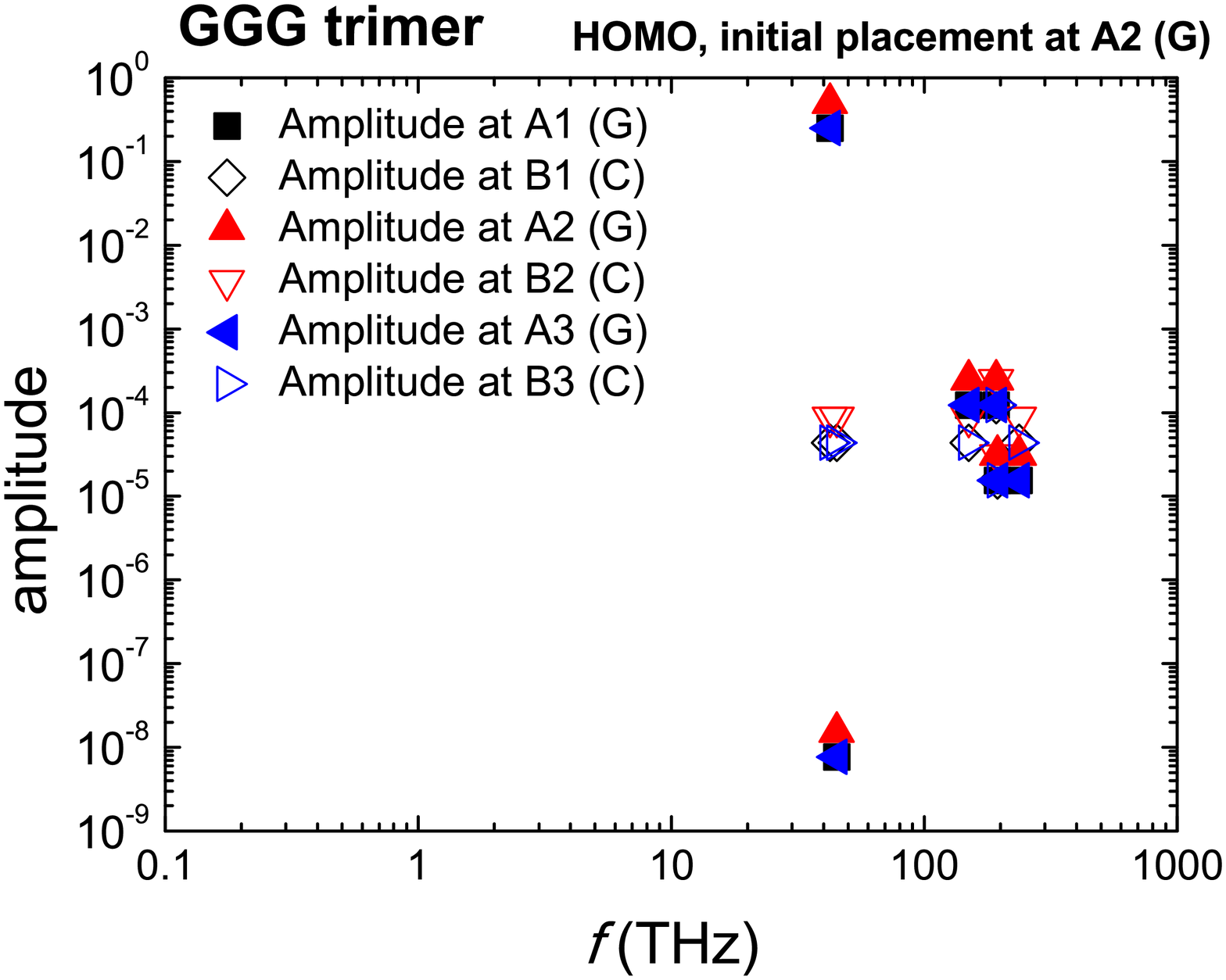}
\includegraphics[width=7.5cm]{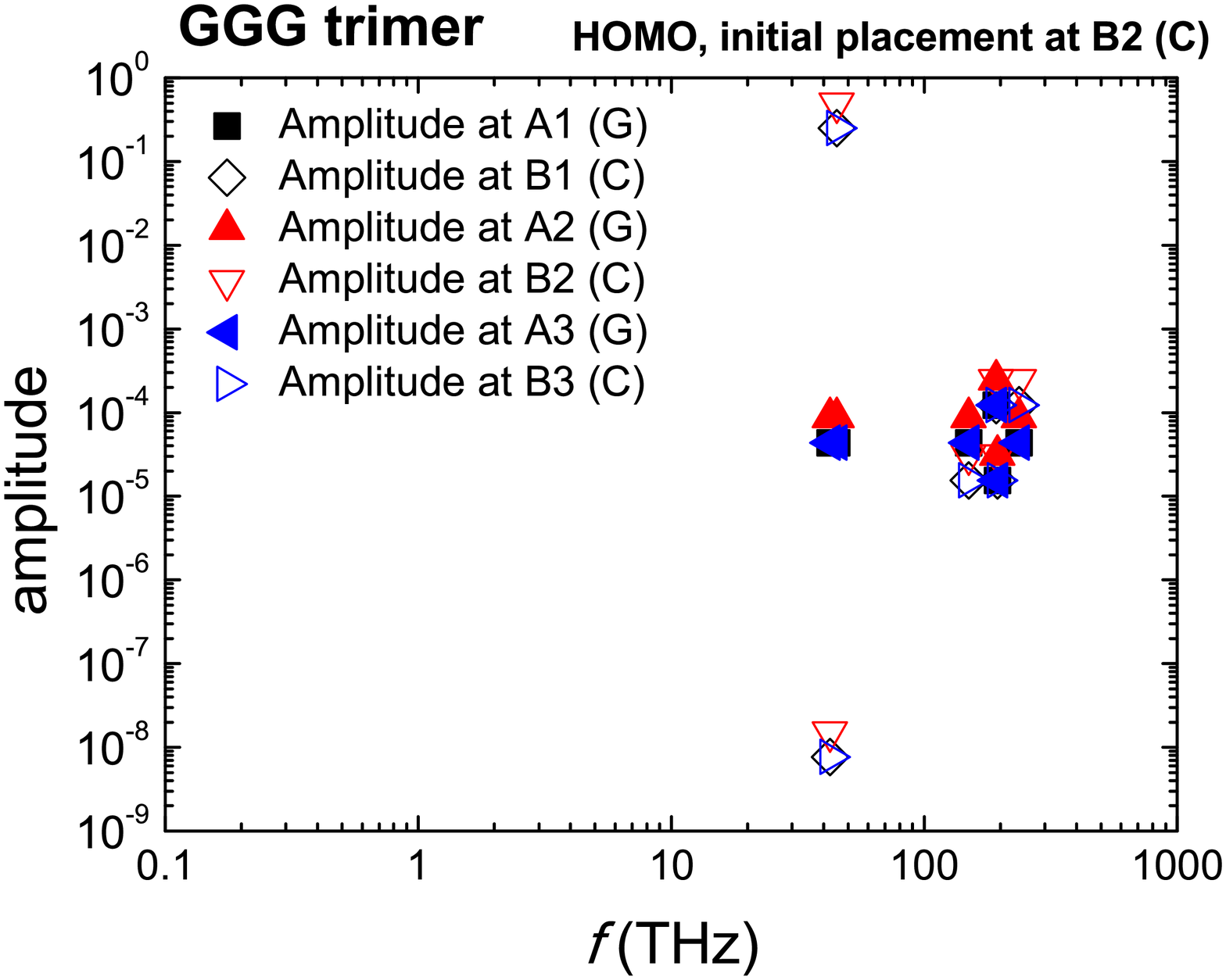}
\includegraphics[width=7.5cm]{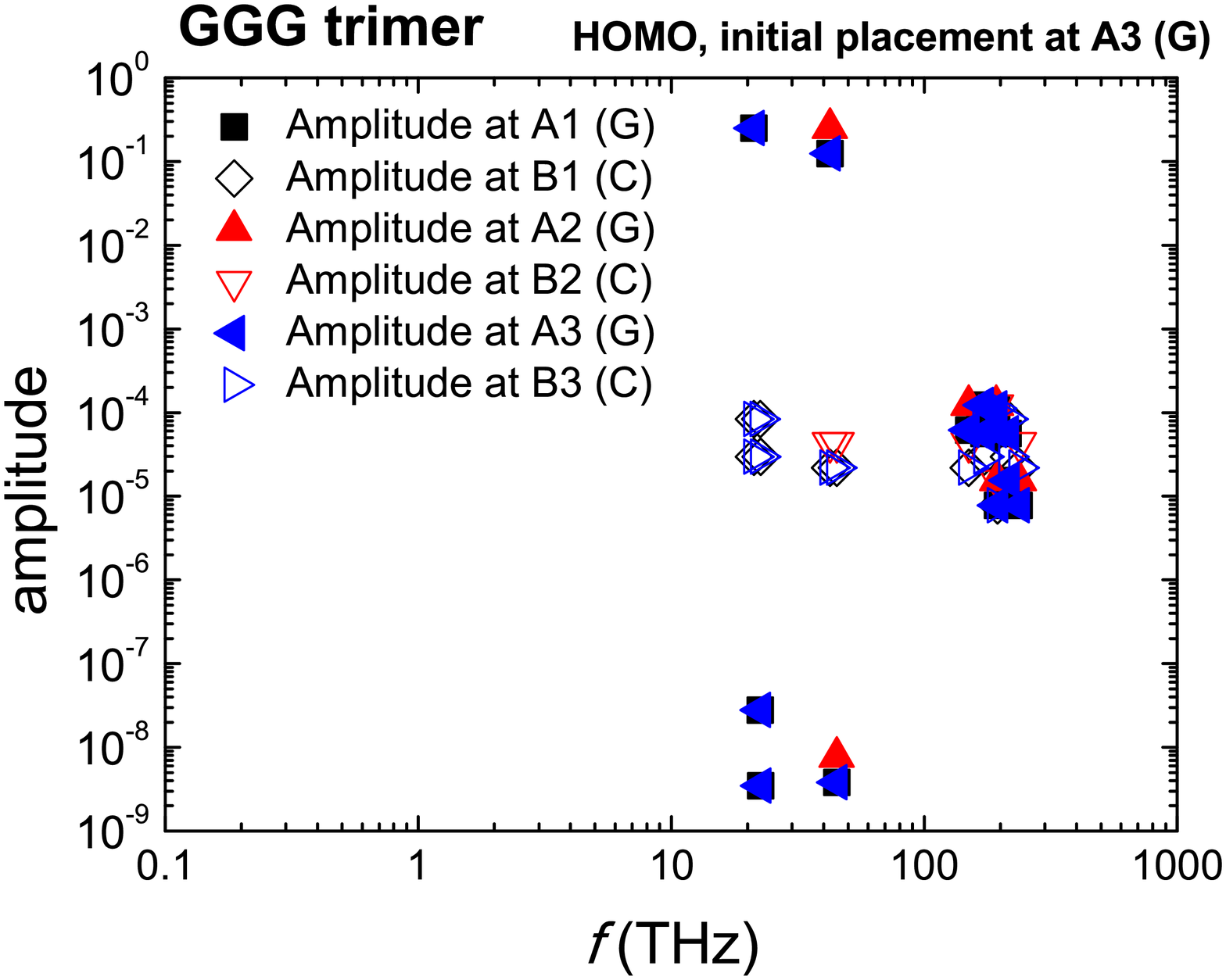}
\includegraphics[width=7.5cm]{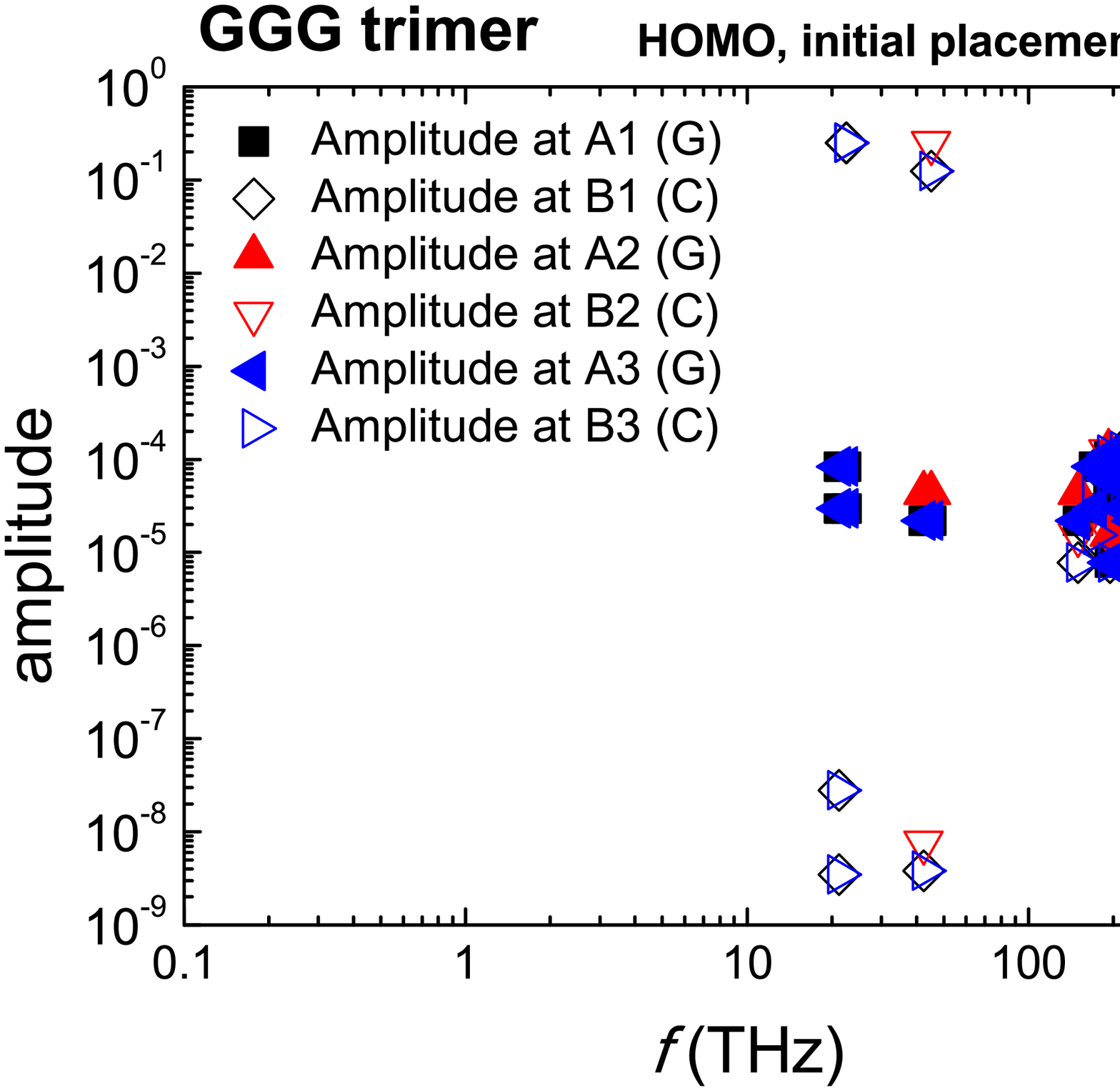}
\caption{Fourier analysis, within TB approach II and HKS parametrization~\cite{HKS:2010-2011}, of hole transfer in the GGG trimer. A hole is placed initially at a base and we depict the frequency spectrum at all bases, A1(G), B1(C), A2(G), B2(C), A3(G), B3(C).}
\label{fig:FourierGGG}
\end{figure}

\begin{figure} [h!]
\centering
\includegraphics[width=7.5cm]{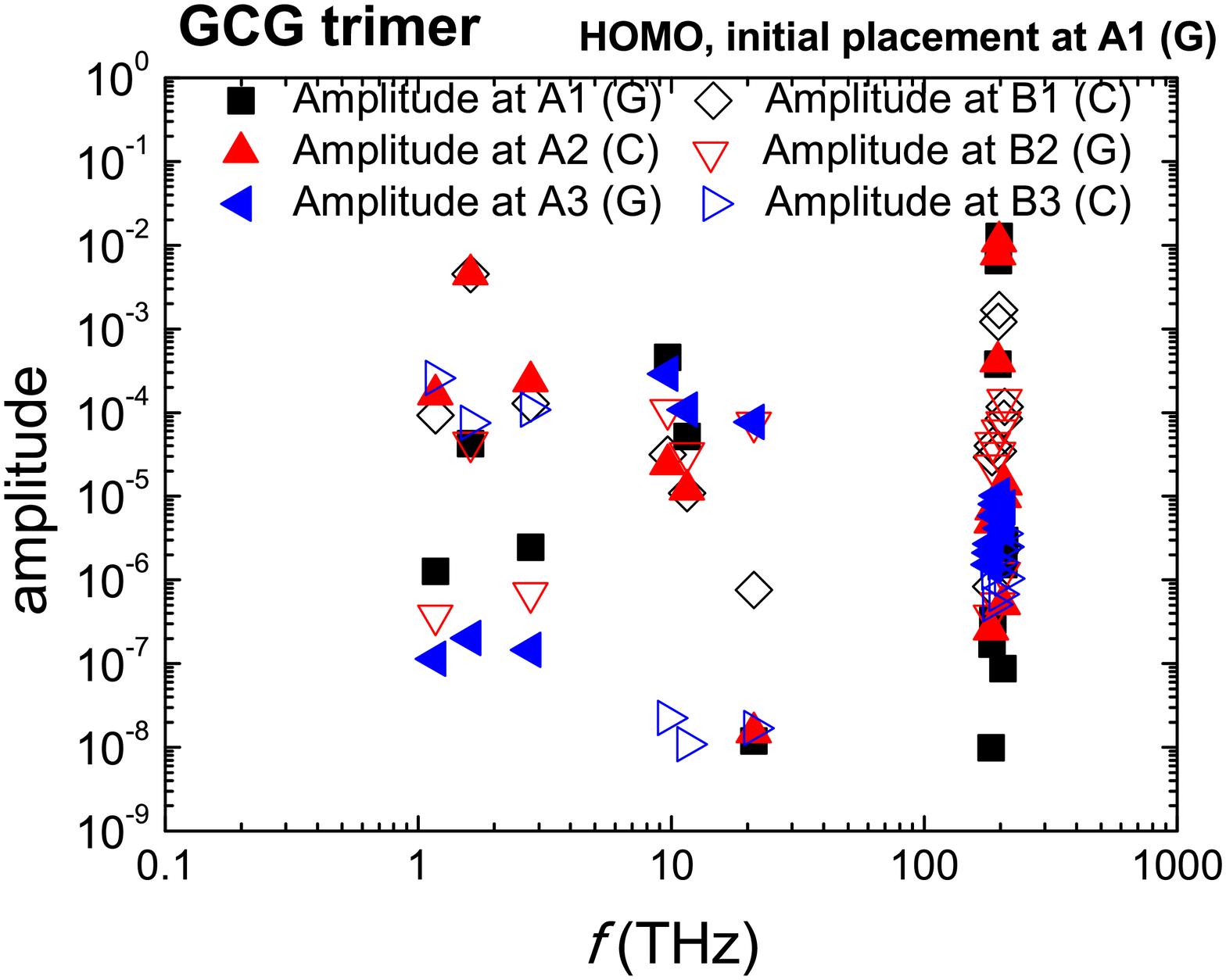}
\includegraphics[width=7.5cm]{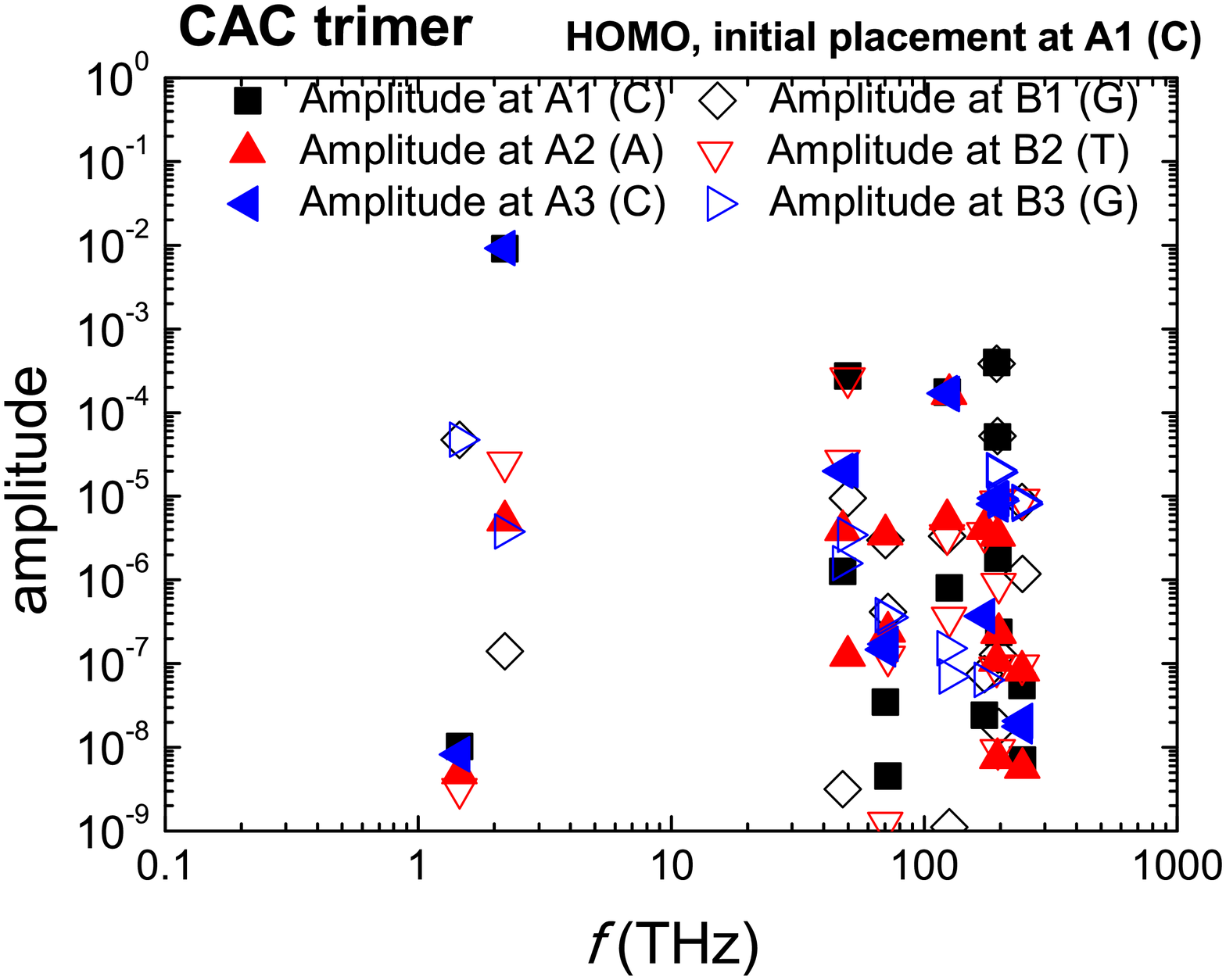}
\includegraphics[width=7.5cm]{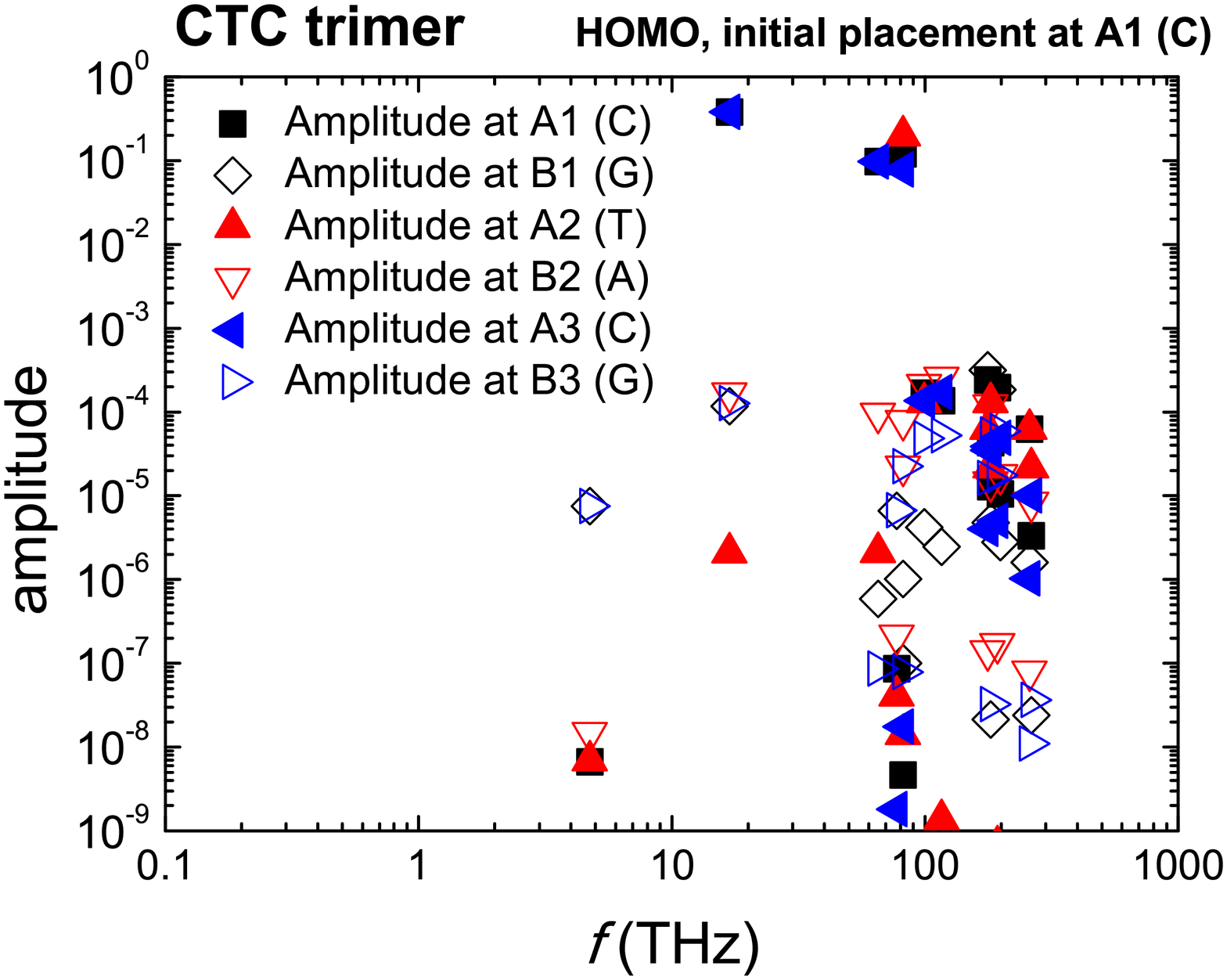}
\caption{Fourier analysis, within TB approach II and HKS parametrization~\cite{HKS:2010-2011}, of the GCG, CAC, CTC trimers for initial placement of a hole at base A1.}
\label{fig:FourierGCGandCACandCTC}
\end{figure}

\end{document}